\documentclass[american,twosided,12pt]{article}
\usepackage{etex}
\usepackage{setspace}
\usepackage{epsfig}
\usepackage{amsmath}
\usepackage{amsfonts}
\usepackage{hyperref}
\usepackage{amssymb}
\usepackage{pgffor}
\usepackage[margin=1.25in]{geometry}
\usepackage{babel}
\usepackage{subfig}
\usepackage{booktabs}
\usepackage{rotating}
\usepackage{diagbox}
\usepackage{tabularx}
\usepackage[authoryear]{natbib}
\usepackage{mathrsfs}
\usepackage{comment}
\usepackage{bm}
\usepackage{tikz}
\usepackage{pgfplots}
\usetikzlibrary{positioning}
\usetikzlibrary{decorations.pathmorphing}
\usetikzlibrary{intersections, pgfplots.fillbetween}
\usetikzlibrary{matrix}
\usepackage{amsmath,fouriernc}
\usepackage{multirow}
\usetikzlibrary{shapes.geometric}
\usepackage{rotating}
\usepackage{pdflscape}
\usepackage{tabulary}
\usepackage{pdfpages}

\usepackage{graphicx,kantlipsum,setspace}
\usepackage{caption}
\usepackage[labelfont=bf]{caption}

\newcommand*{\field}[1]{\mathbb{#1}}

\definecolor{mondrianRed}{rgb}{0.8666,0.1334,0}
\definecolor{mondrianBlue}{rgb}{0.133333, 0.313725, 0.584314}
\definecolor{mondrianYellow}{rgb}{0.980392, 0.788235, 0.00392157}
\definecolor{mondrianGrey}{rgb}{0.756863, 0.784314, 0.788235}
\definecolor{mondrianOrange}{rgb}{0.923529, 0.396078, 0.00196078}
\definecolor{mondrianGreen}{rgb}{0.556863, 0.54902, 0.294118}
\definecolor{mondrianCyan}{rgb}{0.345098, 0.431373, 0.439216}
\definecolor{mondrianWhite}{rgb}{0.976471, 0.976471, 0.976471}
\definecolor{mondrianPurple}{rgb}{0.107843, 0.158824, 0.292157}

\definecolor{goldenyellow}{rgb}{1.0, 0.87, 0.0}
\definecolor{silver}{rgb}{0.75, 0.75, 0.75}

\definecolor {processblue}{cmyk}{0.96,0,0,0}
\usetikzlibrary{arrows}
\input{glyphtounicode}
\DeclareMathAlphabet{\mathpzc}{OT1}{pzc}{m}{it}

\makeatletter
\renewcommand{\section}{\@startsection{section}{1}{0mm}{-1.1\baselineskip}{0.4\baselineskip}{\normalfont\large\centering}}
\renewcommand{\subsection}{\@startsection{subsection}{2}{0mm}{-0.1\baselineskip}{0.4\baselineskip}{\normalfont\bf\flushleft}}
\renewcommand{\@seccntformat}[1]{\csname the#1\endcsname \hspace{+0mm}\large{.}\hspace{+1.9mm}}
\renewcommand{\@seccntformat}[2]{\csname the#1\endcsname \hspace{+0mm}\large{.}\hspace{+1.9mm}}
\makeatother

\newtheorem{definition}{Definition}
\newtheorem{example}{Example}

\newtheorem{proposition}{Proposition}
\newtheorem{remark}{Remark}
\newtheorem{result}{Result}

\newenvironment{support}[1][Support]{\textbf{#1.} }{\,\,\,\,\rule{0.5em}{0.5em}}

\newlength{\extraspace}
\newlength{\extraspaces}
\newcounter{dummy}

\newcommand{\baa}{
\addtocounter{equation}{1} \setcounter{dummy}{\value{equation}}
\setcounter{equation}{0}
\renewcommand{\theequation}{\arabic{dummy}\alph{equation}}
\begin{eqnarray}
\addtolength{\abovedisplayskip}{\extraspaces}
\addtolength{\belowdisplayskip}{\extraspaces}
\addtolength{\abovedisplayshortskip}{\extraspace}
\addtolength{\belowdisplayshortskip}{\extraspace}}

\newcommand{\eaa}{
\end{eqnarray}
\setcounter{equation}{\value{dummy}}
\renewcommand{\theequation}{\arabic{equation}}}

\newcommand{\be}{\begin{equation}
\addtolength{\abovedisplayskip}{\extraspaces}
\addtolength{\belowdisplayskip}{\extraspaces}
\addtolength{\abovedisplayshortskip}{\extraspace}
\addtolength{\belowdisplayshortskip}{\extraspace}}
\newcommand{\ee}{\end{equation}}

\newcommand{\ba}{\begin{eqnarray}
\addtolength{\abovedisplayskip}{\extraspaces}
\addtolength{\belowdisplayskip}{\extraspaces}
\addtolength{\abovedisplayshortskip}{\extraspace}
\addtolength{\belowdisplayshortskip}{\extraspace}}
\newcommand{\ea}{\end{eqnarray}}

\newcommand{\bd}{\begin{displaymath}
\addtolength{\abovedisplayskip}{\extraspaces}
\addtolength{\belowdisplayskip}{\extraspaces}
\addtolength{\abovedisplayshortskip}{\extraspace}
\addtolength{\belowdisplayshortskip}{\extraspace}}
\newcommand{\ed}{\end{displaymath}}

\newcommand{\deel}[2]{{\textstyle{#1 \over #2}}}
\newcommand{\hf}{{\textstyle{1\over 2}}}

\def\inbar{\,\vrule height1.5ex width.4pt depth0pt}
\font\rms=cmr12 at 12pt
\def\ce{\relax\ifmmode\mathchoice
{\hbox{$\inbar\kern-.3em{\rm C}$}} {\hbox{$\inbar\kern-.3em{\rm
C}$}} {\lower.9pt\hbox{\rms $\inbar\kern-.3em{\rm C}$}}
{\lower1.2pt\hbox{\rms $\inbar\kern-.3em{\rm C}$}}
\else{$\inbar\kern-.3em{\rm C}$}\fi}
\font\cmss=cmss12 \font\cmsss=cmss12 at 12pt
\def\ze{\relax\ifmmode\mathchoice
{\hbox{\cmss Z\kern-.4em Z}}{\hbox{\cmss Z\kern-.4em Z}}
{\lower.9pt\hbox{\cmsss Z\kern-.4em Z}} {\lower1.2pt\hbox{\cmsss
Z\kern-.4em Z}}\else{\cmss Z\kern-.4em Z}\fi}

\newcommand{\refsection}[1]{
\vspace{1mm} \pagebreak[3] \addtocounter{section}{1}
\begin{center}
{\large #1}
\end{center}
\nopagebreak
\medskip
\nopagebreak}

\def\thebibliography#1{\refsection{\bf References}
\vspace*{-4mm}\list
 {\relax}{\itemsep=1pt \parsep=0pt
 \usecounter{enumiv}\leftmargin=3em\itemindent=-\leftmargin}%
 \def\newblock{\hskip .11em plus .33em minus .07em}
 \sloppy\clubpenalty4000\widowpenalty4000
 \sfcode`\.=1000\relax}

\newcommand{\startappendix}{
\renewcommand{\thesection}{\Alph{section}}
\setcounter{section}{0}
\renewcommand{\theequation}{\thesection.\arabic{equation}}}

\newcommand{\q}[1]{``#1''}

\begin{document}

\newcolumntype{L}[1]{>{\raggedright\arraybackslash}p{#1}}
\newcolumntype{C}[1]{>{\centering\arraybackslash}p{#1}}
\newcolumntype{R}[1]{>{\raggedleft\arraybackslash}p{#1}}

\setcounter{page}{0}
\thispagestyle{empty}

\begin{center}
{\Huge\bf $\mbox{\bf\em S}$ Equilibrium}\\[3mm]
{\Large\sc A Synthesis of (Behavioral) Game Theory}\\[10mm]
{\large Jacob K. Goeree and Bernardo Garc\'ia-Pola}\footnote{Goeree: AGORA Center for Market Design, UNSW, Sydney, Australia. Garc\'ia-Pola: Department of Economics, Universidad P\'ublica de Navarra, Pamplona, Spain. We gratefully acknowledge funding from the Australian Research Council (DP190103888 and DP220102893). We thank Brett Williams for useful comments. The ``S'' terminology was inspired by Reinhard Selten's work on the role of beliefs in refining Nash equilibria.}\\[5mm]
\today\\[10mm]
{\bf Abstract}
\vspace*{5mm}

\addtolength{\baselineskip}{-1.2mm}

\parbox{15cm}{\addtolength{\baselineskip}{-1.2mm}
$S$ equilibrium synthesizes a century of game-theoretic modeling. $S$-beliefs determine choices as in the refinement literature and level-$k$, without \mbox{anchoring} on Nash \mbox{equilibrium} or imposing ad hoc belief formation. $S$-choices allow for mistakes as in QRE, without imposing rational expectations. $S$ equilibrium is explicitly \mbox{set-valued} to avoid the common practice of selecting the best prediction from an implicitly \mbox{defined} set of unknown, and unaccounted for, size. $S$-equilibrium sets vary with a complexity parameter, offering a trade-off \mbox{between} accuracy and precision \mbox{unlike} in $M$ equilibrium. \mbox{Simple} \mbox{``areametrics''} \mbox{determine} the model's parameter and show that choice sets with a relative size of 5\% capture 58\% of the data. Goodness-of-fit tests applied to data from a broad array of experimental games confirm $S$ \mbox{equilibrium's} ability to predict behavior in and out of sample. In contrast, choice (belief) predictions of level-$k$ and QRE are rejected in most (all) games.}

\end{center}

\vfill

\addtocounter{footnote}{-1}

\newpage

\addtolength{\baselineskip}{0.2mm}

\section{Introduction}

Almost a century ago, \cite{vonNeumann1928} proposed the first solution concept for games. The ``minimax'' solution applies to two-player zero-sum games and entails strategies that minimize the other's maximum payoff. Roughly a quarter century later, \cite{nash1950, nash1951} introduced a new solution concept and proved existence for any finite game. Because of its broad applicability the Nash equilibrium became the predominant solution concept in game theory. Interestingly, when Nash tried to explain his work to von Neumann, the latter interrupted him after a few sentences and jumped to the as yet unstated conclusion and judged \q{That’s trivial, you know, that’s just a fixed-point theorem,} \cite{Nasar1998}.

Perhaps von Neumann's dismissive reaction stemmed from feelings of rivalry but quite possibly he was surprised to see a solution concept defined solely in terms of choices. For von Neumann, game theory was about \q{asking yourself what is the other man going to think I mean to do,} \cite{Bronowski1985}.  Yet beliefs played no role in Nash's solution concept.

\cite{Selten1978} first pointed out that without constraining beliefs, Nash's fixed-point condition could lead to paradoxical outcomes, see the game in Table \ref{chainstore}.

\begin{table}[h]
\vspace*{1mm}
\begin{center}
\begin{tabular}{c|cc}
& $F$ & $A$ \\ \cline{1-3}
\multicolumn{1}{r}{\rule{0pt}{5mm}$N$} & \multicolumn{1}{|c}{2,2} & \multicolumn{1}{c}{2,2}\\
\multicolumn{1}{r}{$E$} & \multicolumn{1}{|c}{0,0} & \multicolumn{1}{c}{3,1}\\
\end{tabular}
\end{center}
\vspace*{-5mm}
\caption{\citeauthor{Selten1978}'s (\citeyear{Selten1978}) Chain-Store Paradox.}\label{chainstore}
\vspace*{-1mm}
\end{table}

\noindent In this game, the Row player is the ``entrant'' and the Column player the ``incumbent.'' It is a Nash equilibrium for the entrant to enter $(E)$ and the incumbent to acquiesce ($A$). But it is also a Nash equilibrium for the entrant not to enter ($N$) and the incumbent to fight $(F)$. The latter outcome relies on the use of the weakly-dominated strategy $F$, which is optimal only if the incumbent believes the entrant will stay out for sure.\footnote{In the extensive-form version of the chain-store paradox, from which Table \ref{chainstore} is derived, the $(N,F)$ equilibrium involves the use of a non-credible threat and is not subgame perfect.} \cite{Selten1975} proposed that equilibrium choices should remain optimal against interior beliefs that put (infinitesimally) small weight on suboptimal strategies (\q{trembles}). For the chain-store paradox, this requirement selects $(E,A)$ as the unique (trembling-hand) \textit{perfect} equilibrium.

$S$ equilibrium takes inspiration from Selten's important insights about the impact of beliefs on choices. It also incorporates features of recent behavioral-game-theory models. Yet, it departs from prior approaches in important ways. An $S$ equilibrium consists of a pair of belief and choice \textit{sets}. Choices are consistent with beliefs in that the best option with the highest expected payoff is most often chosen. Small, but non-infinitesimal, trembles can occur and their size varies with the game's complexity. Beliefs are not necessarily correct but are consistent with choices in that they imply the same best option as observed choices do, i.e. they are consequentially unbiased. Intuitively, there is no need to refine beliefs if doing so does not affect choices.

The motivation for an explicitly set-valued theory is to avoid the common practice of selecting predictions from an implicitly defined set without accounting for its size.  Leading behavioral-game-theory models such as QRE and level-$k$ appear very \mbox{\textit{accurate}}, i.e. they match the bulk of the data, while at the same time being extremely \textit{precise} as they yield point predictions. This predictive success results from picking the best-fitting model from a large set of models. Implicitly, this means selecting the best prediction from a \textit{set} of predictions. Without registering a pre-analysis plan it is impossible to verify what models were sampled or what the (size of the) implied set of predictions was. As a result, the model's precision and its predictive success cannot be properly assessed.

Even if a pre-analyses plan has been registered, i.e. the choice of quantal responses or the distributions of levels is fixed (as are significance levels and number of observations), then there is still a \textit{set} of outcomes that result in non-rejection \mbox{(``acceptance'')} of the model. Also this set is hard to determine and generally ignored when assessing the model's predictive success. This raises a question about current practice: why consider (a slice of) some high-dimensional space of quantal responses or level distributions that implicitly define a set of predictions in the choice simplex? Especially if the theory further requires that beliefs satisfy rational expectations or follows from ad hoc assumptions.  Why not formulate simple choice and belief axioms that define a set of predictions in the choice simplex and a set of associated beliefs that support the predicted choices?

Set-valued theories offer a transparent solution by explicitly modeling the sets to which observed data are compared. \citeauthor{GoereeLouis2021}' (\citeyear{GoereeLouis2021}) $M$ equilibrium is such a theory, but it ignores precision and overemphasizes accuracy as its choice sets were designed to capture all regular QRE. $S$ equilibrium offers a trade-off between accuracy and precision via a single parameter that controls the size of the prediction set. This trade off can be made optimally in the choice simplex. Making this trade-off optimally in the large set of QRE or level-$k$ models, by reverse engineering the models that yield certain choice predictions, is intractable if not impossible.

We find that $S$ equilibrium captures 58\% of the observed choices using only 5\% of the choice simplex. $S$ equilibrium fits the choice data much better than level-$k$ and QRE, both in and out of sample. Observed beliefs are mostly consequentially unbiased (78\%). They refute the rational-expectations assumption underlying QRE and the ad hoc belief model underlying level-$k$ in \textit{all} games reported in this paper.

\subsection{Organization}

The next section details how $S$ equilibrium synthesizes the best features of existing game theory models and why it discards other features. Section \ref{sec:S} defines $S$ equilibrium and shows its choice sets form the roots of a simple function, the $S$ potential. Section \ref{sec:Exp} reports an experiment and compares $S$ equilibrium to leading behavioral-game-theory models. Section \ref{sec:conc} concludes. The Appendices contain proofs, additional results, and instructions.

\section{The Good, the Bad, and a Synthesis}

\citeauthor{Selten1975}'s (\citeyear{Selten1975}) approach to defining robust equilibria underlies virtually all of the refinement literature. We will argue, however, that Selten's definition is not adequate to achieve robustness and that a set-valued solution concept is needed. We then turn to behavioral-game-theory models that allow for sizeable (rather than infinitesimal) trembles and/or disequilibrium beliefs. $S$ equilibrium incorporates mistakes and non-equilibrium beliefs, but avoids the functional form assumptions that characterize existing behavioral models. Finally, we discuss the need to discipline set-valued concepts as not all games, nor all players, are created equal. This discipline is missing from $M$ equilibrium, as recently proposed by \cite{GoereeLouis2021}.

\subsection{Equilibrium Refinement}

\citeauthor{Selten1975} originally defined perfect equilibria as Nash equilibria of perturbed games. A simpler, but equivalent, definition can be found in \citeauthor{vanDamme1996} (\citeyear{vanDamme1996}, Th. 2.2.5).

\vspace*{1mm}

\begin{center}
\parbox{11.5cm}{A Nash equilibrium profile $\sigma$ is perfect if it is the $\varepsilon\downarrow 0$ limit of \color{red}\textbf{a sequence} \color{black} of totally-mixed profiles $\sigma(\varepsilon)$ such that $\sigma$ is a best reply against every element in $\sigma(\varepsilon)$.}
\end{center}

\vspace*{1mm}

\noindent Selten's approach formed the starting point for an entire literature. Virtually all refinement models define a Nash-equilibrium profile $\sigma$ to be robust if there exists a sequence of interior beliefs that converges to $\sigma$ and to which $\sigma$ is a best reply.\footnote{Other selection criteria may restrict the \textit{type} of sequence, e.g. in \citeauthor{GoereeLouis2021}' (\citeyear{GoereeLouis2021}) \textit{profect} equilibrium more costly trembles are less likely and in \citeauthor{Myerson1978}'s (\citeyear{Myerson1978}) \textit{proper} equilibrium they are infinitely less likely. However, both these criteria demand only the existence of one such sequence.} The intuition is that even if the others' choices are subject to some small amount of randomness, the Nash profile $\sigma$ remains the optimal choice. However, merely demanding existence of \color{red}\textbf{a sequence}\color{black}, leaves open the possibility that $\sigma$ is optimal only against exactly one sequence, i.e. against a very specific set of trembles and not to others. In particular, not to \textit{random} trembles.

\begin{figure}[t]
\begin{center}
\begin{tikzpicture}[scale=0.70]
\node[scale=0.6,above,shape=circle,draw] (player3) at (10,3) {\small 1};
\node[scale=0.6,shape=rectangle,draw] (R) at (7.5,1.5) {$R$};
\node[scale=0.6,shape=rectangle,draw] (B) at (10,1.5) {$B$};
\node[scale=0.6,shape=rectangle,draw] (Y) at (12.5,1.5) {$Y$};
\node[scale=0.6,below] (R2) at (5,0) {
\begin{tabular}{c|ccc}
\diagbox[width=7mm,height=7mm,innerleftsep=2pt,innerrightsep=2pt]{\small 2}{\small 3} & $R$ & $B$ & $Y$ \\ \cline{1-4}
\multicolumn{1}{r}{\rule{0pt}{5mm}$R$} & \multicolumn{1}{|c}{$120$} & \multicolumn{1}{c}{$90$} & \multicolumn{1}{c}{$60$} \\
\multicolumn{1}{r}{$B$} & \multicolumn{1}{|c}{$90$} & \multicolumn{1}{c}{$60$} & \multicolumn{1}{c}{$60$} \\
\multicolumn{1}{r}{$Y$} & \multicolumn{1}{|c}{$60$} & \multicolumn{1}{c}{$60$} & \multicolumn{1}{c}{$70$} \\
\end{tabular}};
\node[scale=0.6,below] (B2) at (10,0) {
\begin{tabular}{c|ccc}
\diagbox[width=7mm,height=7mm,innerleftsep=2pt,innerrightsep=2pt]{\small 2}{\small 3} & $R$ & $B$ & $Y$ \\ \cline{1-4}
\multicolumn{1}{r}{\rule{0pt}{5mm}$R$} & \multicolumn{1}{|c}{$25$} & \multicolumn{1}{c}{$25$} & \multicolumn{1}{c}{$55$} \\
\multicolumn{1}{r}{$B$} & \multicolumn{1}{|c}{$25$} & \multicolumn{1}{c}{$25$} & \multicolumn{1}{c}{$75$} \\
\multicolumn{1}{r}{$Y$} & \multicolumn{1}{|c}{$55$} & \multicolumn{1}{c}{$75$} & \multicolumn{1}{c}{$90$} \\
\end{tabular}};
\node[scale=0.6,below] (Y2) at (15,0) {
\begin{tabular}{c|ccc}
\diagbox[width=7mm,height=7mm,innerleftsep=2pt,innerrightsep=2pt]{\small 2}{\small 3} & $R$ & $B$ & $Y$ \\ \cline{1-4}
\multicolumn{1}{r}{\rule{0pt}{5mm}$R$} & \multicolumn{1}{|c}{$5$} & \multicolumn{1}{c}{$20$} & \multicolumn{1}{c}{$60$} \\
\multicolumn{1}{r}{$B$} & \multicolumn{1}{|c}{$20$} & \multicolumn{1}{c}{$5$} & \multicolumn{1}{c}{$85$} \\
\multicolumn{1}{r}{$Y$} & \multicolumn{1}{|c}{$60$} & \multicolumn{1}{c}{$85$} & \multicolumn{1}{c}{$85$} \\
\end{tabular}};
\draw[scale=0.6] (player3) -- (R); \draw (R) -- (R2);
\draw[scale=0.6] (player3) -- (B); \draw (B) -- (B2);
\draw[scale=0.6] (player3) -- (Y); \draw (Y) -- (Y2);
\draw[scale=0.6,dotted] (R) -- (B) -- (Y);

\node at (21.5,.7) {
\begin{tikzpicture}[scale=4.1]
\draw[name path=A1,line width=.5pt,gray] (7/8,1/8*3^.5) -- (1/2,1/2*3^.5) -- (0,0);
\draw[name path=A2,line width=.5pt,gray] (7/8,1/8*3^.5)-- (1,0) -- (0,0);
\filldraw[fill=mondrianRed,draw=mondrianRed,opacity=1] (0,0) -- (0.354809, 0.614547) -- (0.782376, 0.376936) -- (1,0) -- (0,0);

\node[scale=0.8pt] at (-0.05,-0.05) {$R$};
\node[scale=0.8pt] at (1.05,-0.05) {$B$};
\node[scale=0.8pt] at (1/2,1/2*3^.5+0.05) {$Y$};

\draw[name path=B1, domain=0:0.157895, very thick, smooth, variable=\pp, black] plot ({1/18*(12-14*\pp-2^.5*(\pp*(3-19*\pp))^.5)},{3^.5/18*(6-4*\pp+2^.5*(\pp*(3-19*\pp))^.5)});
\draw[name path=B2, domain=0:0.157895, very thick, smooth, variable=\pp, black] plot ({1/18*(12-14*\pp+2^.5*(\pp*(3-19*\pp))^.5)},{3^.5/18*(6-4*\pp-2^.5*(\pp*(3-19*\pp))^.5)});

\draw[very thick, black] (31/57+0.008,17/57*3^.5-0.008) -- (31/57-0.008,17/57*3^.5+0.008);
\fill[black] (2/3,1/3*3^.5) circle(0.7pt);
\node[scale=0.8pt] at (2/3+0.07,1/3*3^.5+0.02) {$\sigma$};
\fill[black] (0,0) circle(0.7pt);

\node at (0.08, 0.034641) [draw,scale=.5,diamond,fill=mondrianOrange]{};
\node[star,star points=10] at (0.19, 0.103923) [draw,scale=.5,fill=mondrianOrange]{};

\end{tikzpicture}};

\end{tikzpicture}

\vspace*{-1mm}
\caption{Game $G_1$ is a symmetric $3\times 3\times 3$ game with two perfect Nash equilibria: $R$ and $\sigma=(0,\deel{1}{3},\deel{2}{3})$. In the belief simplex on the right the red area corresponds to beliefs to which $R$ is a best reply and the black ellipse to beliefs for which $\sigma$ is a best reply. The orange star and diamond show average beliefs and choices respectively.}\label{fig:ellipse}
\end{center}
\vspace*{-7mm}
\end{figure}
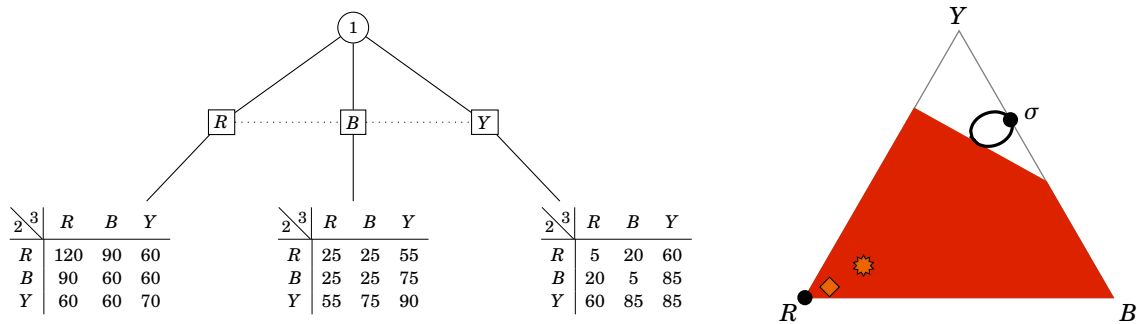

To illustrate, consider the symmetric $3\times3\times 3$ in the left panel of Figure \ref{fig:ellipse}, which has two symmetric Nash equilibria, $R$ and $\sigma=(0,\deel{1}{3},\deel{2}{3})$.\footnote{We only show Row's payoffs. Column's payoffs follow by transposing Row's payoff matrix. Throughout we focus on symmetric equilibria as subjects were matched using a random or ``total strangers'' protocol, which makes it virtually impossible to coordinate on asymmetric equilibria.} Both are perfect. For the pure-strategy equilibrium, $R$, this is obvious as it is strict. For the mixed equilibrium, $\sigma=(0,\deel{1}{3},\deel{2}{3})$, there exists an ellipse inside the simplex that supports it, i.e. for the profiles that satisfy $(2\sigma_R+2\sigma_B-\sigma_Y)^2=2\sigma_R(3\sigma_B-\sigma_R)$ we have $\pi_B(\sigma)=\pi_Y(\sigma)>\pi_R(\sigma)$.  This ellipse is shown by the thick black curve in the belief simplex in the right panel of Figure \ref{fig:ellipse}. Since we focus on symmetric equilibria, players hold the same beliefs about others' play, which allows us to draw beliefs in a single simplex.

While there exists a sequence of beliefs supporting $\sigma$ it restricts players to tremble in a very precise manner. If trembles are random and off the elliptic path, expected payoffs will be strictly ranked and $\sigma$ will not be a best reply. In contrast, $R$ is supported by a \textit{set} of interior beliefs as indicated by the red area in the belief simplex on the right. The orange star and diamond in this simplex show average beliefs and choices respectively (from an experiment described in Section \ref{sec:Exp} below), and confirm that $\sigma$ is irrelevant from an empirical viewpoint.

\subsection{Level-$k$}

A popular model of non-equilibrium beliefs is level-$k$, see \cite{stahl1994,stahl1995} and \cite{nagel1995}.\footnote{See \cite{camerer2004} for a closely related model called \textit{Cognitive Hierarchy} and
\cite{alaoui2016,alaoui2021} for a model called \textit{Endogenous Depth of Reasoning} in which levels are endogenous.} The discrete levels $k=0,1,2,\ldots$ represent a player's strategic sophistication. A naive level-0 player randomizes uniformly over the pure strategies.\footnote{This is the common assumption for the abstract matrix games considered in this paper. In games where some choices are more salient the definition of level-0 play may depend on the game's features.} More sophisticated players with levels $k\geq 1$ believe others are of level $k-1$ and best reply accordingly, i.e. level-1 best replies to random behavior, level-2 best replies to the best reply to random behavior, etc. One concern is that building a best-reply hierarchy on a single belief is unlikely to produce robust results.

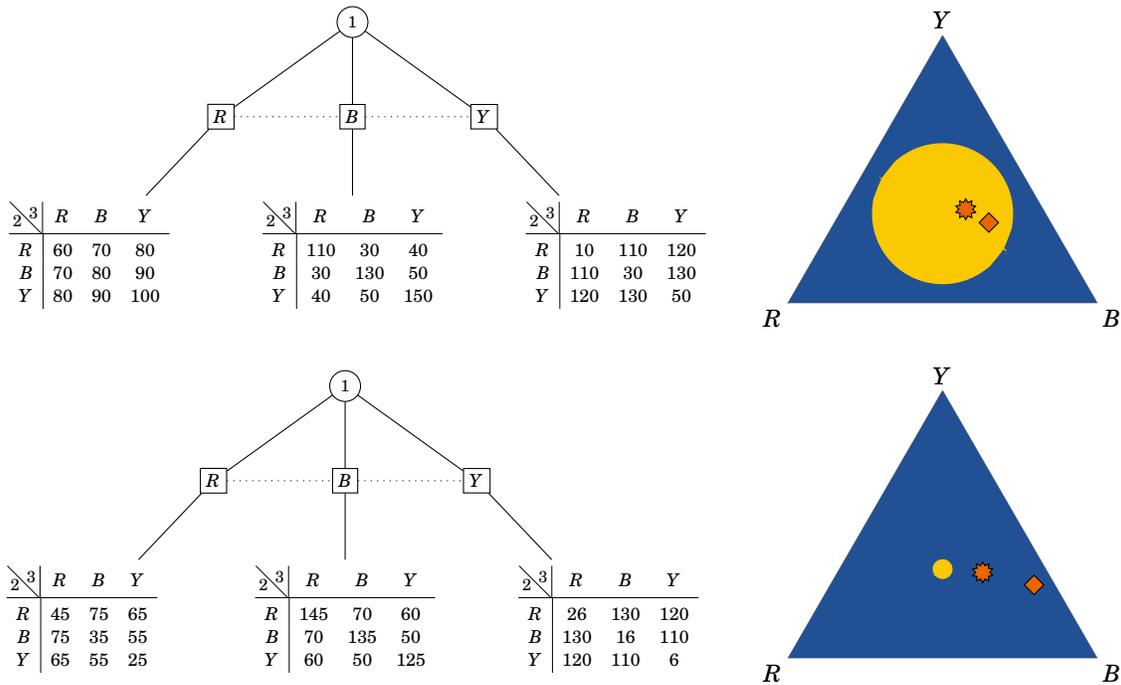
\begin{figure}[t]
\begin{center}
\begin{tikzpicture}[scale=0.70]
\node[scale=0.6,above,shape=circle,draw] (player3) at (10,3) {\small 1};
\node[scale=0.6,shape=rectangle,draw] (R) at (7.5,1.5) {$R$};
\node[scale=0.6,shape=rectangle,draw] (B) at (10,1.5) {$B$};
\node[scale=0.6,shape=rectangle,draw] (Y) at (12.5,1.5) {$Y$};
\node[scale=0.6,below] (R2) at (5,0) {
\begin{tabular}{c|ccc}
\diagbox[width=7mm,height=7mm,innerleftsep=2pt,innerrightsep=2pt]{\small 2}{\small 3} & $R$ & $B$ & $Y$ \\ \cline{1-4}
\multicolumn{1}{r}{\rule{0pt}{5mm}$R$} & \multicolumn{1}{|c}{$60$} & \multicolumn{1}{c}{$70$} & \multicolumn{1}{c}{$80$} \\
\multicolumn{1}{r}{$B$} & \multicolumn{1}{|c}{$70$} & \multicolumn{1}{c}{$80$} & \multicolumn{1}{c}{$90$} \\
\multicolumn{1}{r}{$Y$} & \multicolumn{1}{|c}{$80$} & \multicolumn{1}{c}{$90$} & \multicolumn{1}{c}{$100$} \\
\end{tabular}};
\node[scale=0.6,below] (B2) at (10,0) {
\begin{tabular}{c|ccc}
\diagbox[width=7mm,height=7mm,innerleftsep=2pt,innerrightsep=2pt]{\small 2}{\small 3} & $R$ & $B$ & $Y$ \\ \cline{1-4}
\multicolumn{1}{r}{\rule{0pt}{5mm}$R$} & \multicolumn{1}{|c}{$110$} & \multicolumn{1}{c}{$30$} & \multicolumn{1}{c}{$40$} \\
\multicolumn{1}{r}{$B$} & \multicolumn{1}{|c}{$30$} & \multicolumn{1}{c}{$130$} & \multicolumn{1}{c}{$50$} \\
\multicolumn{1}{r}{$Y$} & \multicolumn{1}{|c}{$40$} & \multicolumn{1}{c}{$50$} & \multicolumn{1}{c}{$150$} \\
\end{tabular}};
\node[scale=0.6,below] (Y2) at (15,0) {
\begin{tabular}{c|ccc}
\diagbox[width=7mm,height=7mm,innerleftsep=2pt,innerrightsep=2pt]{\small 2}{\small 3} & $R$ & $B$ & $Y$ \\ \cline{1-4}
\multicolumn{1}{r}{\rule{0pt}{5mm}$R$} & \multicolumn{1}{|c}{$10$} & \multicolumn{1}{c}{$110$} & \multicolumn{1}{c}{$120$} \\
\multicolumn{1}{r}{$B$} & \multicolumn{1}{|c}{$110$} & \multicolumn{1}{c}{$30$} & \multicolumn{1}{c}{$130$} \\
\multicolumn{1}{r}{$Y$} & \multicolumn{1}{|c}{$120$} & \multicolumn{1}{c}{$130$} & \multicolumn{1}{c}{$50$} \\
\end{tabular}};
\draw[scale=0.6] (player3) -- (R); \draw (R) -- (R2);
\draw[scale=0.6] (player3) -- (B); \draw (B) -- (B2);
\draw[scale=0.6] (player3) -- (Y); \draw (Y) -- (Y2);
\draw[scale=0.6,dotted] (R) -- (B) -- (Y);

\node at (10,-6.25) {
\begin{tikzpicture}[scale=0.70]
\node[scale=0.6,above,shape=circle,draw] (player3) at (10,3) {\small 1};
\node[scale=0.6,shape=rectangle,draw] (R) at (7.5,1.5) {$R$};
\node[scale=0.6,shape=rectangle,draw] (B) at (10,1.5) {$B$};
\node[scale=0.6,shape=rectangle,draw] (Y) at (12.5,1.5) {$Y$};
\node[scale=0.6,below] (R2) at (5,0) {
\begin{tabular}{c|ccc}
\diagbox[width=7mm,height=7mm,innerleftsep=2pt,innerrightsep=2pt]{\small 2}{\small 3} & $R$ & $B$ & $Y$ \\ \cline{1-4}
\multicolumn{1}{r}{\rule{0pt}{5mm}$R$} & \multicolumn{1}{|c}{$45$} & \multicolumn{1}{c}{$75$} & \multicolumn{1}{c}{$65$} \\
\multicolumn{1}{r}{$B$} & \multicolumn{1}{|c}{$75$} & \multicolumn{1}{c}{$35$} & \multicolumn{1}{c}{$55$} \\
\multicolumn{1}{r}{$Y$} & \multicolumn{1}{|c}{$65$} & \multicolumn{1}{c}{$55$} & \multicolumn{1}{c}{$25$} \\
\end{tabular}};
\node[scale=0.6,below] (B2) at (10,0) {
\begin{tabular}{c|ccc}
\diagbox[width=7mm,height=7mm,innerleftsep=2pt,innerrightsep=2pt]{\small 2}{\small 3} & $R$ & $B$ & $Y$ \\ \cline{1-4}
\multicolumn{1}{r}{\rule{0pt}{5mm}$R$} & \multicolumn{1}{|c}{$145$} & \multicolumn{1}{c}{$70$} & \multicolumn{1}{c}{$60$} \\
\multicolumn{1}{r}{$B$} & \multicolumn{1}{|c}{$70$} & \multicolumn{1}{c}{$135$} & \multicolumn{1}{c}{$50$} \\
\multicolumn{1}{r}{$Y$} & \multicolumn{1}{|c}{$60$} & \multicolumn{1}{c}{$50$} & \multicolumn{1}{c}{$125$} \\
\end{tabular}};
\node[scale=0.6,below] (Y2) at (15,0) {
\begin{tabular}{c|ccc}
\diagbox[width=7mm,height=7mm,innerleftsep=2pt,innerrightsep=2pt]{\small 2}{\small 3} & $R$ & $B$ & $Y$ \\ \cline{1-4}
\multicolumn{1}{r}{\rule{0pt}{5mm}$R$} & \multicolumn{1}{|c}{$26$} & \multicolumn{1}{c}{$130$} & \multicolumn{1}{c}{$120$} \\
\multicolumn{1}{r}{$B$} & \multicolumn{1}{|c}{$130$} & \multicolumn{1}{c}{$16$} & \multicolumn{1}{c}{$110$} \\
\multicolumn{1}{r}{$Y$} & \multicolumn{1}{|c}{$120$} & \multicolumn{1}{c}{$110$} & \multicolumn{1}{c}{$6$} \\
\end{tabular}};
\draw[scale=0.6] (player3) -- (R); \draw (R) -- (R2);
\draw[scale=0.6] (player3) -- (B); \draw (B) -- (B2);
\draw[scale=0.6] (player3) -- (Y); \draw (Y) -- (Y2);
\draw[scale=0.6,dotted] (R) -- (B) -- (Y);
\end{tikzpicture}};

\node at (21.2,0.5) {
\begin{tikzpicture}[scale=4.1]
\draw[name path=A,line width=.5pt,gray] (1/2,1/2*3^.5) -- (0,0) -- (1,0) -- (1/2,1/2*3^.5);
\node[scale=0.8pt] at (-0.05,-0.05) {$R$};
\node[scale=0.8pt] at (1.05,-0.05) {$B$};
\node[scale=0.8pt] at (1/2,1/2*3^.5+0.05) {$Y$};

\draw[name path=B1, smooth, domain=0.0611678:0.605499, ultra thin, smooth, variable=\pp, mondrianBlue] plot ({1/12*(3+9*\pp+(-1+18*\pp-27*\pp*\pp)^.5)},{3^.5/12*(3-3*\pp+(-1+18*\pp-27*\pp*\pp)^.5)});
\draw[name path=B2, smooth, domain=0.0611678:0.605499, ultra thin, smooth, variable=\pp, mondrianBlue] plot ({1/12*(3+9*\pp-(-1+18*\pp-27*\pp*\pp)^.5)},{3^.5/12*(3-3*\pp-(-1+18*\pp-27*\pp*\pp)^.5)});
\tikzfillbetween[of=A and B1]{mondrianBlue, opacity=1};
\tikzfillbetween[of=A and B2]{mondrianBlue, opacity=1};
\filldraw[fill=mondrianBlue,draw=mondrianBlue,opacity=1] (0,0) -- (5/8,7/104*3^.5+0.01) -- (5/8,7/104*3^.5-0.01) -- (0,0);
\tikzfillbetween[of=B1 and B2]{mondrianYellow, opacity=1};
\filldraw[fill=mondrianYellow,draw=mondrianYellow,opacity=1] (0.295876, 0.406526) -- (0.704124, 0.170824+0.01) -- (0.704124, 0.170824-0.01) -- (0.295876, 0.406526);
\draw[name path=B1, smooth, samples=50, domain=0.0611678:0.605499, ultra thick, smooth, variable=\pp, mondrianBlue] plot ({1/12*(3+9*\pp+(-1+18*\pp-27*\pp*\pp)^.5)},{3^.5/12*(3-3*\pp+(-1+18*\pp-27*\pp*\pp)^.5)});
\draw[name path=B2, smooth, samples=50, domain=0.0611678:0.605499, ultra thick, smooth, variable=\pp, mondrianBlue] plot ({1/12*(3+9*\pp-(-1+18*\pp-27*\pp*\pp)^.5)},{3^.5/12*(3-3*\pp-(-1+18*\pp-27*\pp*\pp)^.5)});

\node at (0.65, 0.2598) [draw,scale=.5,diamond,fill=mondrianOrange]{};
\node[star,star points=10] at (0.575, 0.3031) [draw,scale=.5,fill=mondrianOrange]{};

\end{tikzpicture}};

\node at (21.2,-6.25) {
\begin{tikzpicture}[scale=4.1]
\draw[name path=A,line width=.5pt,gray] (1/2,1/2*3^.5) -- (0,0) -- (1,0) -- (1/2,1/2*3^.5);
\node[scale=0.8pt] at (-0.05,-0.05) {$R$};
\node[scale=0.8pt] at (1.05,-0.05) {$B$};
\node[scale=0.8pt] at (1/2,1/2*3^.5+0.05) {$Y$};

\draw[name path=B1, smooth, domain=0.29811:0.3685685, ultra thin, smooth, variable=\pp, mondrianYellow] plot ({3/4-3/4*\pp-1/4*(-59/179+2*\pp-3*\pp*\pp)^.5)},{3^.5/716*(179-179*\pp+179^.5*(-59+358*\pp-537*\pp*\pp)^.5)});
\draw[name path=B2, smooth, domain=0.29811:0.3685685, ultra thin, smooth, variable=\pp, mondrianYellow] plot ({3/4-3/4*\pp+1/4*(-59/179+2*\pp-3*\pp*\pp)^.5)},{3^.5/716*(179-179*\pp-179^.5*(-59+358*\pp-537*\pp*\pp)^.5)});
\tikzfillbetween[of=A and B1]{mondrianBlue, opacity=1};
\tikzfillbetween[of=A and B2]{mondrianBlue, opacity=1};
\tikzfillbetween[of=B1 and B2]{mondrianYellow, opacity=1};
\draw[name path=B1, smooth, samples=50, domain=0.29811:0.3685685, ultra thin, smooth, variable=\pp, mondrianYellow] plot ({3/4-3/4*\pp-1/4*(-59/179+2*\pp-3*\pp*\pp)^.5)},{3^.5/716*(179-179*\pp+179^.5*(-59+358*\pp-537*\pp*\pp)^.5)});
\draw[name path=B2, smooth, samples=50, domain=0.29811:0.3685685, ultra thin, smooth, variable=\pp, mondrianYellow] plot ({3/4-3/4*\pp+1/4*(-59/179+2*\pp-3*\pp*\pp)^.5)},{3^.5/716*(179-179*\pp-179^.5*(-59+358*\pp-537*\pp*\pp)^.5)});
\filldraw[fill=mondrianYellow,draw=mondrianYellow,opacity=1] (0.526426, 0.303932) -- (0.473574, 0.273418+0.005) -- (0.473574, 0.273418-0.005) -- (0.526426, 0.303932);

\node at (0.7965, 0.236425) [draw,scale=.5,diamond,fill=mondrianOrange]{};
\node[star,star points=10] at (0.63, 0.277128) [draw,scale=.5,fill=mondrianOrange]{};

\end{tikzpicture}};

\end{tikzpicture}

\end{center}
\vspace*{-5mm}

\caption{The left panels show two symmetric $3\times 3\times 3$ games labeled $G_2$ (top) and $G_3$ (bottom). The blue areas in the belief simplices on the right show beliefs for which $B$ is a best reply and the yellow areas show beliefs for which $Y$ is a best reply. The orange stars and diamonds show average beliefs and choices respectively.}\label{fig:levelk}
\vspace*{-2mm}
\end{figure}

To illustrate, consider the symmetric $3\times 3\times 3$ games in Figure \ref{fig:levelk}. Level-$k$ predicts identical results across these two games: level-0 randomizes, level-1 plays $Y$, and levels 2 and above play $B$. The orange stars in the right panels show average observed beliefs. A statistical test reveals that beliefs do not differ significantly across the two games $(p>0.6)$. However, the set of beliefs for which $Y$ is the best reply is much larger in $G_2$ (top) than in $G_3$ (bottom) and captures more of the observed beliefs. The orange diamonds show average observed choices and confirm that observed play differs significantly across the two games $(p<0.001)$. As in the previous section, belief \textit{sets} are an important determinant of observed play.

\subsection{Quantal Response Equilibrium}

\begin{figure}[t]
\begin{center}

\begin{tikzpicture}

\node[scale=0.85] at (-1,0.35) {
\begin{tabular}{c|ccc|c|c}
$g_1$ & $R$ & $B$ & $Y$ & $\sigma_{obs}$ & $\omega_{obs}$ \\ \cline{1-6}
\multicolumn{1}{r}{\rule{0pt}{5mm}$R$} & \multicolumn{1}{|c}{10} & \multicolumn{1}{c}{120} & \multicolumn{1}{c|}{10} & 0.07 & 0.13\\
\multicolumn{1}{r}{$B$} & \multicolumn{1}{|c}{10} & \multicolumn{1}{c}{100} & \multicolumn{1}{c|}{240} & 0.71 & 0.53\\
\multicolumn{1}{r}{$Y$} & \multicolumn{1}{|c}{10} & \multicolumn{1}{c}{110} & \multicolumn{1}{c|}{120} & 0.22 & 0.34\\
\end{tabular}};

\node at (6,0.43) {
\begin{tikzpicture}[scale=3.5]
\draw[line width=.5pt,gray] (0,0) -- (1,0) -- (1/2,1/2*3^.5) -- (0,0);
\node[scale=0.8pt] at (-0.03,-0.03) {$R$};
\node[scale=0.8pt] at (1.03,-0.03) {$B$};
\node[scale=0.8pt] at (1/2,1/2*3^.5+0.03) {$Y$};

\filldraw[fill=mondrianGrey,draw=mondrianGrey,opacity=1] (1/2,1/6*3^.5) -- (0.888889, 0.06415) -- (24/25,1/25*3^.5) -- (3/4,1/4*3^.5) -- (1/2,1/6*3^.5);

\node at (0.82, 0.190526) [draw,scale=.5,diamond,fill=mondrianOrange]{};
\node[star,star points=10] at (0.7, 0.294449) [draw,scale=.5,fill=mondrianOrange]{};

\begin{axis}[scale=0.175, axis line style={draw=none}, tick style={draw=none}, ticks=none, xmin=0, xmax=1.2, ymin=0, ymax=1]
\addplot[thin, smooth] plot coordinates
            {
                    (0.5,	0.288675)
                    (0.60599,	0.275959)
                    (0.681701,	0.254544)
                    (0.732403,	0.233972)
                    (0.767372,	0.216564)
                    (0.792621,	0.202198)
                    (0.81162,	0.190304)
                    (0.826411,	0.180349)
                    (0.838247,	0.171912)
                    (0.847933,	0.164676)
                    (0.856008,	0.158401)
                    (0.856008,	0.158401)
                    (0.896453,	0.123056)
                    (0.911545,	0.107557)
                    (0.919216,	0.0987294)
                    (0.92365,	0.0929827)
                    (0.926352,	0.0889208)
                    (0.928001,	0.085883)
                    (0.928944,	0.0835146)
                    (0.929381,	0.0816077)
                    (0.92943,	0.0800319)
                    (0.92943,	0.0800319)
                    (0.917524,	0.0717384)
                    (0.888981,	0.0671218)
                    (0.845822,	0.0627798)
                    (0.793274,	0.0583171)
                    (0.738253,	0.0539527)
                    (0.685583,	0.049907)
                    (0.637438,	0.0462729)
                    (0.594348,	0.0430547)
                    (0.556086,	0.0402173)
                    (0, 0)
            };
\end{axis}

\end{tikzpicture}};

\node[scale=0.85] at (-1,0.35-4) {
\begin{tabular}{c|ccc|c|c}
$g_{2}$ & $R$ & $B$ & $Y$ & $\sigma_{obs}$ & $\omega_{obs}$ \\ \cline{1-6}
\multicolumn{1}{r}{\rule{0pt}{5mm}$R$} & \multicolumn{1}{|c}{100} & \multicolumn{1}{c}{80} & \multicolumn{1}{c|}{10} & 0.06 & 0.25 \\
\multicolumn{1}{r}{$B$} & \multicolumn{1}{|c}{100} & \multicolumn{1}{c}{100} & \multicolumn{1}{c|}{20} & 0.88 & 0.52 \\
\multicolumn{1}{r}{$Y$} & \multicolumn{1}{|c}{80} & \multicolumn{1}{c}{100} & \multicolumn{1}{c|}{40} & 0.06 & 0.23 \\
\end{tabular}};

\node at (6,0.43-4) {
\begin{tikzpicture}[scale=3.5]
\draw[line width=.5pt,gray] (0,0) -- (1,0) -- (1/2,1/2*3^.5) -- (0,0);
\node[scale=0.8pt] at (-0.03,-0.03) {$R$};
\node[scale=0.8pt] at (1.03,-0.03) {$B$};
\node[scale=0.8pt] at (1/2,1/2*3^.5+0.03) {$Y$};

\filldraw[fill=mondrianGrey,draw=mondrianGrey,opacity=1] (1/2,1/6*3^.5) -- (3/4,1/4*3^.5) -- (1/2,1/2*3^.5) -- (1/2,1/6*3^.5);

\begin{axis}[scale=0.175, axis line style={draw=none}, tick style={draw=none}, ticks=none, xmin=0, xmax=1.2, ymin=0, ymax=1]
\addplot[thin, smooth] plot coordinates
            {
                (0.5,	0.288675)
                (0.533854,	0.314749)
                (0.564967,	0.359646)
                (0.584349,	0.432397)
                (0.582792,	0.536485)
                (0.563983,	0.649121)
                (0.543665,	0.7341)
                (0.528944,	0.786383)
                (0.519208,	0.817162)
                (0.512822,	0.835492)
                (0.508599,	0.84665)
                (0.505785,	0.853579)
                (0.503898,	0.857955)
            };
\end{axis}

\node at (0.91, 0.0519615) [draw,scale=.5,diamond,fill=mondrianOrange]{};
\node[star,star points=10] at (0.635, 0.199186) [draw,scale=.5,fill=mondrianOrange]{};

\end{tikzpicture}};

\end{tikzpicture}
\end{center}
\vspace*{-6mm}
\caption{The left panels show two $3\times 3$ games labeled $g_1$ (top) and $g_2$ (bottom). The black curves in the choice simplices on the right show the logit-QRE as a function of $\lambda$. The curve starts at the simplex' centroid $(\lambda=0)$ and ends at a Nash equilibrium $(\lambda=\infty)$. The grey areas show $\mathcal{R}$-QRE for all possible regular quantal response functions, $\mathcal{R}$. The orange stars and diamonds show average beliefs and choices respectively.}\label{fig:qre}
\vspace*{-1mm}
\end{figure}
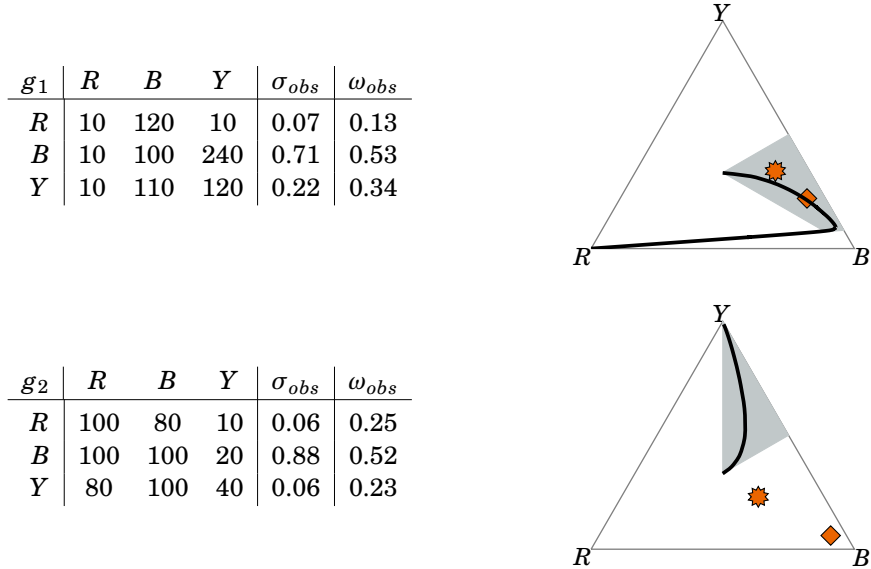

Another departure from \citeauthor{Selten1975}'s (\citeyear{Selten1975}) approach is to allow for sizeable mistakes. The notion of infinitesimal trembles is too restrictive as sizeable deviations are the rule rather than the exception. To illustrate, consider game $g_1$ in the top-left panel of Figure \ref{fig:qre}, for which $R$ is the unique (perfect) Nash equilibrium. Observed choices and beliefs, see the $\sigma_{obs}$ and $\omega_{obs}$ columns, are far from Nash predictions.\footnote{More generally, there is overwhelming evidence from the laboratory that subjects do \textit{not} play Nash equilibrium and that their choices are often far away from Nash-equilibrium predictions. See e.g. \citet{lieberman1960, oneill1987, brown1990, rapoport1992, stahl1994, nagel1995, mckelvey1992, ochs1995, goeree2001, crawford2013, GoereeHoltPalfrey2016, GoereeLouis2021}. This list is far from exhaustive.}

A coherent model of mistakes is \citeauthor{McKelveyPalfrey1995}'s (\citeyear{McKelveyPalfrey1995}) \textit{Quantal Response Equilibrium}. QRE requires the specification of \q{quantal response functions} that map expected payoffs to choice probabilities. Suppose, for instance, that player $i$ has $K_i$ strategies then assuming logistic quantal responses
\begin{equation}\label{logitQRE2}
  \mathcal{L}_{ik}(\pi_i(\sigma_{-i}))\,=\,\frac{\exp(\lambda\pi_{ik}(\sigma_{-i}))}{\sum_{\ell\,=\,1}^{K_i}\exp(\lambda\pi_{i\ell}(\sigma_{-i}))}
\end{equation}
the logit-QRE is defined by $\sigma_{ik}=\mathcal{L}_{ik}(\pi_i(\sigma_{-i}))$ for $i\in N$, $1\leq k\leq K_i$. The logistic formulation in \eqref{logitQRE2} is not the only possibility. Any set of \textit{regular} quantal responses, $\mathcal{R}_i:\field{R}^{K_i}\rightarrow\Sigma_i$, that are interior, continuous, strictly increasing, and monotone in expected payoffs can be used to define an $\mathcal{R}$-QRE: $\sigma=\mathcal{R}(\pi(\sigma))$, where $\mathcal{R}$ denotes the concatenation of players' quantal responses. Note that QRE is a fixed-point model based on rational expectations, i.e. choices on the left match beliefs on the right. Like the Nash equilibrium, QRE is a model of choices not beliefs.

The black curve in the top-right panel of Figure \ref{fig:qre} shows the logit-QRE for game $g_1$ as a function of the rationality parameter $\lambda$. The curve starts at the simplex' centroid when $\lambda=0$ and ends in the unique Nash equilibrium, $R$, when $\lambda=\infty$. The grey area shows the $\mathcal{R}$-QRE for all other possible choices of the regular quantal responses $\mathcal{R}$. For this game, the logit-QRE perfectly captures the observed average choices (indicated by the orange diamond) for some intermediate value of $\lambda$.

However, for game $g_2$ in the bottom-left panel of Figure \ref{fig:qre} no $\mathcal{R}$-QRE can capture observed choices, which are predominantly $B$. The inability of QRE to produce mostly $B$ choices is a direct consequence of the underlying fixed-point assumption that beliefs match choices. If $B$ is (believed to be) most likely then the expected payoff of $Y$ exceeds that of $R$. Hence, $Y$ is (believed to be) more likely chosen than $R$, which, in turn, implies that the expected payoff of $Y$ exceeds that of $B$. But then $B$ cannot (believed to) be more likely. To explain the preponderance of $B$ choices, the rational-expectations assumption underlying QRE has to be relaxed. For instance, for the observed average beliefs, $\omega_{obs}$, the best reply is $B$.

\subsection{Sets of Models versus Models of Sets}
\label{sec:svss}

The inability of QRE and level-$k$ to explain choices in the above examples contrasts with their success in other settings. This success partly results from selecting predictions from a set without accounting for its size. For instance, a QRE is a solution to the fixed-point condition $\sigma=\mathcal{R}(\pi(\sigma))$ for some set of regular quantal responses, $\mathcal{R}$. The choice of $\mathcal{R}$ is made \textit{after} data have been collected to generate the best fit.\footnote{Commonly-used QRE models include the Probit model, e.g. \cite{Zauner1999}, the Luce model, e.g. \cite{GHP2002}, and the Logit model, e.g. \cite{CapraGoereeGomezHolt1999}. QRE models have been estimated with a single rationality parameter, e.g this paper, game-specific rationality parameters, e.g. \cite{McKelveyPalfrey1995}, and player-specific rationality parameters, e.g. \cite{mckelvey2000}.} Likewise, level-$k$ produces point predictions for each level, but the distribution of levels is calibrated to provide the best fit.\footnote{Some papers estimate a general level distribution, e.g. \cite{stahl1994,stahl1995,crawford2001}, while others impose parametric restrictions, e.g. \cite{goeree2017}.} In both examples, the model is cherry picked from a \textit{set of models}. While each model in this set is point valued, collectively they generate a set of predicted choices. This raises questions of falsifiability.

It is easy to show that level-$k$ is non-falsifiable for general level distributions. Consider the symmetric $2\times 2$ game in Table \ref{2by2} and assume identical level distributions for Row and Column. Any observed frequency $p\in[0,1]$ for $A$ can be matched by the level-$k$ model in which level-1 occurs with probability $p$ and level-2 with probability $1-p$. One remedy is to impose parametric restrictions.  A commonly-used parametric form is $p_k=e^{-\tau}\tau^k/k!$, i.e. levels are Poisson distributed. Under this parametrization the chance of $A$ is predicted to lie between $\hf$ and $\deel{5}{8}$ and the model is falsifiable.

\begin{table}[t]
\vspace*{1mm}
\begin{center}
\begin{tabular}{c|cc}
& $A$ & $B$ \\ \cline{1-3}
\multicolumn{1}{r}{\rule{0pt}{5mm}$A$} & \multicolumn{1}{|c}{0,0} & \multicolumn{1}{c}{2,1}\\
\multicolumn{1}{r}{$B$} & \multicolumn{1}{|c}{1,2} & \multicolumn{1}{c}{0,0}\\
\end{tabular}
\end{center}
\vspace*{-5mm}
\caption{A symmetric $2\times 2$ game.}\label{2by2}
\vspace*{-1mm}
\end{table}

What about the set of predicted choices for all regular\footnote{\cite{HaileHortacsuKosenok2008} prove that QRE, as originally defined by \cite{McKelveyPalfrey1995}, is non-falsifiable. \cite{goeree2005} introduce the concept of \textit{regular} QRE and show it is falsifiable.} QRE? Ostensibly, this question is impossible to answer as the set of regular quantal responses, $\mathcal{R}$, is infinite dimensional. And for a typical choice of $\mathcal{R}$ the QRE fixed-point condition can only be solved numerically. Yet, \cite{GoereeLouis2021} show that the union of all regular QRE forms an $M$-equilibrium choice set (see the next section). The latter is falsifiable in generic games and easy to compute. For instance, for the game in Table \ref{2by2}, the symmetric $M$ equilibrium predicts that the chance of $A$ lies between $\hf$ and $\deel{2}{3}$.

While regular QRE is falsifiable, the current practice to select the best-fitting model from a large set of models creates the false impression that the selected model is both very accurate and precise. One solution is to show the set of predicted choices for all regular QRE (and all level-$k$ models), as we do in this paper. However, the resulting set may be too large and contain choice predictions from models that were not considered. But, without registering a pre-analyses plan, it is impossible to verify what models \textit{were} considered and what the implied prediction set was.

\begin{figure}[t]
\vspace*{-4mm}
\begin{center}

\begin{tikzpicture}
\node at (0,0) {\includegraphics[scale=0.3]{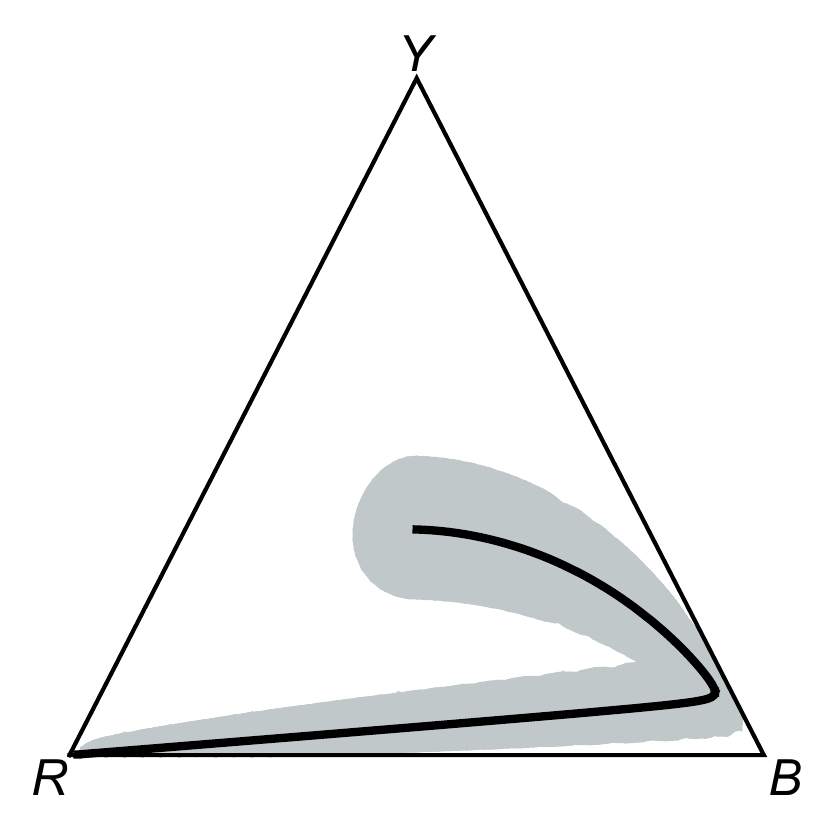}};
\node at (6.6,0) {\includegraphics[scale=0.3]{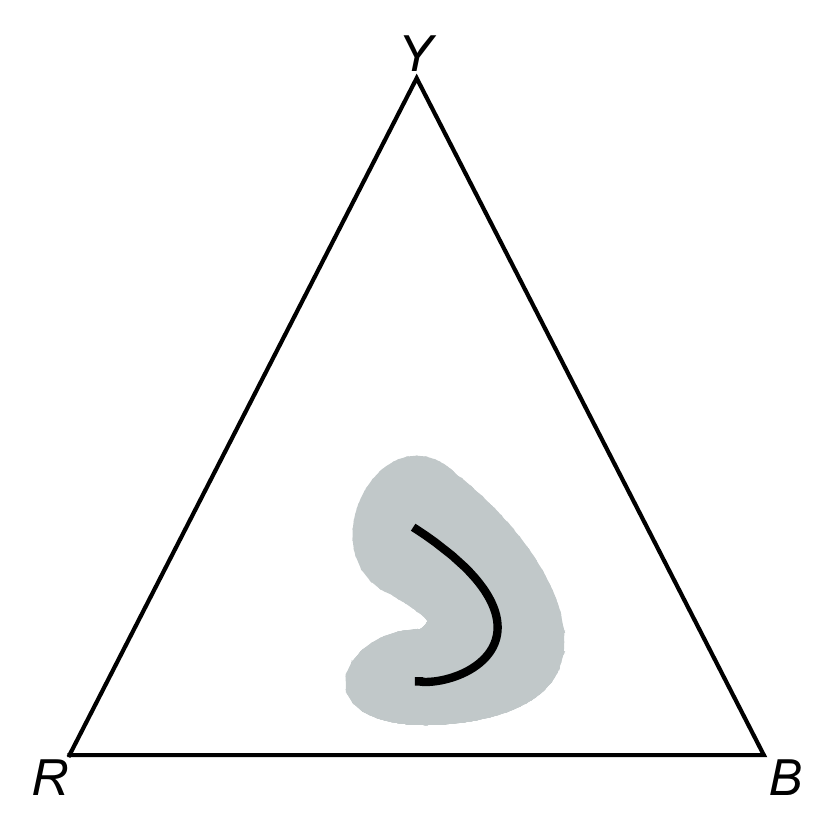}};
\end{tikzpicture}
\vspace*{0mm}
\caption{The black curves show equilibrium correspondences for logit-QRE (left) and level-$k$ model (right) for game $g_1$. The grey areas show choice profiles with a goodness-of-fit $G$ test value below the critical threshold when $\alpha=0.05$ and $N=120$.}\label{fig:logitSet}
\end{center}
\vspace*{-6mm}
\end{figure}

Even with a pre-analyses plan there is a \textit{set} of possible outcomes that results in acceptance (or, rather, non-rejection) of the model. To see this, reconsider game $g_1$ for which the logit-QRE correspondence is shown by the black curve in the left panel of Figure \ref{fig:logitSet}. The grey area that encloses it corresponds to the choice profiles for which a goodness-of-fit test ($G$ test) yields a value below the critical threshold (based on confidence level $\alpha=0.05$ and $N=120$ observations). In other words, if the observed average choice falls anywhere in the grey set then logit-QRE is not rejected.

To summarize, evaluation of QRE involves picking the best option from an implicitly defined set of predictions. To avoid overstating QRE's predictive success the size of this set should be accounted for. This requires computing all regular QRE, as in Figure \ref{fig:qre}. Or, if a parametric form such as logit-QRE has been preregistered, computing all choice profiles that survive a goodness-of-fit test, as in Figure \ref{fig:logitSet}.

The same critique applies to level-$k$. When allowing for general level distributions, level-$k$ typically produces a set of predictions. If one commits to the Poisson distribution then the equilibrium correspondence is one dimensional, but there will be an enclosing set of choice profiles that result in non-rejection of the model. The grey area in the right panel of Figure \ref{fig:logitSet} shows this set for game $g_1$.

A set-valued theory offers a more transparent solution by making explicit the set of predicted choices that the observed data are compared to.

\subsection{$M$ Equilibrium}

\citeauthor{GoereeLouis2021}' (\citeyear{GoereeLouis2021}) \textit{$M$ equilibrium} consists of a pair of choice and belief \textit{sets} $(M^c,M^b)$. Choices are monotone, i.e. options with higher expected payoffs are more likely chosen, and beliefs are consequentially unbiased, i.e. they imply the same choice frequencies as observed choices do. \citeauthor{GoereeLouis2021} show that $M^c$ contains all regular QRE and that each element in $M^b\supseteq M^c$, where the inclusion is typically strict, is a consequentially unbiased belief that supports any of the choices in $M^c$.

A major drawback is that the $M$-equilibrium choice sets can be unrealistically large even in simple games. Consider, for instance, a symmetric $2\times 2$ game in which strategy $A$ pays \$10 and strategy $B$ pays nothing. Presumably, observed play will be close to the Nash equilibrium in which both players choose $A$ with probability one. The $M$-equilibrium choice set, however, contains all profiles in which both players choose $A$ more likely, i.e. with probability one-half or more. The reason is that the $M$-equilibrium choice set contains all regular QRE, including, for instance, the logit QRE with $\lambda=0$, which corresponds to random behavior.

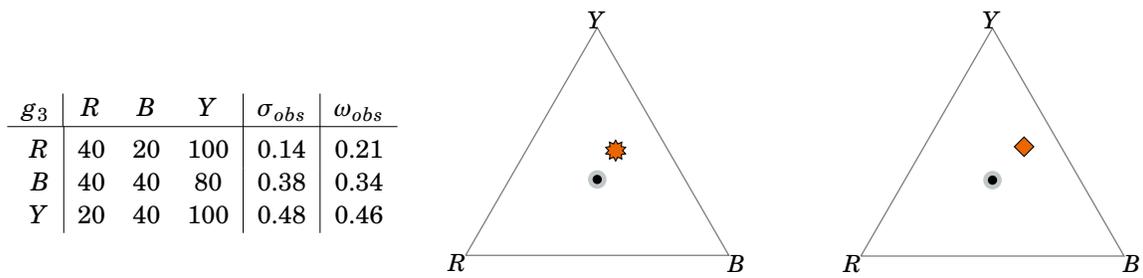
\begin{figure}[t]
\begin{center}

\begin{tikzpicture}[scale=3.5]
\draw[line width=.5pt,gray] (0,0) -- (1,0) -- (1/2,1/2*3^.5) -- (0,0);
\node[scale=0.8pt] at (-0.03,-0.03) {$R$};
\node[scale=0.8pt] at (1.03,-0.03) {$B$};
\node[scale=0.8pt] at (1/2,1/2*3^.5+0.03) {$Y$};

\fill[mondrianGrey] (1/2,1/6*3^.5) circle(1pt); 
\fill[black] (1/2,1/6*3^.5) circle(.5pt); 

\node[star,star points=10] at (0.57, 0.398372) [draw,scale=.5,fill=mondrianOrange]{};

\node at (2,0.43) {
\begin{tikzpicture}[scale=3.5]
\draw[line width=.5pt,gray] (0,0) -- (1,0) -- (1/2,1/2*3^.5) -- (0,0);
\node[scale=0.8pt] at (-0.03,-0.03) {$R$};
\node[scale=0.8pt] at (1.03,-0.03) {$B$};
\node[scale=0.8pt] at (1/2,1/2*3^.5+0.03) {$Y$};

\fill[mondrianGrey] (1/2,1/6*3^.5) circle(1pt); 
\fill[black] (1/2,1/6*3^.5) circle(.5pt); 

\node at (0.62, 0.415692) [draw,scale=.5,diamond,fill=mondrianOrange]{};

\end{tikzpicture}};

\node[scale=0.85] at (-1,0.35) {
\begin{tabular}{c|ccc|c|c}
$g_{3}$ & $R$ & $B$ & $Y$ &$\sigma_{obs}$ & $\omega_{obs}$ \\ \cline{1-6}
\multicolumn{1}{r}{\rule{0pt}{5mm}$R$} & \multicolumn{1}{|c}{40} & \multicolumn{1}{c}{20} & \multicolumn{1}{c|}{100} & 0.14 & 0.21 \\
\multicolumn{1}{r}{$B$} & \multicolumn{1}{|c}{40} & \multicolumn{1}{c}{40} & \multicolumn{1}{c|}{80} & 0.38 & 0.34 \\
\multicolumn{1}{r}{$Y$} & \multicolumn{1}{|c}{20} & \multicolumn{1}{c}{40} & \multicolumn{1}{c|}{100} & 0.48 & 0.46 \\
\end{tabular}};

\end{tikzpicture}
\end{center}
\vspace*{-6mm}
\caption{The left panel shows a symmetric $3\times 3$ game labeled $g_3$. In the belief simplex (middle panel), the grey disk shows the $M$-equilibrium belief set and the orange star the average observed belief. In the choice simplex (right panel), the grey disk shows the $M$-equilibrium choice set and the orange diamond the average observed choice.}\label{fig:windmill}
\vspace*{-2mm}
\end{figure}

Another drawback is that $M$-equilibrium beliefs are required to satisfy monotonicity, i.e. they should imply the same entire ranking of others' expected payoffs as observed choices do. This means that players have to step in others' shoes and calculate expected payoffs for all of their options, which seems unrealistic and restrictive. To illustrate, consider game $g_3$ in the left panel of Figure \ref{fig:windmill}. For this game the set of $M$-equilibrium beliefs consists of a single point. To see this, note that any of the six possible rankings of beliefs, e.g. $\omega_R>\omega_Y>\omega_B$, yields expected payoffs that are ranked differently, e.g. $\pi_R>\pi_B>\pi_Y$. Hence, $M$ equilibrium predicts that the simplex' centroid is the unique belief held by all players. And the only choice profile consistent with this belief is the simplex' centroid itself.\footnote{Together with the previous paragraph this shows that $M$-equilibrium choice sets can be too large and too small.}

In Figure \ref{fig:windmill}, the grey disks at the simplex' centroid show the $M$-equilibrium belief set (middle panel) and the $M$-equilibrium choice set (right panel). The black dots indicate that the simplex' centroid is also the unique prediction for any regular QRE. In the middle panel, the orange star shows the average observed belief and the orange diamond in the right panel shows the average observed choice.  Both the average belief and choice differ significantly from the simplex' centroid ($p<0.005$ and $p<0.001$ respectively). Moreover, they are consistent in that the most frequent choice, $Y$, is the best option given the observed beliefs, $\omega_{obs}$. This is a weaker condition than requiring beliefs to produce the correct ranking of expected payoffs for all options, including inferior ones. It is readily verified that, given $\omega_{obs}$, option $B$ has a lower expected payoff than $R$, yet $B$ is chosen more often than $R$.

\subsection{A Synthesis}

The above results underline that beliefs play an important role in determining robust choices, as first pointed out by \cite{Selten1975}. However:
\begin{itemize}\addtolength{\itemsep}{-2.5mm}
\vspace*{-2mm}
\item[--] robustness should be defined in terms of random trembles not in terms of \textit{a sequence} of trembles as in \citeauthor{Selten1975}'s notion of perfectness (see Figure \ref{fig:ellipse});
\item[--] beliefs cannot be assumed to form a hierarchy of best replies to the simplex' centroid as in level-$k$ (see Figure \ref{fig:levelk});
\item[--] beliefs cannot be anchored on Nash equilibrium as sizeable trembles, not just infinitesimal ones, can occur (see top panel of Figure \ref{fig:qre});
\item[--] beliefs do not follow from a rational-expectations fixed-point condition as in QRE (see bottom panel of Figure \ref{fig:qre});
\item[--] beliefs cannot be expected to induce the correct ranking of all inferior options as in $M$ equilibrium (see Figure \ref{fig:windmill});
\item[--] trembles can be sizeable depending on the game's ``complexity,'' i.e. how difficult it is to identify the best reply (cf. Figure \ref{fig:ellipse} and the top panel of Figure \ref{fig:qre}).
\end{itemize}
Finally, a theory's predictive power can only be properly assessed if both its accuracy and precision are transparent and verifiable. For theories that select a best-fitting model from a large set of models this requires registration of a pre-analyses plan. And even then there is a set of predictions that do not refute the model \mbox{(see Figure \ref{fig:logitSet})}. Set-valued theories offer an alternative by explicitly modeling the sets that the observed data are compared to. We adopt the latter route and synthesize the above insights into four desiderata for an empirically-relevant game theory.
\begin{itemize}\addtolength{\itemsep}{-2.5mm}
\vspace*{0mm}
\item[D1.] The theory is formulated in terms of a pair of choice and belief sets to allow for random deviations in choices and beliefs.
\item[D2.] Choices are consistent with beliefs in that the best option with the highest expected payoff is most frequently chosen.
\item[D3.] Beliefs are consistent with choices in that they imply the same best option as observed choices do.
\item[D4.] The sizes of the choice sets are disciplined by a parameter that facilitates the transparent and optimal trade off between accuracy and precision.
\end{itemize}
The next section presents a theory that satisfies these desiderata.

\section{$S$ Equilibrium}
\label{sec:S}

A finite normal-form game $G$ is a tuple $(N,\{X_{i},\Pi_{i}\}_{i\,\in\,N})$, with $N=\{1,\ldots,n\}$ the set of players, $X_{i}=\{x_{i1},\ldots,x_{iK_{i}}\}$ the set of pure strategies for player $i$, and $\Pi_{i}:X\rightarrow\field{R}$, where $X=\prod_{i=1}^nX_i$, player $i$'s payoff function. Let $\Sigma_{i}$ denote the set of probability distributions over $X_i$ and let $\Sigma=\prod_{i=1}^n\Sigma_i$. Let $\Omega_i=\prod_{j\neq i}\Sigma_i$ denote player $i$'s set of (independent) beliefs and let $\Omega=\prod_{i=1}^n\Omega_i$. We extend player $i$'s payoff function over $\Sigma_i\times\Omega_i$ as follows: for $(\sigma_i,\omega_i)\in\Sigma_i\times\Omega_i$, player $i$'s expected payoff is $\sum_{k=1}^{K_i}\sigma_{ik}\pi_{ik}(\omega_i)$ with $\sigma_{ik}$ the probability player $i$ chooses strategy $x_{ik}$ and $\pi_{ik}(\omega_i)=\sum_{x_{-i}\in X_{-i}}p_i(x_{-i})\Pi_{i}(x_{ik},x_{-i})$, with $p_i(x_{-i})=\prod_{j\neq i}\omega_{ij}(x_j)$, the expected payoff associated with $x_{ik}$. Let $\pi_i(\sigma_{-i})$ denote the vector of expected payoffs when player $i$'s beliefs are correct, i.e. $\omega_i=\sigma_{-i}$. Finally, $S^c_{int}$ and $S^b_{int}$ denote the relative interiors of $S^c\subseteq\Sigma$ and $S^b\subseteq\Omega$ respectively.
\begin{definition}\label{def:S}
For $\varepsilon\in(0,1)$, an $\boldsymbol{S(\varepsilon)}$ \textbf{Equilibrium} of $G$ is a maximal and closed set $S(\varepsilon)=S^c(\varepsilon)\times S^b(\varepsilon)\subseteq\Sigma\times\Omega$ such that, for $1\leq j,k\leq K_i$ and $i\in N$,
\begin{equation}\label{Sdef}
\pi_{ij}(\sigma_{-i})\,<\,\max\nolimits_{k}\pi_{ik}(\sigma_{-i})\,\Longleftrightarrow\,\pi_{ij}(\omega_{i})\,<\,\max\nolimits_{k}\pi_{ik}(\omega_{i})\,\Longrightarrow\,\sigma_{ij}\,<\,\varepsilon\,\max\nolimits_{k}\sigma_{ik}
\end{equation}
for all $\sigma\in S^c_{int}(\varepsilon)$ and $\omega\in S^b_{int}(\varepsilon)$. An $S(\varepsilon)$ equilibrium is \textbf{colorable} if $\Longrightarrow$ in \eqref{Sdef} is sharpened to $\Longleftrightarrow$. Let $\mathcal{S}_\varepsilon(G)$ denote the set of all $S(\varepsilon)$-equilibria of $G$. $S(\varepsilon)\in\mathcal{S}_\varepsilon(G)$ is \textbf{robust} if $\dim(S^c(\varepsilon))\geq\dim(\tilde{S}^c(\varepsilon))$ for all $\tilde{S}(\varepsilon)\in\mathcal{S}_\varepsilon(G)$.
\end{definition}
\begin{remark}{\em Alternatively, we could require $\pi_{ij}(\sigma_{-i})<\max\nolimits_{k}\pi_{ik}(\sigma_{-i})\Rightarrow\sigma_{ij}\,<\,\varepsilon/K_i$ for $\varepsilon\in(0,1)$ in line with \citeauthor{Selten1975}'s (\citeyear{Selten1975}) original definition. For small $\varepsilon$ the resulting sets would be similar, but not so for larger $\varepsilon$. Consider a symmetric $3\times 3$ game like the ones of the previous section and suppose $\pi_R>\pi_B>\pi_Y$. Then $\sigma=(\deel{3}{5},\deel{2}{5},0)$ satisfies \eqref{Sdef} for $\varepsilon\in(\deel{2}{3},1)$ but it does not satisfy the alternative requirement for any $\varepsilon\in(0,1)$.}
\end{remark}
\begin{remark}{\em Equation \eqref{Sdef} does not restrict the choice probabilities of the strategies that tie for the highest expected payoff. \textit{Colorability} means these strategies are equally likely. When there is a unique highest expected payoff in \eqref{Sdef} it allows us to ``color'' the $S(\varepsilon)$-equilibrium sets with the strategy chosen most frequently.}
\end{remark}
\begin{remark}{\em \textit{Robustness} reflects the idea that lower-dimensional choice sets are empirically irrelevant. Obviously, an $S(\varepsilon)$ equilibrium is robust if its choice set is full-dimensional. However, full-dimensionality is not necessary for robustness. For example, for a matching-pennies game the $S(\varepsilon)$-equilibrium choice and belief sets consist of a single profile for any $\varepsilon\in(0,1)$: the unique Nash equilibrium in which both players randomize uniformly. Since payoff ties are matched by ties in the choice probabilities, this lower-dimensional $S$ equilibrium is colorable. It is also robust.}
\end{remark}
\begin{remark}{\em Since \eqref{Sdef} has to hold for all $\sigma\in S^c_{int}(\varepsilon)$ and $\omega\in S^b_{int}(\varepsilon)$, $S(\varepsilon)$ equilibria have a Cartesian product structure, i.e. $S(\varepsilon)=S^c(\varepsilon)\times S^b(\varepsilon)$. Except in two-player games, $S(\varepsilon)$-equilibrium sets are generally not the product of individual players' $S_i(\varepsilon)$-equilibrium sets.  For a two-player game, if $(\sigma_1,\sigma_2)$ and $(\sigma'_1,\sigma'_2)$ belong to some $S^c(\varepsilon)$ then $\sigma_1$ and $\sigma'_1$ generate the same best option for player 2, and $\sigma_2$ and $\sigma'_2$ generate the same best option for player 1. Hence, $(\sigma_1,\sigma'_2)$ and $(\sigma'_1,\sigma_2)$ also belong to $S^c(\varepsilon)$. In a three-player game, however, if $(\sigma_1,\sigma_2,\sigma_3)$ and $(\sigma'_1,\sigma'_2,\sigma'_3)$ belong to some $S^c(\varepsilon)$ then, for instance, $(\sigma_1,\sigma_2,\sigma'_3)$ does \textit{not} necessarily belong to $S^c(\varepsilon)$ as player 1's expected payoffs involve the product of the other two players' choice probabilities.}
\end{remark}
\begin{proposition}\label{prop:exist}
$\mathcal{S}_\varepsilon(G)$ is non-empty for any finite normal-form game $G$ and $\varepsilon\in(0,1)$. If $S(\varepsilon)\in\mathcal{S}_\varepsilon(G)$ then for $\tilde{\varepsilon}\geq\varepsilon$ there exists $\tilde{S}(\tilde{\varepsilon})\in\mathcal{S}_{\tilde{\varepsilon}}(G)$ such that $S(\varepsilon)\subseteq\tilde{S}(\tilde{\varepsilon})$.
\end{proposition}
This existence result follows since any Nash-equilibrium profile $\sigma$ satisfies
\begin{displaymath}
\pi_{ij}(\sigma_{-i})\,<\,\max\nolimits_{k}\pi_{ik}(\sigma_{-i})\,\Longrightarrow\,\sigma_{ij}\,=\,0
\end{displaymath}
and, hence, $\sigma\in S^c(\varepsilon)$ for any $\varepsilon\in(0,1)$. Moreover, any Nash-equilibrium profile $\sigma$ can be supported by correct beliefs, i.e. $\omega_i=\sigma_{-i}$ for $i\in N$. The fact that $S(\varepsilon)$ equilibrium sets are increasing in $\varepsilon$ is a direct consequence of \eqref{Sdef}.
\begin{example}{\em To illustrate the construction of $S(\varepsilon)$-equilibrium choice and belief sets consider Selten's Chain-Store Paradox in Table \ref{chainstore}. Let $p$ denote the probability with which the incumbent fights and $q$ the probability with which the entrant stays out. Furthermore, let $\nu$ denote the incumbent's belief that the entrant stays out and $\omega$ the entrant's belief that the incumbent fights. The choice and belief sets can be summarized by unit squares consisting of the pairs $(p,q)$ and $(\nu,\omega)$, see Figure~\ref{fig:chainstore}.

For any non-degenerate belief the incumbent is worse off fighting so $p<\varepsilon(1-p)$, or, equivalently, $p<\varepsilon/(1+\varepsilon)$. When $\varepsilon<\hf$ we thus have $p<\deel{1}{3}$, which implies that the entrant is better off entering. Hence, $q<\varepsilon(1-q)$, or, equivalently, $q<\varepsilon/(1+\varepsilon)$. To summarize, when $\varepsilon<\hf$ there is only one $S(\varepsilon)$ equilibrium in which $p<\varepsilon/(1+\varepsilon)$ and $q<\varepsilon/(1+\varepsilon)$. The yellow area in the top-left panel of Figure~\ref{fig:chainstore} shows the $S(\varepsilon)$-equilibrium choice set for $\varepsilon=\deel{1}{3}$. The yellow area in the top-right panel shows the corresponding $S(\varepsilon)$-equilibrium belief set. This set consists of all beliefs that imply the same ordering of expected payoff as the choices in the top-left panel do.

When $\varepsilon>\hf$, the incumbent's fight probability, which again satisfies $p<\varepsilon/(1+\varepsilon)$, may be larger than $\deel{1}{3}$. If so then the entrant is better off staying out. Hence, $1-q<\varepsilon q$, or, equivalently, $q>1/(1+\varepsilon)$. This possibility gives rise to a second $S(\varepsilon)$ equilibrium, which is indicated by the blue area in the bottom-left panel of Figure~\ref{fig:chainstore} for the case $\varepsilon=\deel{2}{3}$. The blue area in the bottom-right panel shows the corresponding belief set.

\begin{figure}[t]
\begin{center}
\begin{tikzpicture}

\node at (0,0) {
\begin{tikzpicture}[scale=4]
\draw[line width=.5pt,gray] (0,0) -- (1,0) -- (1,1) -- (0,1) -- (0,0);
\filldraw[fill=mondrianYellow,draw=mondrianYellow,opacity=1] (0,0) -- (1/4,0) -- (1/4,1/4) -- (0,1/4) -- (0,0);
\node[scale=0.6pt] at (-0.03,-0.06) {$0$};
\node[scale=0.6pt] at (-0.03,1) {$1$};
\node[scale=0.6pt] at (1,-0.06) {$1$};
\node[scale=0.8] at (0.8,-0.06) {$p$};
\node[scale=0.8] at (-0.08,0.8) {$q$};
\node[scale=0.8] at (1/4,-0.08) {$\deel{\varepsilon}{1+\varepsilon}$};
\node[scale=0.8] at (-0.1,1/4) {$\deel{\varepsilon}{1+\varepsilon}$};
\end{tikzpicture}};

\node at (5.5,0) {
\begin{tikzpicture}[scale=4]
\draw[line width=.5pt,gray] (0,0) -- (1,0) -- (1,1) -- (0,1) -- (0,0);
\filldraw[fill=mondrianYellow,draw=mondrianYellow,opacity=1] (0,0) -- (1/3,0) -- (1/3,1) -- (0,1) -- (0,0);
\node[scale=0.6pt] at (-0.03,-0.06) {$0$};
\node[scale=0.6pt] at (-0.03,1) {$1$};
\node[scale=0.6pt] at (1,-0.06) {$1$};
\node[scale=0.8] at (0.8,-0.06) {$\omega$};
\node[scale=0.8] at (-0.08,0.8) {$\nu$};
\node[scale=0.8] at (1/3,-0.08) {$\deel{1}{3}$};
\end{tikzpicture}};

\node at (0,-5.5) {
\begin{tikzpicture}[scale=4]
\draw[line width=.5pt,gray] (0,0) -- (1,0) -- (1,1) -- (0,1) -- (0,0);
\filldraw[fill=mondrianYellow,draw=mondrianYellow,opacity=1] (0,0) -- (1/3,0) -- (1/3,2/5) -- (0,2/5) -- (0,0);
\filldraw[fill=mondrianBlue,draw=mondrianBlue,opacity=1] (1/3,3/5) -- (1/3,1) -- (2/5,1) -- (2/5,3/5) -- (1/3,3/5);
\node[scale=0.6pt] at (-0.03,-0.06) {$0$};
\node[scale=0.6pt] at (-0.03,1) {$1$};
\node[scale=0.6pt] at (1,-0.06) {$1$};
\node[scale=0.8] at (0.8,-0.06) {$p$};
\node[scale=0.8] at (-0.08,0.8) {$q$};
\draw[line width=1pt] (1/3,0) -- (1/3,1);
\node[scale=0.8] at (1/3,-0.08) {$\deel{1}{3}$};
\node[scale=0.8] at (2/5+0.03,-0.08) {$\deel{\varepsilon}{1+\varepsilon}$};
\node[scale=0.8] at (-0.1,2/5) {$\deel{\varepsilon}{1+\varepsilon}$};
\node[scale=0.8] at (-0.1,3/5) {$\deel{1}{1+\varepsilon}$};
\end{tikzpicture}};

\node at (5.5,-5.5) {
\begin{tikzpicture}[scale=4]
\draw[line width=.5pt,gray] (0,0) -- (1,0) -- (1,1) -- (0,1) -- (0,0);
\filldraw[fill=mondrianYellow,draw=mondrianYellow,opacity=1] (0,0) -- (1/3,0) -- (1/3,1) -- (0,1) -- (0,0);
\filldraw[fill=mondrianBlue,draw=mondrianBlue,opacity=1] (1/3,0) -- (1/3,1) -- (1,1) -- (1,0) -- (1/3,0);
\node[scale=0.6pt] at (-0.03,-0.06) {$0$};
\node[scale=0.6pt] at (-0.03,1) {$1$};
\node[scale=0.6pt] at (1,-0.06) {$1$};
\node[scale=0.8] at (0.8,-0.06) {$\omega$};
\node[scale=0.8] at (-0.08,0.8) {$\nu$};
\node[scale=0.8] at (1/3,-0.08) {$\deel{1}{3}$};
\draw[line width=1pt] (1/3,0) -- (1/3,1);
\end{tikzpicture}};

\end{tikzpicture}
\vspace*{0mm}
\caption{$S(\varepsilon)$-equilibrium choice sets (left) and belief sets (right) for Selten's Chain-Store Paradox in Table \ref{chainstore}. In the top panels $\varepsilon=1/3$, which results in a unique $S(\varepsilon)$-equilibrium choice set that is full-dimensional, robust, and colorable (see the yellow area).  In the bottom panels $\varepsilon=2/3$, which results in three $S(\varepsilon)$-equilibrium choice sets. Two are full-dimensional, robust, and colorable (see the yellow and blue areas). A third, indicated by the black vertical line, is lower-dimensional and non-colorable.}\label{fig:chainstore}
\end{center}
\vspace*{-5mm}
\end{figure}
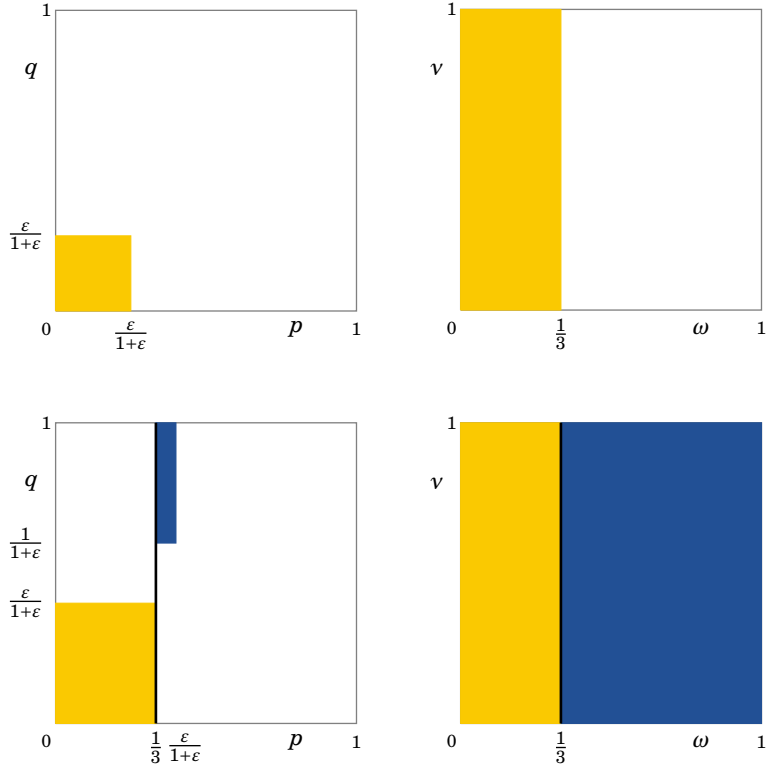

The black vertical line in the bottom panels corresponds to a lower-dimensional $S(\varepsilon)$ equilibrium, which requires $\varepsilon\geq\hf$. It entails $p=\deel{1}{3}$ so that the entrant is indifferent and there is no condition on the entry probability, $q$.  This lower-dimensional $S(\varepsilon)$ equilibrium is not colorable since entrant's expected payoffs match, but the entrant's choice probabilities do not. This
lower-dimensional $S(\varepsilon)$ equilibrium is also not robust since at least one higher-dimensional $S(\varepsilon)$ equilibrium exists.\footnote{Another lower-dimensional, non-colorable, and non-robust $S$ equilibrium is the Nash-equilibrium component $q=1$ and $p\in[\deel{1}{3},1]$.}}$\hfill\blacksquare$
\end{example}

\subsection{Properties of $S$ Equilibrium}

$S$ equilibrium sets are determined by payoff inequalities and the $\varepsilon$ parameter, see Definition \ref{def:S}.
To derive an upper bound for the size of an $S(\varepsilon)$-equilibrium choice set, suppose player $i$'s first strategy is dominant. Player $i$'s $S(\varepsilon)$-equilibrium choice set consists of $\{\sigma_i\in\Sigma_i|\sigma_{ij}<\varepsilon\sigma_{i1}\mbox{ for }j>1\}$, and, hence, the largest possible relative measure is:\footnote{See the proof of Proposition \ref{gen1} in Appendix \ref{app:proofs}.}
\begin{equation}\label{maxArea}
  \overline{\mu}_i(\varepsilon)\,=\,\prod_{k\,=\,1}^{K_i-1} \frac{k\varepsilon}{1+k\varepsilon}
\end{equation}
which falls as $\varepsilon^{K_i-1}$ when $\varepsilon\downarrow 0$ and equals $1/K_i$ when $\varepsilon=1$.

Denote by $\Gamma_i=\field{R}^{|S|}$ the space of payoffs of player $i$ and let $\Gamma=\prod_{i=1}^n\Gamma_i$.
\begin{proposition}\label{gen1}
There exists a generic subset $\mathcal{G}\subset\Gamma$ such that for $G\in\mathcal{G}$:
\begin{itemize}\addtolength{\itemsep}{-2mm}
\vspace*{-2mm}
\item[(i)] A full-dimensional, robust, colorable $S(\varepsilon)$ equilibrium exists for some $\varepsilon\in(0,1)$;
\item[(ii)] The measure of each $S(\varepsilon)$-choice set is bounded by $\prod_{i\in N}\overline{\mu}_i(\varepsilon)$;
\item[(iii)] The measure of the union of $S(\varepsilon)$-choice sets is bounded by $\min_{i\in N}\overline{\mu}_i(\varepsilon)$;
\item[(iv)] An $S(\varepsilon)$-belief set may have full measure.
\end{itemize}
\end{proposition}
In generic games, full-dimensionality, robustness, and colorability all reflect that one strategy has \textit{strictly} higher expected payoffs.  In non-generic games, however, colorability does not imply full dimensionality as the matching-pennies example of Remark 3 shows. Nor does full-dimensionality imply colorability, as we demonstrate below (see game $g_7$ in Figure \ref{fig:beliefSets2}). Since colorable or full-dimensional sets may not exist in general, we apply robustness as a refinement criterium for $S$ equilibria. From an empirical viewpoint, only $S$-equilibria of maximum dimension matter.

\subsection{Relation to Other Concepts}
\label{sec:Spot}

\cite{Selten1975} modeled trembles by restricting strategy sets to interior simplices.
Consider, for $i\in N$, the restricted strategy sets $\mathscr{X}_i=\{\mu_{ik}\}_{k=1}^{K_i}$ where
\begin{equation}\label{triangle}
  \mu_{ik}(\boldsymbol{\varepsilon}_i)\,=\,\bigl(\varepsilon_{i1},\ldots,\varepsilon_{ik-1},1-\sum_{j\,\neq\,k}\varepsilon_{ij},\varepsilon_{ik+1},\ldots,\varepsilon_{iK_i}\bigr)
\end{equation}
and the $\varepsilon_{ik}\geq 0$ are such that the $k$-th entry of $\mu_{ik}$ is largest: $\varepsilon_{i\ell}\leq 1-\sum_{j\neq k}\varepsilon_{ij}$ for $\ell\neq k$. Let $\boldsymbol{\varepsilon}$ denote the concatenation of $(\varepsilon_{i1},\ldots,\varepsilon_{iK_i})$ for $i\in N$. The convex hull $\mathscr{P}_i(\boldsymbol{\varepsilon})=\text{co}(\mathscr{X}_i)$ is an interior simplex with faces parallel to $\Sigma_i$. We call a Nash equilibrium on $\mathscr{P}(\boldsymbol{\varepsilon})=\prod_{i\in N}\mathscr{P}_i(\boldsymbol{\varepsilon})$ an \textit{$\bm{\varepsilon}$-perfect equilibrium} of $G$.

\begin{figure}[t]
\vspace*{-8mm}
\begin{center}
\begin{tikzpicture}
\node at (0,0) {\includegraphics[scale=0.3]{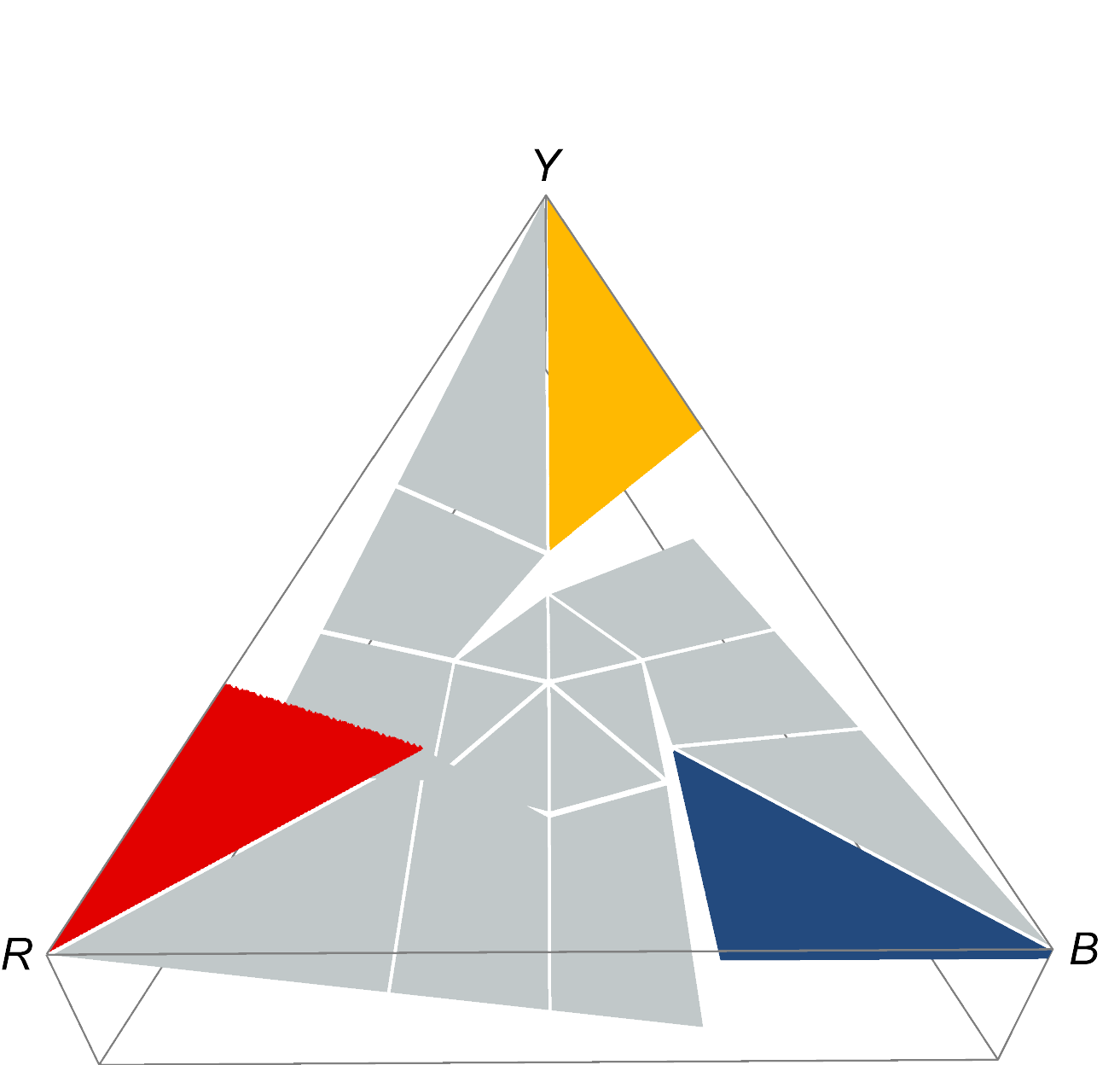}};
\node at (8,0) {\includegraphics[scale=0.3]{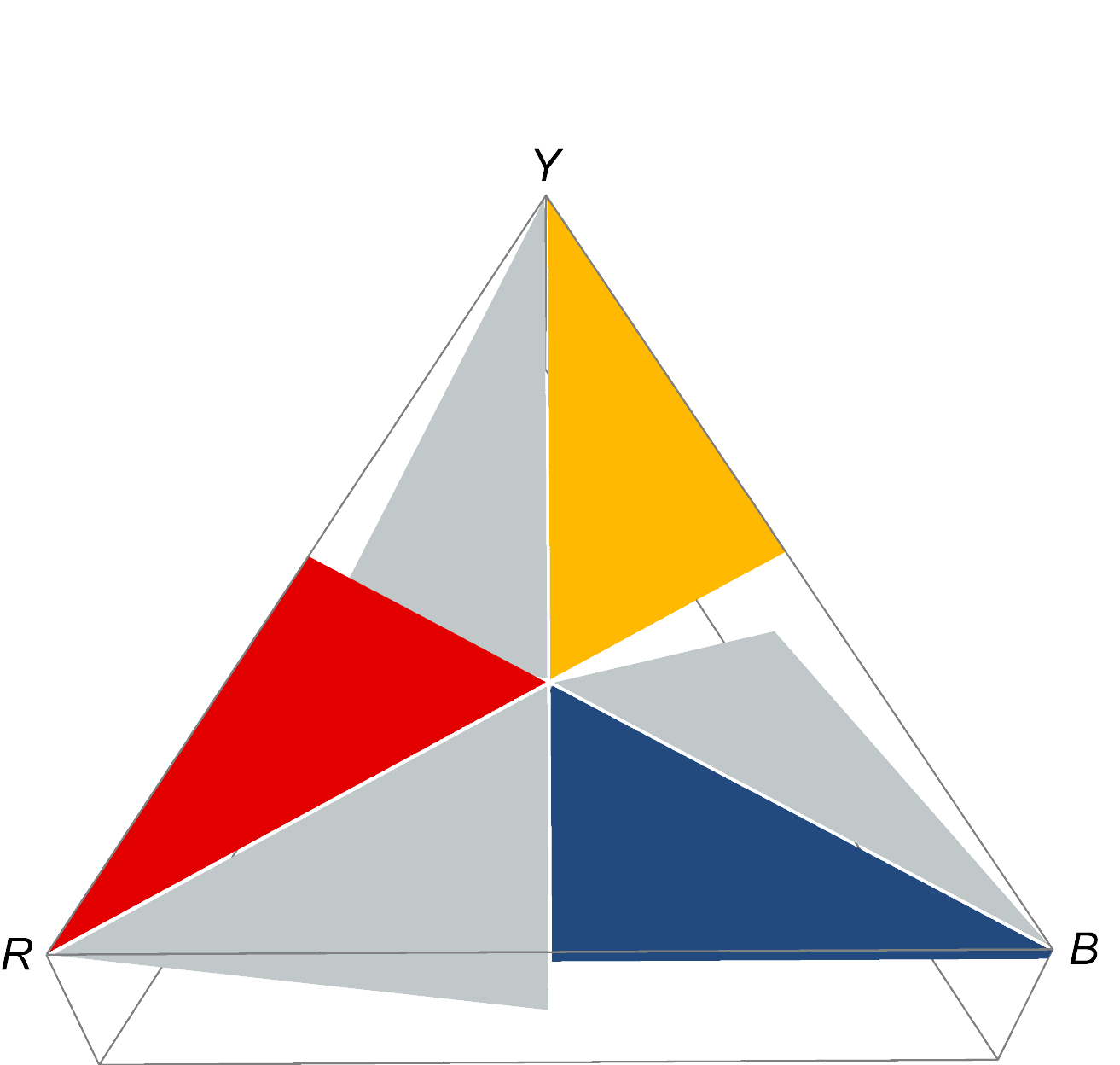}};
\end{tikzpicture}
\vspace*{1mm}
\caption{Graphs of the potential $Y_\varepsilon(\sigma)$ for game $g_3$ when $\varepsilon=1/2$ (left) and $\varepsilon=1$ (right). In both panels the potential vanishes on three robust and full-dimensional sets that can be colored. In the left panel, the relative measure of any $S$-equilibrium set is $1/12$ and in the right panel it is $1/6$.}\label{fig:Spotential}
\end{center}
\vspace*{-5mm}
\end{figure}

We will demonstrate that the union of $S(\varepsilon)$ choice sets is equal to the set of all $\boldsymbol{\varepsilon}$-perfect equilibria. First, we discuss how to compute the $S(\varepsilon)$-equilibrium choice sets.  Since expected payoffs are polynomial in players' beliefs, Definition \ref{def:S} implies that the $S(\varepsilon)$-equilibrium choice sets are defined by finitely many polynomial inequalities. In other words, they fit the definition of \textit{semi-algebraic sets}. This is reassuring as it implies that, in principle, the $S(\varepsilon)$-equilibrium choice and belief sets can be computed using a finite algorithm. Here we show they are, in fact, \textit{simple} to compute because they are the roots of a piecewise polynomial function -- the $S(\varepsilon)$ potential.
\begin{definition}\label{def:potS}
For $\varepsilon\in(0,1)$, the $S(\varepsilon)$ potential, $Y_\varepsilon:\Sigma\rightarrow\field{R}_{\leq 0}$, is given by
\begin{equation}\label{eq:potS}
  Y_\varepsilon(\sigma)\,=\,\sum_{i\,\in\,N}\,\,\,\min_{k\,\in\,\text{\em supp}_\varepsilon(\sigma_{i})}\,(\pi_{ik}(\sigma_{-i}))-\max_{k}(\pi_{ik}(\sigma_{-i}))
\end{equation}
where $\text{\em supp}_\varepsilon(\sigma_i)=\{\ell\,|\,\sigma_{i\ell}\geq\varepsilon\max_k(\sigma_{ik})\}$ is the support of $\sigma_i$.
\end{definition}
This definition assumes expected payoffs are positive. If not, then we add the same constant to all the $\pi_{ik}(\sigma_{-i})$ for $i\in N$, $1\leq k\leq K_i$ to ensure they become positive.
\begin{example}{\em Consider game $g_3$ in Figure \ref{fig:windmill} for which all QRE are located at the simplex' centroid. The left panel of Figure \ref{fig:Spotential} shows the $S(\varepsilon)$ potential $Y_\varepsilon(\sigma)$ for $\varepsilon=\deel{1}{2}$ and the right panel shows $Y_\varepsilon(\sigma)$ for $\varepsilon=1$. The $S$ potential ``kinks'' or has a discontinuity along the simplex' diagonals or when the probability of an inferior option equals $\varepsilon$ times that of the best option. The potential's roots form three full-dimensional choice sets that are colorable and robust.  Their sizes grow with $\varepsilon$, see Proposition \ref{prop:exist}.} $\hfill\blacksquare$
\end{example}
\begin{proposition}\label{rootS}
For any normal-form game $G$:
\begin{itemize}\addtolength{\itemsep}{-2mm}
\vspace*{-2mm}
\item[$(i)$] $\sigma\in S^c(\varepsilon)$ for $\varepsilon\in(0,1)$ if and only if $\sigma$ is a root, whence maximizer, of $Y_\varepsilon(\sigma)$;
\item[$(ii)$] $\sigma\,\in\,\bigcup_{\varepsilon\in(0,1)}\,S^c(\varepsilon)\,$ if and only if $\sigma$ is an $\boldsymbol{\varepsilon}$-perfect equilibrium;
\item[$(iii)$] $\sigma\,\not\in\,\bigcup_{\varepsilon\in(0,1)}\,S^c(\varepsilon)\,$ implies $\sigma$ is not a regular QRE.
\end{itemize}
\end{proposition}
The first property shows that the $S(\varepsilon)$-equilibrium choice sets can readily be computed using a semi-algebraic potential. The second property shows that when $\varepsilon$ limits to one the $S(\varepsilon)$-equilibrium choice sets nest all of \citeauthor{Selten1975}'s (\citeyear{Selten1975}) $\boldsymbol{\varepsilon}$-perfect equilibria. The final property implies that if the choice data cannot be explained by any $S(\varepsilon)$ equilibrium then they cannot be explained by any $\mathcal{R}$-QRE either. This means that $S$ equilibrium can be more accurate than regular QRE but it does \textit{not} mean that it is less precise because the size of its choice set depends on $\varepsilon$ (while the size of the set of all regular QRE is fixed).
The bound in part (ii) of Proposition \ref{gen1} implies that for small $\varepsilon$ the size of any $S(\varepsilon)$-equilibrium choice set falls as
\begin{displaymath}
  \mu(\varepsilon)\,=\,\prod_{i\,=\,1}^N\mu_i(\varepsilon)\,\leq\,\prod_{i\,=\,1}^N\overline{\mu}_i(\varepsilon)\,\sim\,\varepsilon^{\dim(\Sigma)}
\end{displaymath}
For instance, for the chain-store paradox in Table \ref{chainstore} we have $\dim(\Sigma)=2$. The yellow area in the upper-left panel of Figure \ref{fig:chainstore} equals $\varepsilon^2/(1+\varepsilon)^2$ and falls as $\varepsilon^2$.

\section{Experimental Test of $S$ Equilibrium}
\label{sec:Exp}

We report experimental results for the three $3\times 3\times 3$ games in \mbox{Figures \ref{fig:ellipse}--\ref{fig:levelk}} and the ten $3\times 3$ games in Figures \ref{fig:beliefSets}--\ref{fig:beliefSets2}. These games were chosen to address several topics: the role of belief sets for equilibrium selection, the complexity of identifying a unique best reply, QRE's prediction that (almost) identical strategies are played (almost) equally often, and level-$k$'s comparative statics prediction that games with the same hierarchy of best replies yield the same outcomes.

Section \ref{sec:expDesign} details the experimental design and protocol. Section \ref{sec:expResults} analyzes observed choices. Section \ref{sec:areametrics} demonstrates how simple ``areametrics'' based on \citeauthor{selten1991}'s (\citeyear{selten1991}) measure of predictive of success can be used to determine $\varepsilon$ and compare $S$ equilibrium to QRE and level-$k$. Section \ref{sec:Structural} provides structural estimates based on standard likelihood techniques, which are used to determine the ``in-sample'' and ``out-of-sample'' fit of the various models. Section \ref{sec:beliefs} analyzes observed beliefs.

\subsection{Experimental Design and Protocol}
\label{sec:expDesign}

We recruited 141 subjects to participate in 16 sessions. We ran six laboratory sessions using z-Tree, see \cite{Fischbacher2007}, and ten online sessions using Zoom and z-Tree unleashed, see \cite{duch2020}. Each session contained eight or nine subjects, depending on whether the session employed a $3\times 3$ or $3\times 3\times 3$ game.

Each session started with a PowerPoint presentation of the experimental instructions that were read aloud. Then, in each round, subjects' screens displayed the three actions they could take and the corresponding payoffs. Subjects chose an action by selecting the appropriate row. After each round, we rematched participants using a perfect-stranger protocol in games $g_4$--$g_6$ (for a total of seven rounds) and a minimal-repeated-interactions protocol in all other games (for a total of fifteen periods).

We also elicited subjects' beliefs about the opponent’s choice in each round. This was incentivized using a generalization of a method proposed by \cite{wilson2018}, which is an implementation of \citeauthor{hossain2013}'s (\citeyear{hossain2013}) binarized scoring rule (BSR). BSR is incentive compatible for general risk-preferences and thus avoids issues of risk-aversion that plague other scoring rules. The method operationalizes BSR for binary-choice settings in a simple manner.

After each round, subjects were shown their opponents' choices, and the results of the belief elicitation task. They received 60 tokens for a correct guess in the belief elicitation task. To avoid hedging, their payoff in each round was randomly selected to be either their payoff from the game or their payoff from the belief elicitation task. At the end of the experiment, subjects were informed about their total earnings and paid. Participants received AU\$1 for every 80 tokens in games $g_4$--$g_6$ and for every 150 tokens in all other games, with an average payment of approximately AU\$27 including show-up fee (which included payment for some other unrelated tasks).

\subsection{Analyzing Observed Choices}
\label{sec:expResults}

We first apply a simple area-based measure to compare $S$ equilibrium with regular QRE and level-$k$. We next follow up with structural estimates.

\subsubsection{\citeauthor{selten1991}'s (\citeyear{selten1991}) Areametrics}
\label{sec:areametrics}

To estimate $\varepsilon$ we use the measure of predictive success that \cite{selten1991} developed for set-valued theories. This measure of predictive success (MPS) compares the accuracy of a prediction with its precision. The accuracy is given by the ``hit rate,'' which is the relative frequency of correct predictions. The precision equals the ``area size,'' i.e. the relative size of the predicted subset compared to the set of all possible outcomes.
\begin{equation}\label{MPS}
  \text{MPS}\,=\,\text{hit rate } - \text{ area size}
\end{equation}
\cite{selten1991} derives this measure based on plausible axioms and shows that it has desirable properties compared to alternative functional forms.\footnote{See also \cite{selten1982}.}  The MPS takes on values between $-1$ and $1$ with a more successful theory yielding a higher MPS. An example of a theory with zero MPS is one that predicts the entire outcome space. Another trivial theory is a point-valued theory that captures none of the data.

Table \ref{mps_table} displays the MPS results. $S$ equilibrium has a higher MPS than QRE in all but one game and a higher MPS than level-$k$ in ten out of thirteen games. Pooling all games yields an MPS for $S$ equilibrium that is 44\% higher than that of QRE and 73\% higher than that of level-$k$. $S$ equilibrium is as accurate as QRE (i.e. it has virtually the same hit rate), but is much more precise (i.e. it has a four times smaller area size). $S$ equilibrium is both more accurate and precise than level-$k$. Finally, unlike QRE and level-$k$, $S$ equilibrium consistently outperforms random predictions that result in a zero MPS.

\begin{table}[t]
		\setlength\tabcolsep{5pt}
		\begin{center}
		\begin{tabular}{c|c|ccc|ccc|ccc}
			\hline\hline
			\multirow{2}{*}{Game} & $\varepsilon$ & \multicolumn{3}{c|}{hit rate} & \multicolumn{3}{c|}{area size} & \multicolumn{3}{c}{MPS} \\
			& $S$ & $S$ & QRE & level-$k$ & $S$ & QRE & level-$k$ & $S$ & QRE & level-$k$ \\ \hline
			$G_1$ & 0.167 & 0.89 & 1.00 & 0.67 & 0.04 & 0.17 & 0.00 & 0.85 & 0.83 & 0.67\\
			$G_{2}$ & 0.364 & 0.44 & 0.22 & 0.78 & 0.11 & 0.17 & 0.33 & 0.33 & 0.05 & 0.44 \\
			$G_{3}$ & 0.750 & 0.78 & 0.83 & 0.94 & 0.26 & 0.17 & 0.33 & 0.52 & 0.67 & 0.61 \\
			$g_1$ & 0.875 & 0.38 & 0.25 & 0.50 & 0.24 & 0.16 & 0.33 & 0.13 & 0.09 & 0.17\\
			$g_2$ & 0.154 & 0.75 & 0.00 & 0.00 & 0.05 & 0.17 & 0.17 & 0.70 & -0.17 & -0.17 \\
			$g_3$ & 0.071 & 0.19 & 0.00 & 0.00 & 0.01 & 0.00 & 0.00 & 0.17 & 0.00 & 0.00\\
			$g_4$ & 0.167 & 0.88 & 0.88 & 0.62 & 0.04 & 0.17 & 0.00 & 0.83 & 0.71 & 0.62\\
			$g_5$ & 0.750 & 1.00 & 0.88 & 0.88 & 0.31 & 0.21 & 0.33 & 0.69 & 0.67 & 0.54\\
			$g_6$ & 0.167 & 0.88 & 1.00 & 0.62 & 0.04 & 0.17 & 0.00 & 0.84 & 0.83 & 0.62\\
			$g_7$ & 0.154 & 0.50 & 0.00 & 0.00 & 0.06 & 0.00 & 0.00 & 0.44 & 0.00 & 0.00\\
			$g_8$ & 0.154 & 0.75 & 0.88 & 0.12 & 0.03 & 0.17 & 0.00 & 0.72 & 0.71 & 0.12\\
			$g_{9}$& 0.250 & 0.62 & 0.81 & 0.25 & 0.20 & 0.67 & 0.00 & 0.42 & 0.14 & 0.25\\
			$g_{10}$ & 0.154 & 0.88 & 0.94 & 0.06 & 0.09 & 0.58 & 0.00 & 0.78 & 0.35 & 0.06\\
			  \hline
			Pooled & 0.167 & 0.58 & 0.59 & 0.42 & 0.05 & 0.21 & 0.12 & 0.52 & 0.38 & 0.30\\   \hline \hline
		\end{tabular}
    \end{center}
   	\vspace*{-5mm}
	\caption{Measure of predictive success for $S$ equilibrium, regular QRE, and level-$k$.}\label{mps_table}
	\vspace*{-2mm}
\end{table}

\begin{result}\label{mps_result}
$S$ equilibrium outperforms regular QRE and level-$k$ based on the measure of predictive success proposed by \cite{selten1991}.
\end{result}
\begin{support}
A paired Wilcoxon rank-sum test (with continuity correction) shows that the MPS for $S$ equilibrium is significantly greater than that of QRE ($p=0.004$) and of level-$k$ ($p=0.003$).
\end{support}

\subsubsection{Structural Estimation}
\label{sec:Structural}

Here we consider parametric models such as logit-QRE with a single precision parameter, $\lambda$, and level-$k$ in which levels follow a Poisson distribution determined by a single parameter, $\tau$. Parsimony is imposed so that the model's parameter can be used for out-of-sample predictions without overfitting. We also include two models that, in the tradition of the refinement literature, are defined in terms of Nash equilibria of games with restricted strategy sets.

First, Proposition \ref{rootS} shows that the union of $S$-equilibrium choice sets is equal to the set of all $\boldsymbol{\varepsilon}$-perfect equilibria \citep{Selten1975}. A one-dimensional curve in this set is obtained by restricting the $NK$-dimensional vector $\boldsymbol{\varepsilon}$ to a single parameter $\varepsilon$. Then the $\mu_{ik}(\varepsilon)$ are the $K_i$ permutations  of
\begin{equation}\label{eq:perfect}
  \frac{1}{1+(K_i-1)\varepsilon}\,\bigl(1,\varepsilon,\varepsilon,\ldots,\varepsilon\bigr)
\end{equation}
This one-dimensional model, which we refer to as the \textit{$\varepsilon$-perfect model}, is based on interior simplices centered around the simplex' centroid.

Second, \cite{GoereeLouis2021} have shown that the union of $M$-equilibrium choice sets is equal to the set of all $\boldsymbol{\varepsilon}$-proper equilibria \citep{Myerson1978}. A one-dimensional curve in this set is obtained by considering permutahedra with vertices $\mu_{ik}(\varepsilon)$ that are the $K_i!$ permutations of
\begin{equation}\label{eq:proper}
  \frac{1-\varepsilon}{1-\varepsilon^{K_i}}\,\bigl(1,\varepsilon,\varepsilon^2,\ldots,\varepsilon^{K_i-1}\bigr)
\end{equation}
This one-dimensional model, which we refer to as the \textit{$\varepsilon$-proper model}, is based on interior permutahedra centered around the simplex' centroid.

For both models, let $\mathscr{P}_i(\mu_i(\varepsilon))$ denote convex hull of the $\mu_{ik}(\varepsilon)$ and consider the better replies $B\!R_i^{\varepsilon}:\Sigma_{-i}\rightarrow\mathscr{P}_i(\mu_i(\varepsilon))$ defined by\footnote{We use the terminology ``better replies'' rather than best replies because the $B\!R_i^{\varepsilon}$ involve suboptimal strategies in the unrestricted game.}
\begin{equation}\label{epsPandP}
  B\!R_i^{\varepsilon}(\sigma_{-i})\,=\,\text{argmax}_{\sigma_i\,\in\,\mathscr{P}_i(\mu_i(\varepsilon))}\,\langle\sigma_i|\pi_i(\sigma_{-i})\rangle
\end{equation}
with $\mu_i(\varepsilon)$ as in \eqref{eq:perfect} or \eqref{eq:proper}. Let $B\!R^{\varepsilon}$ denote the concatenation of players' better replies then $\sigma\in B\!R^{\varepsilon}(\sigma)$ defines a Nash equilibrium of the game in which player $i$'s mixed strategies are restricted to $\mathscr{P}_i(\mu_i(\varepsilon))$.

\begin{figure}[t]
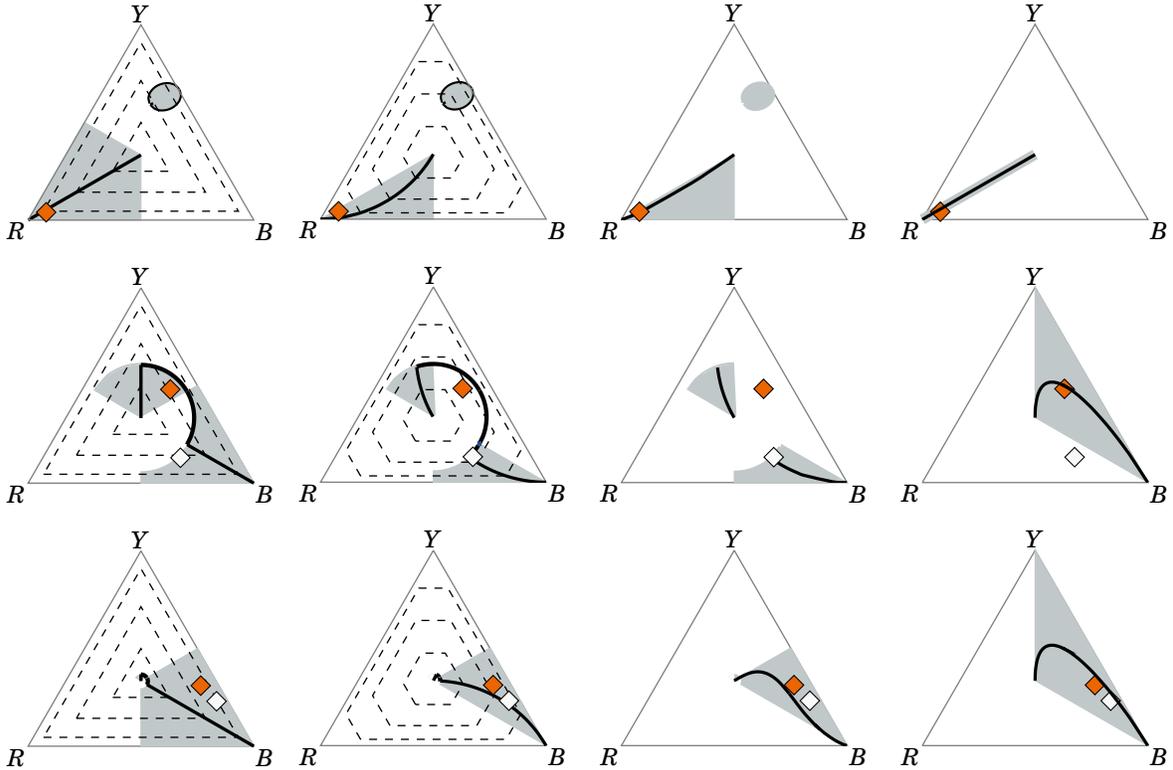

\hspace*{-3mm}
};

\end{tikzpicture}
\vspace*{-6mm}
\caption{The black curves show equilibrium correspondences in games $G_1$--$G_3$ for the $\varepsilon$-perfect model (left), the $\varepsilon$-proper model (middle-left), logit-QRE (middle-right) and Poisson level-$k$ (right). The vertices of the interior simplices in the left panel are permutations of $\mu(\varepsilon)=(1,\varepsilon,\varepsilon)/(1+2\varepsilon)$ for $\varepsilon=1/2$, $1/5$, and $1/20$. The vertices of the permutohedra in the middle-left panel are permutations of $\mu(\varepsilon)=(1,\varepsilon,\varepsilon^2)/(1+\varepsilon+\varepsilon^2)$ for $\varepsilon=2/3$, $2/5$, and $1/5$. The grey areas show predictions of more general versions of the models, i.e. all $\boldsymbol{\varepsilon}$-perfect equilibria (left), all $\boldsymbol{\varepsilon}$-proper equilibria (middle-left), all regular QRE (middle-right), and level-$k$ for arbitrary level distributions (right). The orange and white diamonds indicate the average observed choice in a session.}\label{epsEqgraphs}
\vspace*{-2mm}
\end{figure}

The $\varepsilon$-perfect and $\varepsilon$-proper better responses are interior and reflect random behavior when $\varepsilon=1$ and rational behavior when $\varepsilon=0$, i.e. $\varepsilon$ plays a similar role as $\lambda$ does for logit-QRE. The similarities between the various models are illustrated in Figure \ref{epsEqgraphs}, which shows the equilibrium correspondences for games $G_1$--$G_3$. (Similar graphs for games $g_1$--$g_{10}$ can be found in Appendix B.) All models yield a one-dimensional (black) curve that starts at the simplex' centroid and ends at a Nash equilibrium.\footnote{This is generally true for the $\varepsilon$-perfect model, the $\varepsilon$-proper model, and logit-QRE, but not necessarily for the level-$k$ model, see Figure \ref{fig:logitSet}.}
Figure \ref{epsEqgraphs} also highlights that the predictions of each model belong to a larger set of predictions generated by all possible parameterizations of the model. Specifically, the grey areas correspond to all $\boldsymbol{\varepsilon}$-perfect equilibria (left), all $\boldsymbol{\varepsilon}$-proper equilibria (middle-left), all regular QRE (middle-right), and to level-$k$ for arbitrary level distributions (right). Note that the set of QRE is identical to the set of $\boldsymbol{\varepsilon}$-proper equilibria, see \cite{GoereeLouis2021}, and a subset of the set of $\boldsymbol{\varepsilon}$-perfect equilibria, see Proposition \ref{rootS}. As such, there are no QRE predictions not already implied by the earlier models of \cite{Selten1975} and \cite{Myerson1978}.

\begin{table}[t]
	\begin{center}
		\setlength\tabcolsep{4pt}
		\hspace*{-2mm}
		\begin{tabular}{c|c|c|c|c|c|c|c|c|c|c|c}
			\hline\hline
			\multirow{2}{*}{Game} & \multirow{2}{*}{\#Obs} & \multicolumn{2}{c|}{$S$} & \multicolumn{2}{c|}{$\varepsilon$-perfect} & \multicolumn{2}{c|}{$\varepsilon$-proper} & \multicolumn{2}{c|}{logit QRE} & \multicolumn{2}{c}{level-$k$} \\
			& & $\varepsilon$  & $\overline{G}$ & $\varepsilon$  & $\overline{G}$ & $\varepsilon$ & $\overline{G}$  & $\lambda$ & $\overline{G}$ & $\tau $ & $\overline{G}$ \\ \hline
			$G_1$ & 135 & 0.07 & \textbf{0.00} & 0.053 & \textbf{0.08} & 0.122 & \textbf{0.81} & 0.159 & \textbf{0.00} & 2.11 & \textbf{0.08} \\
			$G_{2}$ & 270 & 0.80 & \textbf{0.00} & 0.110 & \textbf{0.86} & 0.338 & \textbf{0.35} & 0.999 & 3.96 & 1.19 & 6.41 \\
			$G_{3}$ & 270 & 0.52 & \textbf{0.00} & 0.254 & 2.97 & 0.350 & \textbf{0.45} & 0.035 & 1.07 & 2.42 & \textbf{0.79} \\
			$g_1$ & 120 & 0.31 & \textbf{0.00} & 0.206 & \textbf{0.94} & 0.318 & \textbf{0.01} & 0.063 & \textbf{0.00} & 0.51 & 3.71 \\
			$g_2$ & 240 & 0.12 & \textbf{0.03} & 0.066 & 1.23 & 1.000 & 25.0 & 0.033 & 23.1 & 0.31 & 23.0 \\
			$g_3$ & 240 & 1.00 & \textbf{0.35} & 0.370 & 1.24 & $[0,1]$ & 5.66 & $[0,\infty]$ & 5.66 & $[0,\infty]$ & 5.66 \\
			$g_4$ & 56 & 0.06 & \textbf{0.00} & 0.049 & \textbf{0.02} & 0.130 & \textbf{0.82} & 0.058 & \textbf{0.75} & 2.21 & \textbf{0.02} \\
			$g_5$ & 56 & 0.26 & \textbf{0.00} & 0.027 & \textbf{0.03} & 0.235 & \textbf{0.01} & 0.041 & 2.07 & 1.21 & \textbf{0.94} \\
			$g_6$ & 56 & 0.08 & \textbf{0.00} & 0.049 & \textbf{0.21} & 0.099 & \textbf{0.07} & 0.025 & \textbf{0.14} & 2.21 & \textbf{0.21} \\
			$g_7$ & 120 & 0.42 & \textbf{0.02} & 0.062 & \textbf{0.02} & 0.274 & \textbf{0.02} & 0.000 & 7.65 & 0.00 & 7.65 \\
			$g_8$ & 120 & 0.23 & \textbf{0.00} & 0.159 & \textbf{0.66} & 0.269 & \textbf{0.08} & 0.096 & 4.60 & 1.35 & \textbf{0.87} \\
			$g_{9}$ & 240 & 0.60 & \textbf{0.00} & 0.123 & 1.68 & 0.354 & 4.70 & 0.291 & 6.97 & 0.46 & 8.94 \\
			$g_{10}$ & 240 & 0.16 & \textbf{0.00} & 0.063 & 1.82 & 0.133 & 2.00 & 0.050 & 1.92 & 0.00 & 26.0 \\
			\hline\hline
		\end{tabular}
	\end{center}
	\vspace*{-5mm}
	\caption{Estimated model parameters and normalized goodness-of-fit measures. The bold numbers indicate instances when the model is not rejected ($\overline{G}\leq 1$).}\label{estFit}
	\vspace*{-2mm}
\end{table}

Table \ref{estFit} shows the estimated parameters for the different models in each of the games. The table also reports a normalized goodness-of-fit measure. Let $\mathcal{L}_{max}$ denote the best-possible likelihood given observed choices. The $G$ statistic:
\begin{displaymath}
  G\,=\,2\log\,\bigl(\mathcal{L}_{max}\,/\mathcal{L}\bigr)
\end{displaymath}
has a $\chi^2$ distribution. The normalized $\overline{G}$ measure shown in Table \ref{estFit} follows by dividing the $G$ statistic by $\chi^2_{0.01,d}$, which is the critical threshold for confidence level $\alpha=0.01$ and $d=2$ or $d=4$ degrees of freedom.\footnote{In games $G_1$, $g_1$, $g_2$, and $g_4$--$g_8$ there is a single choice average that consists of three numbers that sum to one so there are 2 degrees of freedom. For the other games, we conducted two sessions and allow the different groups to settle on different equilibria so the degrees of freedom double.} The bold numbers in Table \ref{estFit} indicate cases when the model is not rejected ($\overline{G}\leq 1$). $S$ equilibrium is not rejected in any game. Logit-QRE is rejected in more than two-third of the games and level-$k$ is rejected in a majority of the games. The $\varepsilon$-perfect and $\varepsilon$-proper models perform better than logit-QRE and level-$k$, but are still rejected in roughly one-third of the games.

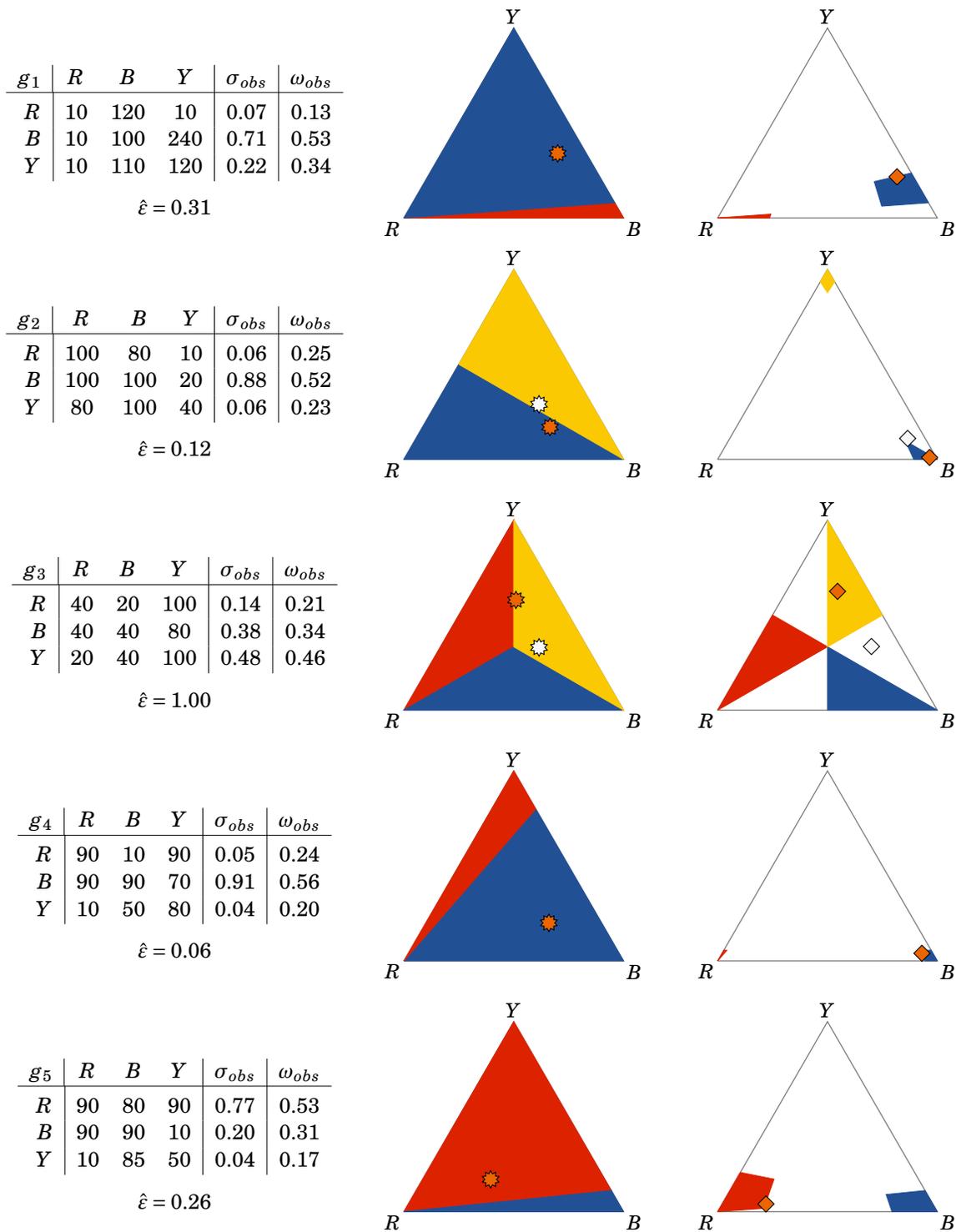
\begin{figure}[p]
\begin{center}
\begin{tikzpicture}

\node[scale=0.85] at (-1.4,0.35) {
\begin{tabular}{c|ccc|c|c}
$g_1$ & $R$ & $B$ & $Y$ & $\sigma_{obs}$ & $\omega_{obs}$ \\ \cline{1-6}
\multicolumn{1}{r}{\rule{0pt}{5mm}$R$} & \multicolumn{1}{|c}{10} & \multicolumn{1}{c}{120} & \multicolumn{1}{c|}{10} & 0.07 & 0.13\\
\multicolumn{1}{r}{$B$} & \multicolumn{1}{|c}{10} & \multicolumn{1}{c}{100} & \multicolumn{1}{c|}{240} & 0.71 & 0.53\\
\multicolumn{1}{r}{$Y$} & \multicolumn{1}{|c}{10} & \multicolumn{1}{c}{110} & \multicolumn{1}{c|}{120} & 0.22 & 0.34\\
\end{tabular}};

\node[scale=0.85] at (-1.4,-1) {$\hat{\varepsilon}=0.31$};

\node at (4,0.35) {
\begin{tikzpicture}[scale=3.5]
\draw[line width=.5pt,gray] (0,0) -- (1,0) -- (1/2,1/2*3^.5) -- (0,0);
\node[scale=0.8pt] at (-0.05,-0.05) {$R$};
\node[scale=0.8pt] at (1.05,-0.05) {$B$};
\node[scale=0.8pt] at (1/2,1/2*3^.5+0.05) {$Y$};
\filldraw[fill=mondrianRed,draw=mondrianRed,opacity=1] (0,0) -- (1,0) -- (24/25,1/25*3^.5) -- (0,0);
\filldraw[fill=mondrianBlue,draw=mondrianBlue,opacity=1] (0,0) -- (24/25,1/25*3^.5) -- (1/2,1/2*3^.5) -- (0,0);

\node[star,star points=10] at (0.7000000, 0.2944486) [draw,scale=.5,fill=mondrianOrange]{};

\end{tikzpicture}};

\node at (9,0.35) {
\begin{tikzpicture}[scale=3.5]
\draw[line width=.5pt,gray] (0,0) -- (1,0) -- (1/2,1/2*3^.5) -- (0,0);
\node[scale=0.8pt] at (-0.05,-0.05) {$R$};
\node[scale=0.8pt] at (1.05,-0.05) {$B$};
\node[scale=0.8pt] at (1/2,1/2*3^.5+0.05) {$Y$};
\filldraw[fill=mondrianRed,draw=mondrianRed,opacity=1] (0,0) -- (0.2366412, 0.) -- (0.24195122 ,0.01746133) -- (0,0);
\filldraw[fill=mondrianBlue,draw=mondrianBlue,opacity=1] (0.96, 0.069282) -- (0.74696545, 0.05390757) -- (0.7129630, 0.1657209) -- (0.8816794, 0.20493735) -- (0.96, 0.069282);

\node at (0.816667, 0.187639) [draw,scale=.5,diamond,fill=mondrianOrange]{};

\end{tikzpicture}};

\node[scale=0.85] at (-1.4,-3.5) {
\begin{tabular}{c|ccc|c|c}
$g_{2}$ & $R$ & $B$ & $Y$ & $\sigma_{obs}$ & $\omega_{obs}$ \\ \cline{1-6}
\multicolumn{1}{r}{\rule{0pt}{5mm}$R$} & \multicolumn{1}{|c}{100} & \multicolumn{1}{c}{80} & \multicolumn{1}{c|}{10} & 0.06 & 0.25 \\
\multicolumn{1}{r}{$B$} & \multicolumn{1}{|c}{100} & \multicolumn{1}{c}{100} & \multicolumn{1}{c|}{20} & 0.88 & 0.52 \\
\multicolumn{1}{r}{$Y$} & \multicolumn{1}{|c}{80} & \multicolumn{1}{c}{100} & \multicolumn{1}{c|}{40} & 0.06 & 0.23 \\
\end{tabular}};

\node[scale=0.85] at (-1.4,-4.85) {$\hat{\varepsilon}=0.12$};

\node at (4,-3.5) {
\begin{tikzpicture}[scale=3.5]
\draw[line width=.5pt,gray] (0,0) -- (1,0) -- (1/2,1/2*3^.5) -- (0,0);
\node[scale=0.8pt] at (-0.05,-0.05) {$R$};
\node[scale=0.8pt] at (1.05,-0.05) {$B$};
\node[scale=0.8pt] at (1/2,1/2*3^.5+0.05) {$Y$};
\filldraw[fill=mondrianBlue,draw=mondrianBlue,opacity=1] (1,0) -- (1/4,1/4*3^.5) -- (0,0) -- (1,0);
\filldraw[fill=mondrianYellow,draw=mondrianYellow,opacity=1] (1,0) -- (1/4,1/4*3^.5) -- (1/2,1/2*3^.5) -- (1,0);

\node[star,star points=10] at (0.6150000, 0.2511474) [draw,scale=.5,fill=mondrianWhite]{};
\node[star,star points=10] at (0.6650000, 0.1472243) [draw,scale=.5,fill=mondrianOrange]{};

\end{tikzpicture}};

\node at (9,-3.5) {
\begin{tikzpicture}[scale=3.5]
\draw[line width=.5pt,gray] (0,0) -- (1,0) -- (1/2,1/2*3^.5) -- (0,0);
\node[scale=0.8pt] at (-0.05,-0.05) {$R$};
\node[scale=0.8pt] at (1.05,-0.05) {$B$};
\node[scale=0.8pt] at (1/2,1/2*3^.5+0.05) {$Y$};
\filldraw[fill=mondrianBlue,draw=mondrianBlue,opacity=1] (1,0) -- (0.8928571, 0.0000000) -- (0.85483871, 0.08380891) -- (1,0);
\filldraw[fill=mondrianYellow,draw=mondrianYellow,opacity=1] (0.5, 0.8660254) -- (0.4672897, 0.8093695) -- (0.5000000, 0.7596714) -- (0.5327103, 0.8093695) -- (0.5, 0.866025);

\node at (0.865, 0.0952628) [draw,scale=.5,diamond,fill=mondrianWhite]{};
\node at (0.965, 0.00866025) [draw,scale=.5,diamond,fill=mondrianOrange]{};

\end{tikzpicture}};

\node[scale=0.85] at (-1.4,-7.5) {
\begin{tabular}{c|ccc|c|c}
$g_{3}$ & $R$ & $B$ & $Y$ &$\sigma_{obs}$ & $\omega_{obs}$ \\ \cline{1-6}
\multicolumn{1}{r}{\rule{0pt}{5mm}$R$} & \multicolumn{1}{|c}{40} & \multicolumn{1}{c}{20} & \multicolumn{1}{c|}{100} & 0.14 & 0.21 \\
\multicolumn{1}{r}{$B$} & \multicolumn{1}{|c}{40} & \multicolumn{1}{c}{40} & \multicolumn{1}{c|}{80} & 0.38 & 0.34 \\
\multicolumn{1}{r}{$Y$} & \multicolumn{1}{|c}{20} & \multicolumn{1}{c}{40} & \multicolumn{1}{c|}{100} & 0.48 & 0.46 \\
\end{tabular}};

\node[scale=0.85] at (-1.4,-8.85) {$\hat{\varepsilon}=1.00$};

\node at (4,-7.5) {
\begin{tikzpicture}[scale=3.5]
\draw[line width=.5pt,gray] (0,0) -- (1,0) -- (1/2,1/2*3^.5) -- (0,0);
\node[scale=0.8pt] at (-0.05,-0.05) {$R$};
\node[scale=0.8pt] at (1.05,-0.05) {$B$};
\node[scale=0.8pt] at (1/2,1/2*3^.5+0.05) {$Y$};
\filldraw[fill=mondrianRed,draw=mondrianRed,opacity=1] (0,0) --  (1/2,1/6*3^.5) -- (1/2,1/2*3^.5) -- (0,0);
\filldraw[fill=mondrianBlue,draw=mondrianBlue,opacity=1] (1,0) -- (0,0) -- (1/2,1/6*3^.5) -- (1,0);
\filldraw[fill=mondrianYellow,draw=mondrianYellow,opacity=1] (1/2,1/2*3^.5) -- (1,0) -- (1/2,1/6*3^.5) -- (1/2,1/2*3^.5);

\node[star,star points=10] at (0.5100000, 0.5022947) [draw,scale=.5,fill=mondrianOrange]{};
\node[star,star points=10] at (0.6150000, 0.2857884) [draw,scale=.5,fill=mondrianWhite]{};

\end{tikzpicture}};

\node at (9,-7.5) {
\begin{tikzpicture}[scale=3.5]
\draw[line width=.5pt,gray] (0,0) -- (1,0) -- (1/2,1/2*3^.5) -- (0,0);
\node[scale=0.8pt] at (-0.05,-0.05) {$R$};
\node[scale=0.8pt] at (1.05,-0.05) {$B$};
\node[scale=0.8pt] at (1/2,1/2*3^.5+0.05) {$Y$};
\filldraw[fill=mondrianRed,draw=mondrianRed,opacity=1] (0,0) --  (0.2500000, 0.4330127) -- (0.5,0.2886751) -- (0,0);
\filldraw[fill=mondrianBlue,draw=mondrianBlue,opacity=1] (1,0) -- (0.50, 0.0) -- (0.5, 0.2886751) -- (1,0);
\filldraw[fill=mondrianYellow,draw=mondrianYellow,opacity=1] (0.5000000, 0.8660254) -- (0.5000000 ,0.2886751) -- (0.7500000, 0.4330127) -- (0.5000000, 0.8660254);

\node at (0.545833, 0.541266) [draw,scale=.5,diamond,fill=mondrianOrange]{};
\node at (0.7, 0.29) [draw,scale=.5,diamond,fill=mondrianWhite]{};

\end{tikzpicture}};

\node[scale=0.85] at (-1.4,-11.5) {
\begin{tabular}{c|ccc|c|c}
$g_4$ & $R$ & $B$ & $Y$ & $\sigma_{obs}$ & $\omega_{obs}$ \\ \cline{1-6}
\multicolumn{1}{r}{\rule{0pt}{5mm}$R$} & \multicolumn{1}{|c}{90} & \multicolumn{1}{c}{10} & \multicolumn{1}{c|}{90} & 0.05 & 0.24\\
\multicolumn{1}{r}{$B$} & \multicolumn{1}{|c}{90} & \multicolumn{1}{c}{90} & \multicolumn{1}{c|}{70} & 0.91 & 0.56\\
\multicolumn{1}{r}{$Y$} & \multicolumn{1}{|c}{10} & \multicolumn{1}{c}{50} & \multicolumn{1}{c|}{80} & 0.04 & 0.20\\
\end{tabular}};

\node[scale=0.85] at (-1.4,-12.85) {$\hat{\varepsilon}=0.06$};

\node at (4,-11.5) {
\begin{tikzpicture}[scale=3.5]
\draw[line width=.5pt,gray] (0,0) -- (1,0) -- (1/2,1/2*3^.5) -- (0,0);
\node[scale=0.8pt] at (-0.05,-0.05) {$R$};
\node[scale=0.8pt] at (1.05,-0.05) {$B$};
\node[scale=0.8pt] at (1/2,1/2*3^.5+0.05) {$Y$};
\filldraw[fill=mondrianBlue,draw=mondrianBlue,opacity=1] (0,0) -- (1,0) -- (3/5,2/5*3^.5) -- (0,0);
\filldraw[fill=mondrianRed,draw=mondrianRed,opacity=1] (0,0) -- (3/5,2/5*3^.5) -- (1/2,1/2*3^.5) -- (0,0);

\node[star,star points=10] at (0.6600000, 0.1732051) [draw,scale=.5,fill=mondrianOrange]{};
\end{tikzpicture}};

\node at (9,-11.5) {
\begin{tikzpicture}[scale=3.5]
\draw[line width=.5pt,gray] (0,0) -- (1,0) -- (1/2,1/2*3^.5) -- (0,0);
\node[scale=0.8pt] at (-0.05,-0.05) {$R$};
\node[scale=0.8pt] at (1.05,-0.05) {$B$};
\node[scale=0.8pt] at (1/2,1/2*3^.5+0.05) {$Y$};
\filldraw[fill=mondrianRed,draw=mondrianRed,opacity=1] (0,0) -- (0.04186045, 0.04833626) -- (0.0283019,0 0.04902033) -- (0,0);
\filldraw[fill=mondrianBlue,draw=mondrianBlue,opacity=1] (1,0) -- (0.9433962, 0.) -- (0.91964286, 0.04639422) -- (0.97169811, 0.04902031) -- (1,0);

\node at (0.93,0.034641) [draw,scale=.5,diamond,fill=mondrianOrange]{};
\end{tikzpicture}};

\node[scale=0.85] at (-1.4,-15.5) {
\begin{tabular}{c|ccc|c|c}
$g_5$ & $R$ & $B$ & $Y$ &$\sigma_{obs}$ & $\omega_{obs}$ \\ \cline{1-6}
\multicolumn{1}{r}{\rule{0pt}{5mm}$R$} & \multicolumn{1}{|c}{90} & \multicolumn{1}{c}{80} & \multicolumn{1}{c|}{90} & 0.77 & 0.53\\
\multicolumn{1}{r}{$B$} & \multicolumn{1}{|c}{90} & \multicolumn{1}{c}{90} & \multicolumn{1}{c|}{10} & 0.20 & 0.31\\
\multicolumn{1}{r}{$Y$} & \multicolumn{1}{|c}{10} & \multicolumn{1}{c}{85} & \multicolumn{1}{c|}{50} & 0.04 & 0.17\\
\end{tabular}};

\node[scale=0.85] at (-1.4,-16.85) {$\hat{\varepsilon}=0.26$};

\node at (4,-15.5) {
\begin{tikzpicture}[scale=3.5]
\draw[line width=.5pt,gray] (0,0) -- (1,0) -- (1/2,1/2*3^.5) -- (0,0);
\node[scale=0.8pt] at (-0.05,-0.05) {$R$};
\node[scale=0.8pt] at (1.05,-0.05) {$B$};
\node[scale=0.8pt] at (1/2,1/2*3^.5+0.05) {$Y$};
\filldraw[fill=mondrianRed,draw=mondrianRed,opacity=1] (0,0) -- (17/18,1/18*3^.5) -- (1/2,1/2*3^.5) -- (0,0);
\filldraw[fill=mondrianBlue,draw=mondrianBlue,opacity=1] (0,0) -- (16/17,0) -- (17/18,1/18*3^.5) -- (0,0);
\filldraw[fill=mondrianBlue,draw=mondrianBlue,opacity=1] (1,0) -- (16/17,0) -- (17/18,1/18*3^.5) -- (1,0);

\node[star,star points=10] at (0.3950000, 0.1472243) [draw,scale=.5,fill=mondrianOrange]{};
\end{tikzpicture}};

\node at (9,-15.5) {
\begin{tikzpicture}[scale=3.5]
\draw[line width=.5pt,gray] (0,0) -- (1,0) -- (1/2,1/2*3^.5) -- (0,0);
\node[scale=0.8pt] at (-0.05,-0.05) {$R$};
\node[scale=0.8pt] at (1.05,-0.05) {$B$};
\node[scale=0.8pt] at (1/2,1/2*3^.5+0.05) {$Y$};
\filldraw[fill=mondrianRed,draw=mondrianRed,opacity=1] (0,0) -- (0.20986460, 0.01507594) -- (0.2565789, 0.1481359) -- (0.1031746, 0.1787037) -- (0,0);
\filldraw[fill=mondrianBlue,draw=mondrianBlue,opacity=1] (1,0) -- (0.7936508 , 0.) -- (0.76714805, 0.07816113) -- (0.94444445, 0.09622504) -- (1,0);

\node at (0.22,0.035) [draw,scale=.5,diamond,fill=mondrianOrange]{};
\end{tikzpicture}};

\end{tikzpicture}
\end{center}
\vspace*{-5mm}
\caption{The middle panels show $S(\varepsilon)$-equilibrium belief sets and the right panels show $S(\varepsilon)$-equilibrium choice sets for the estimated $\hat{\varepsilon}$ parameter in the left panel.
The star(s) in the middle panels indicate average beliefs and the diamond(s) in the right panels indicate average choices.}\label{fig:beliefSets}
\end{figure}


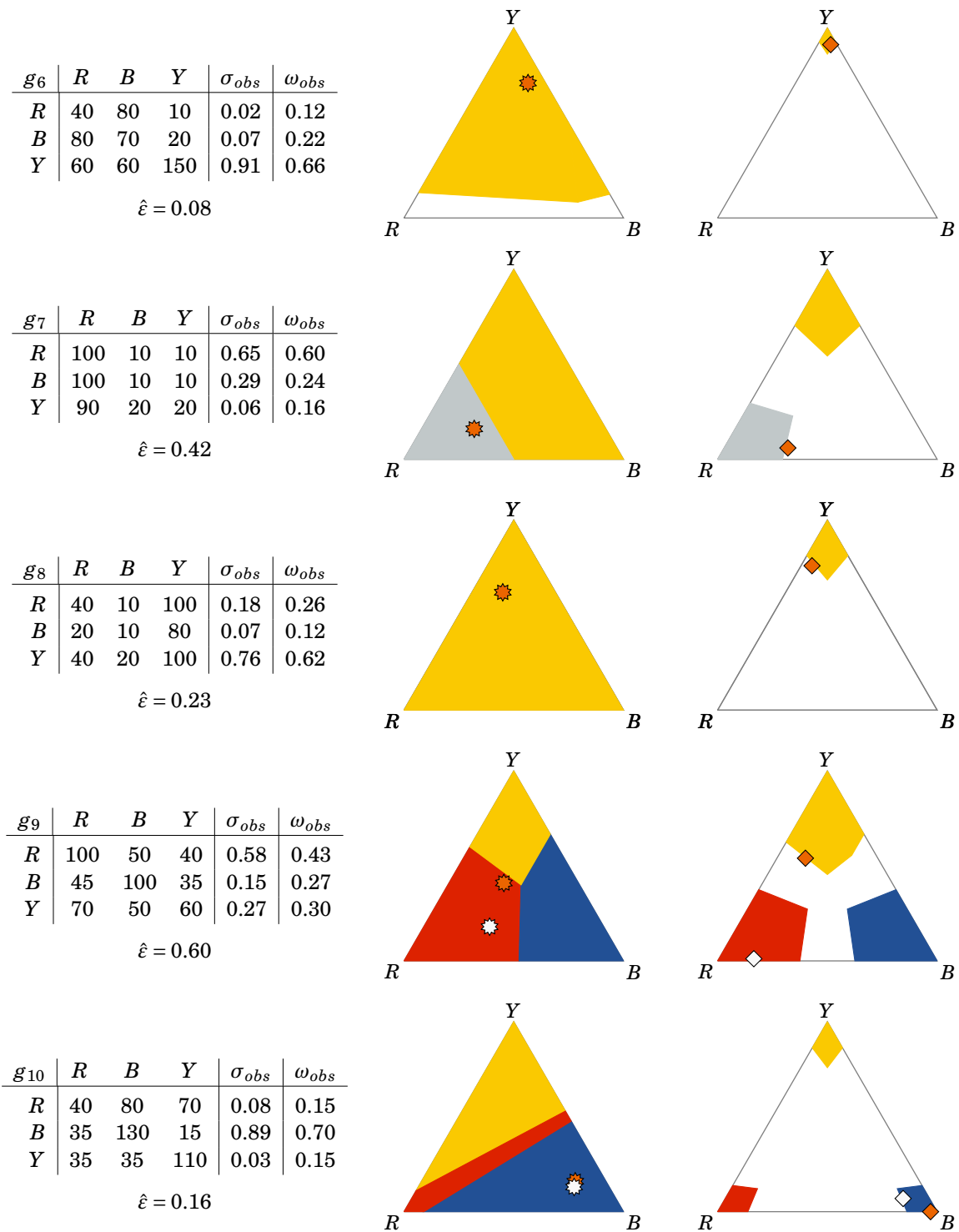
\begin{figure}[p]
\begin{center}

\begin{tikzpicture}
\node[scale=0.85] at (-1.4,0.35) {
\begin{tabular}{c|ccc|c|c}
$g_6$ & $R$ & $B$ & $Y$ & $\sigma_{obs}$ & $\omega_{obs}$ \\ \cline{1-6}
\multicolumn{1}{r}{\rule{0pt}{5mm}$R$} & \multicolumn{1}{|c}{40} & \multicolumn{1}{c}{80} & \multicolumn{1}{c|}{10} & 0.02 & 0.12\\
\multicolumn{1}{r}{$B$} & \multicolumn{1}{|c}{80} & \multicolumn{1}{c}{70} & \multicolumn{1}{c|}{20} & 0.07 & 0.22\\
\multicolumn{1}{r}{$Y$} & \multicolumn{1}{|c}{60} & \multicolumn{1}{c}{60} & \multicolumn{1}{c|}{150} & 0.91 & 0.66\\
\end{tabular}};

\node[scale=0.85] at (-1.4,-1) {$\hat{\varepsilon}=0.08$};

\node at (4,0.35) {
\begin{tikzpicture}[scale=3.5]
\draw[line width=.5pt,gray] (0,0) -- (1,0) -- (1/2,1/2*3^.5) -- (0,0);
\node[scale=0.8pt] at (-0.05,-0.05) {$R$};
\node[scale=0.8pt] at (1.05,-0.05) {$B$};
\node[scale=0.8pt] at (1/2,1/2*3^.5+0.05) {$Y$};
\filldraw[fill=mondrianYellow,draw=mondrianYellow,opacity=1] (1/15,1/15*3^.5) -- (19/24,1/24*3^.5) -- (3/4,1/4*3^.5) -- (1/2,1/2*3^.5) -- (1/15,1/15*3^.5);
\filldraw[fill=mondrianYellow,draw=mondrianYellow,opacity=1] (19/24,1/24*3^.5) -- (3/4,1/4*3^.5) -- (15/16,1/16*3^.5) -- (19/24,1/24*3^.5);

\node[star,star points=10] at (0.565000, 0.614878) [draw,scale=.5,fill=mondrianOrange]{};

\end{tikzpicture}};

\node at (9,0.35) {
\begin{tikzpicture}[scale=3.5]
\draw[line width=.5pt,gray] (0,0) -- (1,0) -- (1/2,1/2*3^.5) -- (0,0);
\node[scale=0.8pt] at (-0.05,-0.05) {$R$};
\node[scale=0.8pt] at (1.05,-0.05) {$B$};
\node[scale=0.8pt] at (1/2,1/2*3^.5+0.05) {$Y$};
\filldraw[fill=mondrianYellow,draw=mondrianYellow,opacity=1] (0.5, 0.866025) -- (0.4629630, 0.8018754) -- (0.5, 0.7465736) -- (0.5370370, 0.8018754) -- (0.5, 0.866025);

\node at (0.515, 0.79) [draw,scale=.5,diamond,fill=mondrianOrange]{};
\end{tikzpicture}};

\node[scale=0.85] at (-1.4,-3.5) {
\begin{tabular}{c|ccc|c|c}
$g_7$ & $R$ & $B$ & $Y$ & $\sigma_{obs}$ & $\omega_{obs}$\\ \hline
\multicolumn{1}{r}{\rule{0pt}{5mm}$R$} & \multicolumn{1}{|c}{$100$} & \multicolumn{1}{c}{$10$} & \multicolumn{1}{c|}{$10$} & 0.65 & 0.60\\
\multicolumn{1}{r}{$B$} & \multicolumn{1}{|c}{$100$} & \multicolumn{1}{c}{$10$} & \multicolumn{1}{c|}{$10$} & 0.29 & 0.24 \\
\multicolumn{1}{r}{$Y$} & \multicolumn{1}{|c}{$90$} & \multicolumn{1}{c}{$20$} & \multicolumn{1}{c|}{$20$} & 0.06 & 0.16\\
\end{tabular}};

\node[scale=0.85] at (-1.4,-4.85) {$\hat{\varepsilon}=0.42$};

\node at (4,-3.5) {
\begin{tikzpicture}[scale=3.5]
\draw[line width=.5pt,gray] (0,0) -- (1,0) -- (1/2,1/2*3^.5) -- (0,0);
\node[scale=0.8pt] at (-0.05,-0.05) {$R$};
\node[scale=0.8pt] at (1.05,-0.05) {$B$};
\node[scale=0.8pt] at (1/2,1/2*3^.5+0.05) {$Y$};
\filldraw[fill=mondrianYellow,draw=mondrianYellow,opacity=1] (1/4,1/4*3^.5) -- (1/2,0) -- (1,0) -- (1/2,1/2*3^.5) -- (1/4,1/4*3^.5);
\filldraw[fill=mondrianGrey,draw=mondrianGrey,opacity=1] (0,0) -- (1/2,0)  -- (1/4,1/4*3^.5) -- (0,0);

\node[star,star points=10] at (0.32,0.139) [draw,scale=.5,fill=mondrianOrange]{};
\end{tikzpicture}};

\node at (9,-3.5) {
\begin{tikzpicture}[scale=3.5]
\draw[line width=.5pt,gray] (0,0) -- (1,0) -- (1/2,1/2*3^.5) -- (0,0);
\node[scale=0.8pt] at (-0.05,-0.05) {$R$};
\node[scale=0.8pt] at (1.05,-0.05) {$B$};
\node[scale=0.8pt] at (1/2,1/2*3^.5+0.05) {$Y$};
\filldraw[fill=mondrianYellow,draw=mondrianYellow,opacity=1] (1/2,1/2*3^.5) -- (0.3521127, 0.6098770) -- (0.5, 0.470666) -- (0.6478873, 0.6098770) -- (1/2,1/2*3^.5);
\filldraw[fill=mondrianGrey,draw=mondrianGrey,opacity=1] (0,0) -- (0.2957746, 0.)  -- (0.3423913, 0.1976797) -- (0.1478873, 0.2561484) -- (0,0);

\node at (0.32,0.052) [draw,scale=.5,diamond,fill=mondrianOrange]{};
\end{tikzpicture}};

\node[scale=0.85] at (-1.4,-7.5) {
\begin{tabular}{c|ccc|c|c}
$g_8$ & $R$ & $B$ & $Y$ & $\sigma_{obs}$ & $\omega_{obs}$ \\ \hline
\multicolumn{1}{r}{\rule{0pt}{5mm}$R$} & \multicolumn{1}{|c}{$40$} & \multicolumn{1}{c}{$10$} & \multicolumn{1}{c|}{$100$} & 0.18 & 0.26\\
\multicolumn{1}{r}{$B$} & \multicolumn{1}{|c}{$20$} & \multicolumn{1}{c}{$10$} & \multicolumn{1}{c|}{$80$} & 0.07 & 0.12 \\
\multicolumn{1}{r}{$Y$} & \multicolumn{1}{|c}{$40$} & \multicolumn{1}{c}{$20$} & \multicolumn{1}{c|}{$100$} &0.76 & 0.62\\
\end{tabular}};

\node[scale=0.85] at (-1.4,-8.85) {$\hat{\varepsilon}=0.23$};

\node at (4,-7.5) {
\begin{tikzpicture}[scale=3.5]
\draw[line width=.5pt,gray] (0,0) -- (1,0) -- (1/2,1/2*3^.5) -- (0,0);
\node[scale=0.8pt] at (-0.05,-0.05) {$R$};
\node[scale=0.8pt] at (1.05,-0.05) {$B$};
\node[scale=0.8pt] at (1/2,1/2*3^.5+0.05) {$Y$};

\draw[line width=.5pt,gray] (0,0) -- (1,0) -- (1/2,1/2*3^.5) -- (0,0);
\node[scale=0.8pt] at (-0.05,-0.05) {$R$};
\node[scale=0.8pt] at (1.05,-0.05) {$B$};
\node[scale=0.8pt] at (1/2,1/2*3^.5+0.05) {$Y$};
\filldraw[fill=mondrianYellow,draw=mondrianYellow,opacity=1] (0,0) -- (1,0) -- (1/2,1/2*3^.5) -- (0,0);

\node[star,star points=10] at (0.45,0.537) [draw,scale=.5,fill=mondrianOrange]{};
\end{tikzpicture}};

\node at (9,-7.5) {
\begin{tikzpicture}[scale=3.5]
\draw[line width=.5pt,gray] (0,0) -- (1,0) -- (1/2,1/2*3^.5) -- (0,0);
\node[scale=0.8pt] at (-0.05,-0.05) {$R$};
\node[scale=0.8pt] at (1.05,-0.05) {$B$};
\node[scale=0.8pt] at (1/2,1/2*3^.5+0.05) {$Y$};

\draw[line width=.5pt,gray] (0,0) -- (1,0) -- (1/2,1/2*3^.5) -- (0,0);
\node[scale=0.8pt] at (-0.05,-0.05) {$R$};
\node[scale=0.8pt] at (1.05,-0.05) {$B$};
\node[scale=0.8pt] at (1/2,1/2*3^.5+0.05) {$Y$};

\filldraw[fill=mondrianYellow,draw=mondrianYellow,opacity=1] (1/2,1/2*3^.5) -- (0.4065041, 0.7040857) -- (0.5, 0.5931681) -- (0.5934959, 0.7040857) -- (1/2,1/2*3^.5);

\node at (0.43,0.658) [draw,scale=.5,diamond,fill=mondrianOrange]{};
\end{tikzpicture}};

\node[scale=0.85] at (-1.4,-11.5) {
\begin{tabular}{c|ccc|c|c}
$g_{9}$ & $R$ & $B$ & $Y$ & $\sigma_{obs}$ & $\omega_{obs}$ \\ \cline{1-6}
\multicolumn{1}{r}{\rule{0pt}{5mm}$R$} & \multicolumn{1}{|c}{100} & \multicolumn{1}{c}{50} & \multicolumn{1}{c|}{40} & 0.58 & 0.43 \\
\multicolumn{1}{r}{$B$} & \multicolumn{1}{|c}{45} & \multicolumn{1}{c}{100} & \multicolumn{1}{c|}{35} & 0.15 & 0.27 \\
\multicolumn{1}{r}{$Y$} & \multicolumn{1}{|c}{70} & \multicolumn{1}{c}{50} & \multicolumn{1}{c|}{60} & 0.27 & 0.30 \\
\end{tabular}};

\node[scale=0.85] at (-1.4,-12.85) {$\hat{\varepsilon}=0.60$};

\node at (4,-11.5) {
\begin{tikzpicture}[scale=3.5]
\draw[line width=.5pt,gray] (0,0) -- (1,0) -- (1/2,1/2*3^.5) -- (0,0);
\node[scale=0.8pt] at (-0.05,-0.05) {$R$};
\node[scale=0.8pt] at (1.05,-0.05) {$B$};
\node[scale=0.8pt] at (1/2,1/2*3^.5+0.05) {$Y$};
\filldraw[fill=mondrianRed,draw=mondrianRed,opacity=1] (0,0) -- (11/21,0) -- (8/15,3/15*3^.5) -- (3/10,3/10*3^.5) -- (0,0);
\filldraw[fill=mondrianBlue,draw=mondrianBlue,opacity=1] (1,0) -- (11/21,0) -- (8/15,3/15*3^.5) -- (2/3,1/3*3^.5) -- (1,0);
\filldraw[fill=mondrianYellow,draw=mondrianYellow,opacity=1] (1/2,1/2*3^.5) -- (3/10,3/10*3^.5) -- (8/15,3/15*3^.5) -- (2/3,1/3*3^.5) -- (1/2,1/2*3^.5);

\node[star,star points=10] at (0.4550000, 0.3550704) [draw,scale=.5,fill=mondrianOrange]{};
\node[star,star points=10] at (0.3900000, 0.1558846) [draw,scale=.5,fill=mondrianWhite]{};

\end{tikzpicture}};

\node at (9,-11.5) {
\begin{tikzpicture}[scale=3.5]
\draw[line width=.5pt,gray] (0,0) -- (1,0) -- (1/2,1/2*3^.5) -- (0,0);
\node[scale=0.8pt] at (-0.05,-0.05) {$R$};
\node[scale=0.8pt] at (1.05,-0.05) {$B$};
\node[scale=0.8pt] at (1/2,1/2*3^.5+0.05) {$Y$};
\filldraw[fill=mondrianRed,draw=mondrianRed,opacity=1] (0,0) -- (0.1875000, 0.3247595) -- (0.4090909, 0.2361887) -- (0.375,0) -- (0,0);
\filldraw[fill=mondrianBlue,draw=mondrianBlue,opacity=1] (1,0) -- (0.625,0) -- (0.5909091, 0.2361887) -- (0.8125000, 0.3247595) -- (1,0);
\filldraw[fill=mondrianYellow,draw=mondrianYellow,opacity=1] (1/2,1/2*3^.5) -- (0.3125000, 0.54126592) -- (0.5000000, 0.3936479) -- (0.6111111, 0.4811253) -- (2/3,1/3*3^.5) -- (1/2,1/2*3^.5);

\node at (0.4, 0.468) [draw,scale=.5,diamond,fill=mondrianOrange]{};
\node at (0.165, 0.01) [draw,scale=.5,diamond,fill=mondrianWhite]{};

\end{tikzpicture}};

\node[scale=0.85] at (-1.4,-15.5) {
\begin{tabular}{c|ccc|c|c}
$g_{10}$ & $R$ & $B$ & $Y$ & $\sigma_{obs}$ & $\omega_{obs}$ \\ \cline{1-6}
\multicolumn{1}{r}{\rule{0pt}{5mm}$R$} & \multicolumn{1}{|c}{40} & \multicolumn{1}{c}{80} & \multicolumn{1}{c|}{70} & 0.08 & 0.15 \\
\multicolumn{1}{r}{$B$} & \multicolumn{1}{|c}{35} & \multicolumn{1}{c}{130} & \multicolumn{1}{c|}{15} & 0.89  & 0.70 \\
\multicolumn{1}{r}{$Y$} & \multicolumn{1}{|c}{35} & \multicolumn{1}{c}{35} & \multicolumn{1}{c|}{110} & 0.03  & 0.15 \\
\end{tabular}};

\node[scale=0.85] at (-1.4,-16.85) {$\hat{\varepsilon}=0.16$};

\node at (4,-15.5) {
\begin{tikzpicture}[scale=3.5]
\draw[line width=.5pt,gray] (0,0) -- (1,0) -- (1/2,1/2*3^.5) -- (0,0);
\node[scale=0.8pt] at (-0.05,-0.05) {$R$};
\node[scale=0.8pt] at (1.05,-0.05) {$B$};
\node[scale=0.8pt] at (1/2,1/2*3^.5+0.05) {$Y$};
\filldraw[fill=mondrianYellow,draw=mondrianYellow,opacity=1] (1/18,1/18*3^.5) -- (25/34,9/34*3^.5) -- (1/2,1/2*3^.5) -- (1/18,1/18*3^.5);
\filldraw[fill=mondrianBlue,draw=mondrianBlue,opacity=1] (1,0) -- (1/11,0) -- (16/21,5/21*3^.5) -- (1,0);
\filldraw[fill=mondrianRed,draw=mondrianRed,opacity=1] (0,0) -- (1/11,0) -- (16/21,5/21*3^.5) --  (25/34,9/34*3^.5) -- (1/18,1/18*3^.5) -- (0,0);

\node[star,star points=10] at (0.7800000, 0.1385641) [draw,scale=.5,fill=mondrianOrange]{};
\node[star,star points=10] at (0.7750000, 0.1125833) [draw,scale=.5,fill=mondrianWhite]{};

\end{tikzpicture}};

\node at (9,-15.5) {
\begin{tikzpicture}[scale=3.5]
\draw[line width=.5pt,gray] (0,0) -- (1,0) -- (1/2,1/2*3^.5) -- (0,0);
\node[scale=0.8pt] at (-0.05,-0.05) {$R$};
\node[scale=0.8pt] at (1.05,-0.05) {$B$};
\node[scale=0.8pt] at (1/2,1/2*3^.5+0.05) {$Y$};
\filldraw[fill=mondrianBlue,draw=mondrianBlue,opacity=1] (1,0) -- (0.862069 , 0.) -- (0.8181818, 0.1049728) -- (0.9310345, 0.1194518) -- (1,0);
\filldraw[fill=mondrianYellow,draw=mondrianYellow,opacity=1] (0.5, 0.866025) -- (0.4310345, 0.7465736) -- (0.5, 0.6560799) -- (0.5689655, 0.7465736) -- (0.5, 0.866025);
\filldraw[fill=mondrianRed,draw=mondrianRed,opacity=1] (0,0) -- (0.06896552, 0.11945178) -- (0.1818182, 0.1049728) -- (0.137931,0) -- (0,0);

\node at (0.97, 0) [draw,scale=.5,diamond,fill=mondrianOrange]{};
\node at (0.845, 0.06) [draw,scale=.5,diamond,fill=mondrianWhite]{};
\end{tikzpicture}};

\end{tikzpicture}
\end{center}
\vspace*{-5mm}
\caption{The middle panels show $S(\varepsilon)$-equilibrium belief sets and the right panels show $S(\varepsilon)$-equilibrium choice sets for the estimated $\hat{\varepsilon}$ parameter in the left panel.
The star(s) in the middle panels indicate average beliefs and the diamond(s) in the right panels indicate average choices.}\label{fig:beliefSets2}
\end{figure}

The right panels of Figures \ref{fig:beliefSets} and \ref{fig:beliefSets2} show the resulting $S$-equilibrium choice sets for the estimated $\varepsilon$ parameters. The orange diamond shows the average choice in a session (and the white diamond shows the average choice in the second session, if there was one). The $S$-equilibrium choice sets catch all but one session average. In several games, $S$ equilibrium is perfectly accurate while being very precise as reflected by the small choice sets.

Our findings highlight the role of belief sets for equilibrium selection. Consider game $g_4$ of Figure \ref{fig:beliefSets} for which $R$ is a perfect equilibrium. To support $R$ the belief must be that the more costly mistake $Y$ is more likely than the less costly mistake $B$. This is \mbox{\citeauthor{Myerson1978}'s} (\citeyear{Myerson1978}) critique of perfect equilibria and may explain why the predominant choice (91\%) is the proper equilibrium $B$. But in game $g_5$ the perfect equilibrium $R$ is chosen more frequently (77\%) than the proper equilibrium $B$. And in game $g_1$, observed choices are far away from the unique perfect and proper equilibrium $R$. While the data are puzzling for classical refinement they have an intuitive explanation in terms of belief sets.\footnote{Another shortcoming of classical refinement models is that they often do not select. Game $g_6$ has three symmetric Nash equilibria: $Y$, $\sigma=(\deel{1}{6},\deel{3}{4},\deel{1}{12})$, and $\sigma=(\deel{1}{5},\deel{4}{5},0)$, all of which are proper. While the various refinement models are silent about which of the three equilibria should be played, the data unequivocally favor (91\%) the $Y$ equilibrium with the largest belief set. (See Figure \ref{fig:ellipse} for a similar result.)} The observed choices for these games, see Figure \ref{fig:beliefSets}, belong to a set of the same color as observed beliefs, \textit{irrespective} of whether this set contains a unique, perfect, or proper equilibrium, or no equilibrium at all!

Game $g_3$ is the only game for which one of the session averages does not belong to an $S$-equilibrium choice set. All models fit the choice data of this game poorly, see Table \ref{estFit}, and the estimated parameter for $S$-equilibrium is $\hat{\varepsilon}=1$. This high estimate reflects the complexity of identifying a unique best reply in this game. By design there are two best replies for any choice of the opponent, which hampers convergence to equilibrium.

Games $g_7$ and $g_8$ test QRE's prediction that (almost) identical strategies are played (almost) equally often. In game $g_7$, strategies $R$ and $B$ are clones. Any QRE predicts they are equally likely and less likely than $Y$, see the second row of Figure \ref{appCalcGraph2}. In game $g_8$, strategies $R$ and $Y$ are almost clones and logit-QRE predicts they are almost equally likely, see the third row of Figure \ref{appCalcGraph2}, even though $R$ is weakly dominated. In both games, QRE predictions are refuted by the observed choices.

Finally, games $g_9$ and $g_{10}$ test level-$k$'s comparative statics prediction that games with the same hierarchy of best replies to the simplex' centroid yield the same outcomes, irrespective of whether the simplex' centroid belongs to a small or large belief set. In both games, level-1's beliefs belong to the red belief set and $R$ is the best reply to itself. Hence, level-$k$ predicts the same behavior in these two games: all levels $k\geq 1$ choose $R$. In game $g_{9}$, the white diamond in the fourth row of Figure \ref{fig:beliefSets2} shows that $R$ is a frequent choice in one of the sessions. However, in game $g_{10}$, choices and beliefs fall into the blue $S$-equilibrium choice and belief sets, see the bottom row of Figure \ref{fig:beliefSets2}. As in Figure \ref{fig:levelk}, the level-$k$ model fails when level-1's beliefs belong to a small $S$-equilibrium belief set.

\begin{figure}[t]
	\begin{center}
		\begin{tikzpicture}[scale=0.5]
			\begin{axis}[
				width=20cm,
				height=13.5cm,
				line width=2,
				grid=major, 
				tick label style={font=\normalsize},
				legend style={nodes={scale=0.9, transform shape}},
				legend pos=south west,
				label style={font=\normalsize},
				grid style={white},
				xlabel={Games in sample},
				ylabel={Average G},
				legend style={at={(1,1)}, anchor=north east,  draw=none, fill=none},
				xtick={2,3,4,5,6,7,8,9,10,11,12},
				ytick={0,0.1,0.2,0.3,0.4,0.5,0.6,0.7,0.8,0.9,1.0},
				xmin=1, xmax=14,
				ymin=0, ymax=1,
				]
				\addplot[mondrianRed,mark=diamond] coordinates
				{(2, 0.07819365)(3, 0.03045529)(4, 0.0157397)(5, 0.01110544)(6, 0.009391625)(7, 0.008453947)
					(8, 0.007683521)(9, 0.006944227)(10, 0.006160412)(11, 0.005194005 )(12, 0.003622551)};
				\addlegendentry{$S$}

				\addplot[mondrianYellow,mark=*] coordinates
				{(2, 0.29377176)(3, 0.26399755 )(4, 0.2403098)(5, 0.22397800)(6,0.213220187 )(7, 0.205140196)(8,0.198843829 )
					(9, 0.194140411 )(10, 0.189755299)(11, 0.184355046)(12, 0.175033215)};
				\addlegendentry{$\varepsilon$-perfect}

				\addplot[mondrianCyan,mark=triangle] coordinates
				{(2, 0.50219534 )(3, 0.48987121)(4, 0.4663351)(5, 0.44382788)(6,0.430770025 )(7, 0.424176785 )(8, 0.420736568)
					(9,0.421519871 )(10, 0.428155429)(11, 0.437575744 )(12,0.435914734 )};
				\addlegendentry{$\varepsilon$-proper}

				\addplot[black,mark=square] coordinates
				{(2, 0.87029463 )(3, 0.81903995  )(4, 0.7760716  )(5, 0.74772888 )(6, 0.730577104 )(7, 0.726937674)(8, 0.732538953)
					(9, 0.739654432 )(10, 0.740945453 )(11, 0.728796091 )(12, 0.662549404)};
				\addlegendentry{logit-QRE}

				\addplot[mondrianBlue,mark=otimes] coordinates
				{(2, 0.92669060)(3,  0.88702786)(4, 0.8607175)(5, 0.84166012)(6, 0.827532011)(7, 0.816712635)(8, 0.808209874)
					(9, 0.801415800)(10, 0.795173939)(11, 0.788052437)(12, 0.771826884)};
				\addlegendentry{level-$k$}
					
			\end{axis}
		\end{tikzpicture}
	\end{center}
	\vspace*{-4mm}
	\caption{Out-of-sample results for the different models by number of in-sample games (the curves show the average over all combinations of in-sample games).}\label{outofsample}
	\vspace*{-2mm}
\end{figure}
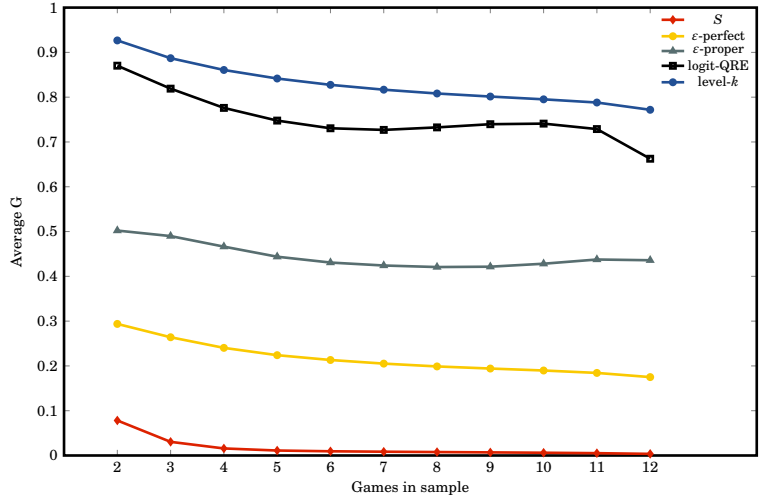

We end this section with formal statistical tests that compare the various models. For model $m$, let $\text{Fit}_{m}(x)$ denote the fraction of the observations for which $\overline{G}_m\le x$, where
$\overline{G}_m$ are the numbers listed in Table \ref{estFit}. The $\text{Fit}_m(x)$ define cumulative distribution functions, see Figure \ref{fitCumulative}, which can be compared using Kolmogorov--Smirnov tests. Let $\succ^{*}$ denote significance at the 1\% confidence level.

\begin{result}\label{res:fit} $\text{\em Fit}_S\,\succ^{*}\,\text{\em Fit}_{\varepsilon\text{-perfect}}\,\succ^{*}\,\text{\em Fit}_{\varepsilon\text{-proper}}\,\succ^{*}\,
  \text{\em Fit}_{\text{logit}}\,\succ^{*}\,\text{\em Fit}_{\text{level-k}}$.
\end{result}
\vspace*{-2mm}

\noindent The superior fit of $S$ equilibrium is not caused by over-fitting. Figure \ref{outofsample} shows results for out-of-sample tests. The $x$-axis displays the number of in-sample games that were selected, which varies between 2 and 13.  We first estimate the various models using only the in-sample games and then apply the estimated parameters to predict behavior in the out-of-sample games. The curves in Figure \ref{outofsample} correspond to the average $G$ of the models for every possible combination of in-sample games.
\begin{result}\label{res:OSFit}
  $\text{\em OSFit}_S\,\succ^{*}\,\text{\em OSFit}_{\varepsilon\text{-perfect}}\,\succ^{*}\,\text{\em OSFit}_{\varepsilon\text{-proper}}\,\succ^{*}\,
  \text{\em OSFit}_{\text{logit}}\,\succ^{*}\,\text{\em OSFit}_{\text{level-k}}$.
\end{result}
\vspace*{-2mm}

\noindent Support for Results \ref{res:fit} and \ref{res:OSFit} can be found in Appendix B.

\subsection{Analyzing Observed Beliefs}
\label{sec:beliefs}

This section evaluates the belief predictions of level-$k$, QRE, and $S$ equilibrium. Level-$k$ is a non-equilibrium model built on a hierarchy of beliefs that reflect different levels of sophistication. This hierarchy starts with level-$1$'s belief that others are level-0 and randomize uniformly over their pure strategies.\footnote{There is an alternative specification in which level-$0$ selects a salient strategy. However, uniform random choice is the specification used in normal-form games with neutral framing (as in our study). The motivation for this specification is to model a hypothetical player that ``selects a strategy at random without forming beliefs,'' \cite{nagel1995}, and ``plays unpredictably,'' \cite{stahl1994}.} The belief of level-$k$ for $k>1$ is then given by the composition of $k-1$ best replies to the simplex' centroid. The second column in Table \ref{tab:stats} shows that the percentage of level-$k$ beliefs is small in any of the games. Pooling over all thirteen games, only 3.7\% of all elicited beliefs are level-$k$ beliefs.

Another aspect of the level-$k$ model is that choices are best replies to beliefs. Columns 3--5 in Table \ref{tab:stats} list the percentages of best, second-best, and third-best replies in each of the games. While the best reply is most frequent, a substantial portion of the replies are second or third best. Pooling results from all thirteen games shows that 63.8\% of the choices are best replies, 25.6\% are second-best replies, 10.6\% are third-best replies.

\begin{result}\label{level-k-non-para}
In all thirteen games, subjects better respond rather than best respond and their beliefs differ from level-$k$ beliefs.
\end{result}

\begin{table}[t]
\setlength\tabcolsep{4pt}
\begin{center}
\begin{tabular}{c|c|c|c|c|c|c|c|c}
\hline\hline
\multirow{2}{*}{Game} & \% level-$k$  & \multicolumn{3}{c|}{\% best reply} & \multirow{2}{*}{$\sigma_{obs}$} & \multirow{2}{*}{$\omega_{obs}$} & $\sigma\,=\,\omega$ & \% CU  \\
 & beliefs & 1st & 2nd & 3rd &                                 &                                 & $p$-value & beliefs  \\ \hline
 $G_1$   & 22.2 & 91.1 & 5.9 & 3.0 & (.90,.06,.04) & (.75,.13,.12) & $<.01$ & 96.3  \\
 $G_{2}$ & 0.0  & 37.0 & 20.0 & 43.0 & (.20,.50,.30) & (.25,.40,.35) & $<.01$ & 65.6  \\
 $G_{3}$ & 0.7  & 56.7 & 36.7 & 6.7 & (.07,.66,.27) & (.22,.46,.32) & $<.01$ & 97.8  \\
 $g_{1}$ & 0.0  & 69.2 & 22.5 & 8.3 & (.07,.71,.22) & (.13,.53,.34) & $<.01$ & 98.3  \\
 $g_{2}$ & 7.1  & 54.2 & 40.8 & 5.0 &  (.06,.88,.06) & (.25,.52,.23) & $<.01$ & 100  \\
 $g_3$   & 0.0  & 45.8 & 43.8 & 10.4 & (.14,.38,.48) & (.21,.34,.46) & $.02$ & 58.8  \\
$g_{4}$ & 8.9  & 89.3 & 3.6 & 7.1 & (.05,.91,.04) & (.24,.56,.20) & $<.01$ & 98.2  \\
$g_{5}$ &1.8  & 66.1 & 32.1 & 1.8 & (.77,.20,.04) & (.53,.31,.17) & $<.01$ & 82.1  \\
$g_{6}$ &14.3  & 89.3 & 5.4 & 5.4 & (.02,.07,.91) & (.12,.22,.66) & $<.01$ & 98.2  \\
$g_7$   & 0.0  & 62.5 & 37.5 & 0.0 & (.65,.29,.06) & (.60,.24,.16) & $<.01$ & 60.0  \\
$g_8$   & 5.8  & 77.5 & 15.8 & 6.7 & (.18,.07,.76) & (.26,.12,.62) & $<.01$ & 100  \\
$g_{9}$ & 2.9  & 70.8 & 22.1 & 7.1 & (.58,.15,.27) & (.43,.27,.30) & $<.01$ & 66.7  \\
$g_{10}$ & 0.8 & 86.2 & 9.2 & 4.6 & (.08,.89,.03) & (.15,.70,.15) & $<.01$ & 95.0  \\ \hline
Pooled & 3.7 & 63.8 & 25.6 & 10.6 & & & $<.01$ & 78.2\\
 \hline\hline
\end{tabular}
\end{center}
\vspace*{-3mm}
\caption{Statistics for each of the thirteen games that support Results \ref{level-k-non-para}, \ref{QRE-non-para}, and \ref{CU}.}\label{tab:stats}
\end{table}

\noindent What about QRE beliefs?  The fixed-point equations that define QRE rest on a rational-expectations assumption that beliefs match choices. Columns 6 and 7 of Table \ref{tab:stats} show the average choice and belief respectively for each of the thirteen games. The $p$-values for the test that they are equal are listed in Column 8.
\begin{result}\label{QRE-non-para}
In all thirteen games, the average belief differs from the average choice and the rational-expectations assumption underlying QRE beliefs is rejected.
\end{result}
Finally, what about the consequential unbiasedness assumption that underlies $S$ equilibrium? Comparing the middle and right panels of Figures \ref{fig:beliefSets} and \ref{fig:beliefSets2} reveals that the average belief almost always belongs to a set of the same color as the set that the average choice belongs to, which suggests beliefs are consequentially unbiased. The final column of Table \ref{tab:stats} shows the percentage of individual beliefs that yield the same best option as the observed choice, $\sigma_{obs}$, that is listed in the sixth column. In all games, the majority of beliefs are consequentially unbiased. And in a majority of the games, virtually all beliefs (>90\%) are consequentially unbiased.
\begin{result}\label{CU}
Pooling over all thirteen games shows that the vast majority of beliefs (78\%) are consequentially unbiased.
\end{result}

\section{Conclusions}
\label{sec:conc}

\citeauthor{Selten1975}'s (\citeyear{Selten1975}) seminal contribution highlights the importance of beliefs in determining robust choices. A perfect equilibrium requires that players' choices remain optimal to a sequence of beliefs that entail small and vanishing mistakes \mbox{(``trembles'')}. Selten's ansatz set off an entire literature on equilibrium refinement. Yet, it is insufficient to guarantee robustness as it does not consider random mistakes. In hindsight, the idea that belief \textit{sets}, rather than infinitesimal belief paths, determine robust choices seems intuitive if not obvious, see Figures \ref{fig:ellipse}--\ref{fig:levelk} and \ref{fig:beliefSets}--\ref{fig:beliefSets2}.

A set-valued approach to behavioral game theory has several other advantages.

\subsection{Explicitly Set-Valued Theories versus Implicitly Set-Valued Theories}

An important methodological insight of this paper is that behavioral-game-theory models such as level-$k$ and QRE pick their predictions from an implicitly defined set. QRE requires the specification of quantal responses that map expected payoffs to choice probabilities. Level-$k$ requires the specification of a distribution function for the levels. In both cases, selecting a particular element from an infinite-dimensional function space yields a point prediction in the choice simplex. The common practice of selecting a set of elements from this function space, defines, in a roundabout way, a set of predictions in the choice simplex. Without registering a pre-analysis plan it is impossible to verify what this set is. Moreover, its size is typically hard to compute and generally ignored when reporting the model's predictive success.

This raises the question ``why generate choice predictions using elements from an infinite-dimensional function space?'' Especially since the resulting choice predictions are only supported by beliefs that satisfy rational expectations or follow from some ad hoc rule -- both possibilities are rejected in \textit{all} games reported in this paper, see Results \ref{level-k-non-para} and \ref{QRE-non-para}.

$S$ equilibrium offers a more transparent approach by explicitly defining a set of predictions in the choice simplex, which are supported by sets of consequentially unbiased beliefs. $S$ equilibrium allows for an optimal tradeoff between accuracy and precision, unlike \citeauthor{GoereeLouis2021}' (\citeyear{GoereeLouis2021}) $M$ equilibrium. It outperforms QRE and level-$k$ based on \citeauthor{selten1991}'s (\citeyear{selten1991}) ``areametrics,'' see Result \ref{mps_result}, and standard likelihood techniques, see Table \ref{estFit}. Moreover, the vast majority of observed beliefs (78\%) are consequentially unbiased, see Result \ref{CU}.

\subsection{Parametric Descendants and Comparative Statics}

One knee-jerk reaction is that set-valued theories do not offer comparative statics predictions. In contrast, logit-QRE is often touted for correctly predicting the direction of change when game parameters are varied.  This distinction is superficial as logit-QRE is simply a slice of some infinite-dimensional function space. One can similarly slice the \textit{finite}-dimensional space of $\boldsymbol{\varepsilon}$-perfect equilibria to obtain the $\varepsilon$-perfect model of the previous section.

Indeed, the latter is preferred to logit-QRE for three reasons. First, it fits the choice data better both in and out of sample, see Results \ref{res:fit} and \ref{res:OSFit}. Second, it is a semi-algebraic model that can be computed analytically, allowing for quantitative, not just qualitative, comparative statics. Third, $\varepsilon$-perfect choices form a one-dimensional subset in an $S$-equilibrium choice set. As a result, they are supported by a set of consequentially unbiased beliefs and do not require rational expectations.

\subsection{Across-Subject and Within-Subject Heterogeneity}

In the symmetric games we consider, logit-QRE defines a symmetric Bayes-Nash equilibrium that results in homogeneous behavior. In contrast, observed choices display substantial heterogeneity. Level-$k$ captures across-subject heterogeneity by assuming their levels of sophistication differ. But, assuming levels remain the same across games, it cannot explain within-subject heterogeneity (and neither can QRE).

$S$-equilibrium naturally accommodates within-subject and across-subject heterogeneity. Its choice sets may contain ``higher level'' subjects that never tremble as well as ``lower level'' subjects that occasionally do. A subject's ``level'' does not have to be constant across games. $S$ equilibrium accomplishes this by virtue of being a set-valued theory, without any ad hoc modeling or functional restrictions.

\subsection{Simplicity and Computability}

$S$ equilibrium is governed by simple choice and belief axioms. The best option with the highest expected payoff is most likely chosen and the chance of a mistake is determined by a complexity parameter, $\varepsilon$. Beliefs imply the same best option as observed choices do. Besides its simple formulation, $S$ equilibrium is easy to compute. This is surprising since $S$-equilibrium choice sets consist of infinitely many $\boldsymbol{\varepsilon}$-perfect equilibria, each of which is hard to compute (as it requires solving a system of fixed-point equations on an interior simplex). Yet, to determine the collection of them, no fixed-point equations need to be solved. Proposition \ref{rootS} shows that $S(\varepsilon)$-equilibrium choice sets are simply the maximizers and roots of a single function, the $S(\varepsilon)$ potential.

\newpage
\addtolength{\baselineskip}{0.0mm}
\vspace*{-1mm}
\bibliography{references}
\bibliographystyle{chicago}

\newpage

\startappendix

\setcounter{equation}{0}

\section{Proofs}
\label{app:proofs}

For $v,w\in\field{R}^{K_i}$ let $\langle v|w\rangle=\sum_{k=1}^{K_i}v_kw_k$ denote the usual inner product. Let $\text{supp}_\varepsilon(v)=\{j|v_{j}\geq\varepsilon\max_k(v_k)\}$ and $\text{argmax}(v)=\lim_{\varepsilon\uparrow1}\text{supp}_\varepsilon(v)=\{j|v_{j}\geq\max_k(v_k)\}$.

\medskip

\noindent\textbf{Proof of Proposition 2.}
Let $\sigma^u$ denote the profile where all players randomize uniformly over their available strategies, i.e. $\sigma^u_i$ is the centroid of $\Sigma_i$ for $i\in N$. Let $\mathcal{G}\subset\Gamma$ denote the set of normal-form games that satisfy:
\begin{itemize}\addtolength{\itemsep}{-2mm}
\vspace*{-2mm}
\item[1.] For $i\in N$ and $1\leq j<k\leq K_i$, $\pi_{ij}(\sigma^u)\neq\pi_{ik}(\sigma^u)$.
\item[2.] For $i\in N$, the set of profiles $\sigma_{-i}$ that make player $i$ indifferent between two choices has measure zero in $\Sigma_{-i}$.
\end{itemize}
\vspace*{-3mm}

\noindent The complement of $\mathcal{G}$ is defined by equalities among the payoff parameters. Hence, this complement is closed and of lower dimension than $\Gamma$, and $\mathcal{G}$ is generic.

By continuity of expected payoffs there exists, for $i\in N$, an open ball around $\sigma^u_i$ such that one of the strategies has the highest expected payoff.  For $i\in N$, $k=1,\ldots,K_i$, define the ``primary'' sets $\Sigma_{ik}=\{\sigma_i|\sigma_{ik}=\max_\ell(\sigma_{i\ell})\}$ with $\bigcup_k\Sigma_{ik}=\Sigma_i$. For $i\in N$, all primary sets meet at $\sigma^u_i$ so there exists a full-dimensional, and, hence, robust $S_i(\varepsilon)$-equilibrium choice set in one of the primary sets when $\varepsilon\uparrow1$. Choice profiles in the interior of this choice set have a unique maximum element as do the associated expected payoffs, i.e. they are colorable. This establishes (i).

To show (ii), note that in generic games $\text{argmax}(\pi_i(\sigma_{-i}))$ is single valued for $i\in N$ and almost all $\sigma\in\Sigma$. Hence, for $i\in N$, $\text{supp}(\sigma_i)$ is single valued and constant on an $S_i(\varepsilon)$-choice set and this choice set is contained in one of the $\Sigma_{ik}$, say, in $\Sigma_{i1}$. An upper bound for the size of this set follows by assuming strategy 1 yields the highest expected payoff for all $\sigma_{-i}$, in which case
\begin{displaymath}
  \mu_i(\varepsilon)\,=\,\int_{\Sigma_i}\mathbf{1}(\sigma_{i2}<\varepsilon\sigma_{i1},\ldots,\sigma_{iK_i}<\varepsilon\sigma_{i1})\,=\,\prod_{k\,=\,1}^{K_i-1}\frac{\varepsilon}{1+k\varepsilon}
\end{displaymath}
The size of $\Sigma_i$ is $1/(K_i-1)!$, so the relative size of player $i$'s $S(\varepsilon)$-choice set is at most $\overline{\mu}_i(\varepsilon)$. The size of an $S(\varepsilon)$-choice set can thus not exceed $\prod_{i=1}^n\overline{\mu}_i(\varepsilon)$.

To show (iii), assume wlog that $K_1=\max_{i}K_i$. In generic games $\text{argmax}(\pi_1(\sigma_{-1}))$ is single valued for almost all $\sigma_{-1}\in\Sigma_{-1}$. Hence, $\{\sigma\in\Sigma\,|\,\text{supp}(\sigma_1)\,\subseteq\,\text{argmax}(\pi_1(\sigma_{-1}))\}$ has measure at most $\overline{\mu}_1(\varepsilon)$. A fortiori, the measure of the union of the $S(\varepsilon)$-choice sets $\{\sigma\in\Sigma\,|\,\text{supp}(\sigma_i)\,\subseteq\,\text{argmax}(\pi_i(\sigma_{-i}))\,\forall\,i\in N\}$ is at most $\overline{\mu}_1(\varepsilon)$.

Property (iv) holds for games in which all players have a dominant strategy. $\hspace*{5mm}\blacksquare$

\smallskip

\noindent\textbf{Proof of Proposition 3.}
To prove (i), let $\sigma$ be a root of $Y_\varepsilon(\sigma)$ then, for $i\in N$, either $\pi_{ij}=\max_k(\pi_{ik})$ or $\sigma_{ij}<\varepsilon\max_k(\sigma_{ik})$. This means $\sigma$ belongs to $S^c(\varepsilon)$. Conversely, if $\sigma\in S^c(\varepsilon)$ then, for $i\in N$, either $\pi_{ij}=\max_k(\pi_{ik})$ or $\sigma_{ij}<\varepsilon\max_k(\sigma_{ik})$, which means  $j\not\in\text{supp}_\varepsilon(\sigma_i)$. Hence, $\text{supp}_\varepsilon(\sigma_i)$ contains only strategies that yield the highest expected payoff and
$\min_{k\in\text{supp}_\varepsilon(\sigma_i)}(\pi_{ik})=\max_k(\pi_{ik})$ for all $i\in N$, i.e. $\sigma$ is a root of $Y_\varepsilon(\sigma)$.

To prove (ii) we will show that the set of all $\boldsymbol{\varepsilon}$-perfect equilibria is identical to the set of roots of
\begin{displaymath}
  Y(\sigma)\,=\,\lim_{\varepsilon\uparrow1}Y_\varepsilon(\sigma)\,=\,\sum_{i\,\in\,N}\,\,\,\,\min_{k\,\in\,\text{argmax}(\sigma_{i})}\pi_{ik}(\sigma_{-i})-\max_{k}\,\,\pi_{ik}(\sigma_{-i})
\end{displaymath}
Since \mbox{$Y(\sigma)\leq0$} for all $\sigma\in\Sigma$, a root of $Y(\sigma)$ is a maximizer. Suppose $\sigma$ is a Nash equilibrium on the interior simplex defined by some $\boldsymbol{\varepsilon}$. For any $k\in\text{argmax}(\sigma_i)$ we must have $k\in\text{argmax}(\pi_{i})$. Hence, $\sigma$ is a root of $Y(\sigma)$. Conversely, suppose $\sigma$ is a root of $Y(\sigma)$ and assume, for $i\in N$, that $\sigma_{i1}\geq\sigma_{i1}\geq\cdots\geq\sigma_{iK_i}$ (without loss of generality as we can relabel strategies). Consider the interior simplex with vertices $\mu_{ik}=\sigma_i+(\sigma_{i1}-\sigma_{iK_i})(e_k-e_1)$ for $k=1,\ldots,K_i$ where $e_k$ is the $k$-th unit vector. If $\text{argmax}(\sigma_i)=\{1,\ldots,K_i\}$ then this simplex consists of a single profile, $\sigma=(1/K_i,\ldots,1/K_i)$, which is thus a Nash equilibrium on this simplex. If $\text{argmax}(\sigma_i)=\{1,\ldots,\hat{K}_i\}$ for some $\hat{K}_i<K_i$ then $\pi_{ij}(\sigma_{-i})=\max_k\pi_{ik}(\sigma_{-i})$ for $j=1,\ldots,\hat{K}_i$ since $\sigma$ is a root of $Y(\sigma)$. Since $\langle\mu_{i1}|\pi_i(\sigma_{-i})\rangle=\langle\mu_{ij}|\pi_i(\sigma_{-i})\rangle$ for $j\leq\hat{K}_i$ and $\langle\mu_{i1}|\pi_i(\sigma_{-i})\rangle>\langle\mu_{ij}|\pi_i(\sigma_{-i})\rangle$ for $j>\hat{K}_i$ it follows that $\sigma=\mu_{i1}$ is a Nash equilibrium on the interior simplex with vertices $\mu_{ik}$.

To prove (iii), recall from \cite{GoereeLouis2021} that the set of all regular QRE is contained in the union of all $\boldsymbol{\varepsilon}$-proper equilibria, which in turn is contained in the union of all $\varepsilon$-perfect equilibria. $\hspace*{5mm}\blacksquare$

\section{Additional Results}
\label{app:est}

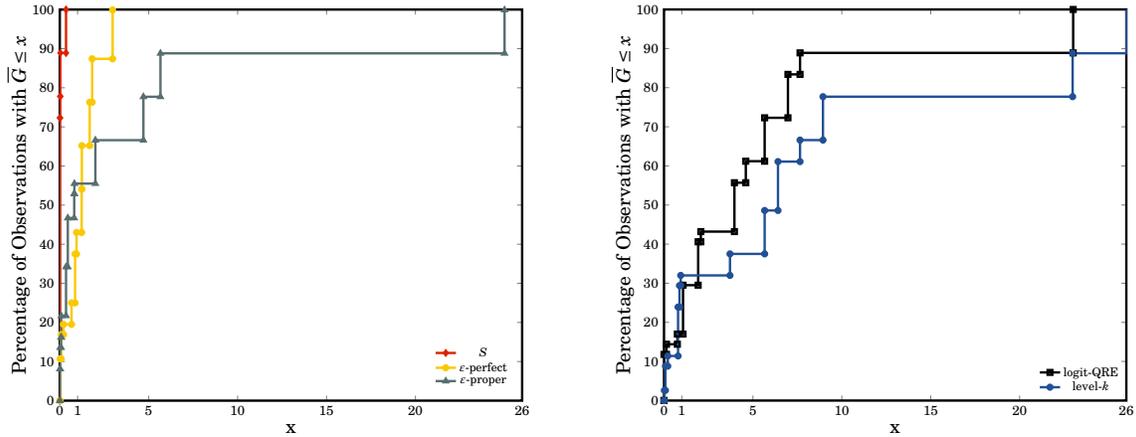
\begin{figure}[h]
	\begin{center}
		\begin{tikzpicture}[scale=0.45]
			\begin{axis}[
				width=\linewidth,
				line width=2,
				grid=major, 
				tick label style={font=\normalsize},
				legend style={nodes={scale=0.9, transform shape}},
				legend pos=south west,
				label style={font=\Large},
				grid style={white},
				xlabel={x},
				ylabel={Percentage of Observations with $\overline{G}\leq x$},
				legend style={at={(0.8,0.15)}, anchor=north west,  draw=none, fill=none},
				ytick={0,10,20,30,40,50,60,70,80,90,100},
				xtick={0,1,5,10,15,20,26},
				ymin=0, ymax=100,
				xmin=0, xmax=26,
				]
				
				\addplot[mondrianRed,mark=diamond] coordinates
				{				      (0,     0)
					(0,     0)
					(0,  72.3)
					(0.02,  72.3)
					(0.02,  77.8)				
					(0.03,  77.8)
					(0.03,  88.9)
					(0.35,  88.9)
					(0.35, 100.0) };
				\addlegendentry{$S$}
				\addplot[mondrianYellow,mark=*] coordinates
				{				
(      0,    0)
(      0.02,    0)
(0.02, 10.7)
( 0.08, 10.7)
(0.08, 16.9)
(0.21, 16.9)
(0.21, 19.5)
(0.66, 19.5)
(0.66, 25.0)
(0.86, 25.0)
(0.86, 37.5)
(0.94, 37.5)
(0.94, 43.0)
(1.23, 43.0)
(1.23, 54.1)
(1.24, 54.1)
(1.24, 65.2)
(1.68, 65.2)
(1.68, 76.3)
(1.82, 76.3)
(1.82, 87.4)
(2.97, 87.4)
(2.97, 99.9)
					};
				\addlegendentry{$\varepsilon$-perfect}
				
				\addplot[mondrianCyan,mark=triangle] coordinates
				{		
				       (0,    0)
				    (0.01,    0)
				    (0.01,  8.1)
				    (0.02,  8.1)
				    (0.02, 13.6)
				    (0.07, 13.6)
				    (0.07, 16.2)
				    (0.08, 16.2)
				    (0.08, 21.7)
				   (0.35, 21.7)
				   (0.35, 34.2)
				   (0.45, 34.2)
				   (0.45, 46.7)
				   (0.81, 46.7)
				   (0.81, 52.9)
				    (0.82, 52.9)
				    (0.82, 55.5)
				   (2, 55.5)
				   (2, 66.6)
				  (4.7, 66.6)
				  (4.7, 77.7)
				     (5.66, 77.7)
				     (5.66, 88.8)
				  (25.01, 88.8)
				  (25.01, 99.9)
				};
				\addlegendentry{$\varepsilon$-proper}
			\end{axis}
			\node at (24,5.37) {
				\begin{tikzpicture}[scale=0.45]
					\begin{axis}[
						width=\linewidth,
						line width=2,
						grid=major, 
						tick label style={font=\normalsize},
						legend style={nodes={scale=0.9, transform shape}},
						legend pos=south west,
						label style={font=\Large},
						grid style={white},
        				xlabel={x},
		          		ylabel={Percentage of Observations with $\overline{G}\leq x$},
						legend style={at={(0.8,0.1)}, anchor=north west,  draw=none, fill=none},
						ytick={0,10,20,30,40,50,60,70,80,90,100},
						xtick={0,1,5,10,15,20,26},
						ymin=0, ymax=100,
						xmin=0, xmax=26,
						]
						
						\addplot[black,mark=square] coordinates
						{
						      (0,    0)
						      (0,    0)
						      (0 , 11.8)
						   (0.14,  11.8)
						   (0.14,  14.4)
						   (0.75,  14.4)
						   (0.75,  17.0)
						   (1.07,  17.0)
						   (1.07,  29.5)
						  (1.92,  29.5)
						  (1.92,  40.6)
						  (2.07,  40.6)
						  (2.07,  43.2)
						 (3.96,  43.2)
						 (3.96,  55.7)
						 (4.6 , 55.7)
						 (4.6,  61.2)
						 (5.66,  61.2)
						 (5.66,  72.3)
						 (6.97,  72.3)
						 (6.97,  83.4)
						 (7.65,  83.4)
						 (7.65,  88.9)
						 (23.01,  88.9)
						 (23.01, 100.0)
						};
						\addlegendentry{logit-QRE}
						\addplot[mondrianBlue,mark=otimes] coordinates
						{			
						       (0,    0)
						     (0.02,    0)
						     (0.02,  2.6)
						    (0.08,  2.6)
						    (0.08,  8.8)
						    (0.21,  8.8)
						    (0.21, 11.4)
						     (0.79, 11.4)
						     (0.79, 23.9)
						   (0.87, 23.9)
						   (0.87, 29.4)
						   (0.94, 29.4)
						   (0.94, 32.0)
						   (3.71, 32.0)
						   (3.71, 37.5)
						  (5.66, 37.5)
						  (5.66, 48.6)
						  (6.41, 48.6)
						  (6.41, 61.1)
						  (7.65, 61.1)
						  (7.65, 66.6)
						  (8.94, 66.6)
						  (8.94, 77.7)
						 (22.98, 77.7)
						 (22.98, 88.8)
						 (26.01, 88.8)
						 (26.01, 100)
						};
						\addlegendentry{level-$k$}
					\end{axis}
			\end{tikzpicture}};
			
		\end{tikzpicture}
	\end{center}
	\vspace*{-6mm}
	\caption{The cumulative distributions $\text{Fit}_m(x)$ for the different models. To avoid an overly cluttered picture, the left panel shows $S$ equilibrium and the $\varepsilon$-perfect and $\varepsilon$-proper models, and the right panel shows logit-QRE and level-$k$.}\label{fitCumulative}
	\vspace*{0mm}
\end{figure}

Support for Result 2 is provided in Table \ref{tab:stats2}, which reports Kolmogorov--Smirnov $D$ statistics for the difference in the cumulative Fit distributions shown in Figure \ref{fitCumulative}. With a total of 2163 observations the reported $D$ statistics strongly support Result \ref{res:fit}.

\medskip

\begin{table}[h]
\begin{center}
\begin{tabular}{l|c|c|c|c|c|c|c}
\hline\hline
KS $D$-statistic & $S$ & $\varepsilon$-perfect & $\varepsilon$-proper & logit-QRE & level-$k$ \\ \hline
$S$  & & 0.80& 0.75& 0.86& 0.89\\
$\varepsilon$-perfect & 0.80&& 0.33& 0.58& 0.68\\
$\varepsilon$-proper  & 0.75& 0.33&& 0.39& 0.40\\
logit-QRE & 0.86& 0.58 &0.39 &&0.24\\
level-$k$ &0.89& 0.68& 0.40& 0.24& \\ \hline\hline
\end{tabular}
\end{center}
\vspace*{-3mm}
\caption{Kolmogorov--Smirnov $D$ statistics for the difference in cumulative Fit distributions that are shown in Figure \ref{fitCumulative}.}\label{tab:stats2}
\end{table}

\medskip
\medskip

\noindent Table \ref{tab:stats3} lists percentages that a model provides better out-of-sample fit than other models.  Based on 16,368 comparisons it provides strong support for Result \ref{res:OSFit}.

\medskip

\begin{table}[h]
\begin{center}
\begin{tabular}{l|c|c|c|c|c|c|c}
\hline\hline
\% better OSFit & $S$ & $\varepsilon$-perfect & $\varepsilon$-proper & logit-QRE & level-$k$ \\ \hline
$S$ && 99.7& 99.8& 99.9& 99.9 \\
$\varepsilon$-perfect & 0.3&  & 74.2& 97.1& 98.2 \\
$\varepsilon$-proper  & 0.2& 25.8& &  84.7&  88.5\\
logit-QRE & 0.1&  2.9& 15.3&   &  68.4\\
level-$k$ & 0.1&  1.8&  11.5&  31.6&    \\ \hline\hline
\end{tabular}
\end{center}
\vspace*{-4mm}
\caption{Percentage that the model listed in the first column provides better out-of-sample fit than the model listed in the first row (16,368 comparisons).}\label{tab:stats3}
\vspace*{-14mm}
\end{table}


\begin{figure}[p]
\hspace*{-3mm}
};

\end{tikzpicture}
\vspace*{-7mm}
\caption{The black curves show the equilibrium correspondences for the $\varepsilon$-perfect model (left), the $\varepsilon$-proper model (middle-left), logit-QRE (middle-right), and level-$k$ (right) in games $g_1$--$g_5$. The grey areas show predictions of more general versions of these models, i.e. all $\boldsymbol{\varepsilon}$-perfect equilibria (left), all $\boldsymbol{\varepsilon}$-proper equilibria (middle-left), all regular QRE (middle-right), and level-$k$ for arbitrary level distributions (right). The diamond(s) indicate(s) the average observed choice in a session.}\label{appCalcGraph1}
\vspace*{-5mm}

\end{figure}


\begin{figure}[p]
\hspace*{-3mm}
};

\end{tikzpicture}
\vspace*{-7mm}
\caption{The black curves show the equilibrium correspondences for the $\varepsilon$-perfect model (left), the $\varepsilon$-proper model (middle-left), logit-QRE (middle-right), and level-$k$ (right) in games $g_1$--$g_5$. The grey areas show predictions of more general versions of these models, i.e. all $\boldsymbol{\varepsilon}$-perfect equilibria (left), all $\boldsymbol{\varepsilon}$-proper equilibria (middle-left), all regular QRE (middle-right), and level-$k$ for arbitrary level distributions (right). The diamond(s) indicate(s) the average observed choice in a session.}\label{appCalcGraph2}
\vspace*{-5mm}

\end{figure}

\newpage



\section*{Supplementary material}

# C. Instructions

**WELCOME**

Thank you for participating in today's session.

We need everybody to stay with us.

The whole experiment will be conducted through your computer. Please turn off your phone and be away of distractions during the session.

---

**General instructions**

- During the experiment you can earn points. At the end of the experiment you will receive **1$ for every 150 points/tokens** that you earned.

- The amount earned may be different for each participant and depends on his/her decisions, the decisions of other participants and luck.

- In addition, you will receive 10 $ as a show-up fee.

---

**The games**

- The experiment consists of a series of different games. You will play each game for 15 periods.

- In each period you will be randomly matched with another participant.

---

**On your screen you will see a table like the one below:**

*(This is just an example. The numbers in the tables in the experiment will be different.)*

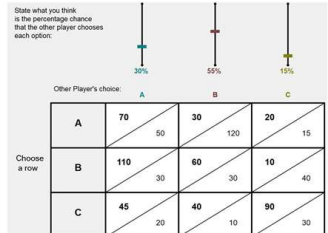

Each player simultaneously selects A, B or C by clicking the corresponding row. Each player wins the points indicated in the cell that corresponds to these choices. In the example above, if you choose C and the other player chooses B, you win 40 points and the other player wins 10 points.

---

**Belief task:**

In each period you are also asked to state your beliefs about the percentage chance of the other player A, B or C.

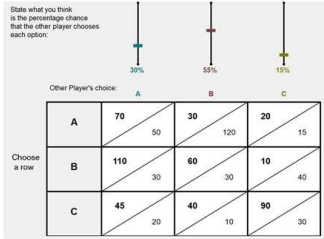

---

- On the screen, above the game you will see three bars like in the picture above. You can indicate what you believe to be the percentage chance for the other player to choose A, B and C by clicking and dragging the corresponding bar. The sum must always be equal to 100%.

- Notice that the other player, exactly like yourself, sees the table with the points he/she may win depending on both of your choices, as well as the points you may win.

- Each time, the computer chooses randomly one of the three bars and on that bar it chooses randomly again two points. If your stated belief in that bar is closer to the other player's actual choice than at least one of these two points, then you may win 60 points
- Examples:

- *To maximize your chance of winning these points you should state your true beliefs about the other player's percentage chance of making each choice.*

### Notice:

- You may change your choice as many times as you like (both for the game and the beliefs) When you are ready, press "Submit".

- For the first period of the game you have to wait 30'' before you are able to submit your choices. Use that time to familiarize yourself with the new game table.

- Please submit your choices within a reasonable time period. All participants will need to submit their choices before the experiment can move on to the next period. You will have a clock at the right to indicate you whether you are expending too much time.

### Notice:

- After submitting your choices, you will be informed about the other player's choice and your possible point earnings from the game. You will also be informed about the bar and points selected by the computer and whether you may earn points from the beliefs task.

- To determine your earnings in each period, the computer will randomly choose either the game or the belief task (both with an equal probability) and you will earn the points you were awarded in the corresponding task.

### Now

- We will start the experiment now. Please, if you have questions aks them.

- When you finish please wait for instructions.



\end{document}